\documentclass[11pt]{article}

\usepackage[final]{acl}

\usepackage{times}
\usepackage{latexsym}
\usepackage{amssymb}
\usepackage{amsmath}
\usepackage[T1]{fontenc}

\usepackage[utf8]{inputenc}

\usepackage{newunicodechar}
\newunicodechar{，}{,}

\usepackage{microtype}

\usepackage{inconsolata}

\usepackage{graphicx}
\usepackage{subcaption}

\usepackage{times}
\usepackage{latexsym}

\usepackage[T1]{fontenc}
\usepackage[utf8]{inputenc}

\usepackage{microtype}

\usepackage{inconsolata}

\usepackage{graphicx}
\usepackage{bibentry}

\usepackage{helvet}
\usepackage{courier}

\usepackage{natbib}
\usepackage{caption}
\usepackage{algorithm}
\usepackage{algorithmic}
\usepackage{color}
\usepackage{colortbl}

\usepackage{tabularx, array}
\newcolumntype{C}[1]{>{\centering\arraybackslash}p{#1}}
\usepackage{hyphenat}
\newcolumntype{Y}{>{\centering\arraybackslash}X}
\usepackage[most]{tcolorbox}
\tcbuselibrary{breakable}
\usepackage{fancyvrb}
\tcbuselibrary{breakable,skins}
\tcbuselibrary{listings, breakable}
\usepackage{enumitem}
\usepackage{fvextra}
\usepackage{booktabs}
\usepackage[table]{xcolor}
\usepackage{pifont}
\usepackage{multirow}
\usepackage{tikz}
\usetikzlibrary{shapes.geometric, arrows.meta, positioning, fit, backgrounds, calc, shadows}

\definecolor{genblue}{HTML}{E6F4FE}
\definecolor{evalorange}{HTML}{FFF3EB}
\definecolor{darkblue}{HTML}{005E99}
\definecolor{darkred}{HTML}{8c2824}
\definecolor{agentpurple}{HTML}{4E2399}

\DefineVerbatimEnvironment{Verbatim}{Verbatim}{breaklines=true, breakanywhere=true}
\definecolor{lightblue}{rgb}{0.9, 0.95, 1}

\definecolor{ysdarkpurple}{HTML}{4E2399}
\definecolor{ysshallowpurple}{HTML}{E6DBFF}
\definecolor{ysdarkred}{HTML}{8c2824}
\definecolor{ysshallowred}{HTML}{F8D7D7}
\definecolor{ysdarkblue}{HTML}{005E99}
\definecolor{ysshallowblue}{HTML}{CCEBFF}
\definecolor{ysdarkgrey}{HTML}{333333}
\definecolor{ysshallowgrey}{HTML}{E5E5E5}

\newtcbinputlisting{\promptbox}[2][]{
  enhanced,
  breakable,
  colback=ysshallowblue,
  colframe=ysdarkblue,
  fonttitle=\bfseries,
  title=#2,
  listing only,
  listing options={
    language=YAML,
    basicstyle=\ttfamily\small,
    breaklines=true,
    breakatwhitespace=true,
    postbreak=\mbox{\textcolor{red}{$\hookrightarrow$}\space},
    showstringspaces=false,
    upquote=true,
  },
  #1
}

\definecolor{ColorGrok}{HTML}{FFFDE7}
\definecolor{ColorPplx}{HTML}{EFFDFE}
\definecolor{ColorOpenAI}{HTML}{F2F2F2}
\definecolor{ColorGemini}{HTML}{E6F4FE}
\definecolor{ColorClaude}{HTML}{FFF3EB}
\definecolor{SectionHeaderColor}{HTML}{FFFFFF}
\colorlet{DarkerColorClaude}{ColorClaude!95!black}
\colorlet{DarkerColorPplx}{ColorPplx!95!black}
\colorlet{DarkerColorGemini}{ColorGemini!95!black}
\colorlet{DarkerColorOpenAI}{ColorOpenAI!95!black}
\colorlet{DarkerColorGrok}{ColorGrok!95!black}
\graphicspath{{figures/}}

\usepackage{newfloat}
\usepackage{listings}
\DeclareCaptionStyle{ruled}{labelfont=normalfont,labelsep=colon,strut=off}
\floatstyle{ruled}
\newfloat{listing}{tb}{lst}{}
\floatname{listing}{Listing}

\title{Diversity Collapse in Multi-Agent LLM Systems: Structural Coupling and
Collective Failure in Open-Ended Idea Generation}

\author{
  \begin{tabular}{cccc}
    Nuo Chen\textsuperscript{1} & Yicheng Tong\textsuperscript{1} & Yuzhe Yang\textsuperscript{2}  & Yufei He\textsuperscript{1} \\
    Xueyi Zhang\textsuperscript{2} & Qingyun Zou\textsuperscript{1} & Qian Wang\textsuperscript{1} & Bingsheng He\textsuperscript{1} \\[0.4em]
  \end{tabular}
     \\[0.2em] \textsuperscript{1}National University of Singapore
    \quad
    \textsuperscript{2}The Chinese University of Hong Kong, Shenzhen
}

\begin{document}

\maketitle

\begingroup
\renewcommand\thefootnote{}
\footnotetext{\hspace{\footnotesep} nuochen@comp.nus.edu.sg, dcsheb@nus.edu.sg}
\endgroup

\begin{abstract}

Multi-agent systems (MAS) are increasingly used for open-ended idea generation, driven by the expectation that collective interaction will broaden the exploration diversity. However, when and why such collaboration truly expands the solution space remains unclear. We present a systematic empirical study of diversity in MAS-based ideation across three bottom-up levels: model intelligence, agent cognition, and system dynamics. At the model level, we identify a compute efficiency paradox, where stronger, highly aligned models yield diminishing marginal diversity despite higher per-sample quality. At the cognition level, authority-driven dynamics suppress semantic diversity compared to junior-dominated groups. At the system level, group-size scaling yields diminishing returns and dense communication topologies accelerate premature convergence. We characterize these outcomes as \textit{collective failures} emerging from \textit{structural coupling}, a process where interaction inadvertently contracts agent exploration and triggers \textit{diversity collapse}. Our analysis shows that this collapse arises primarily from the interaction structure rather than inherent model insufficiency, highlighting the importance of preserving independence and disagreement when designing MAS for creative tasks. Our code is available at \url{https://github.com/Xtra-Computing/MAS_Diversity}.

\end{abstract}

\section{Introduction}

\begin{figure*}
    \centering
    \includegraphics[width=0.95\linewidth]{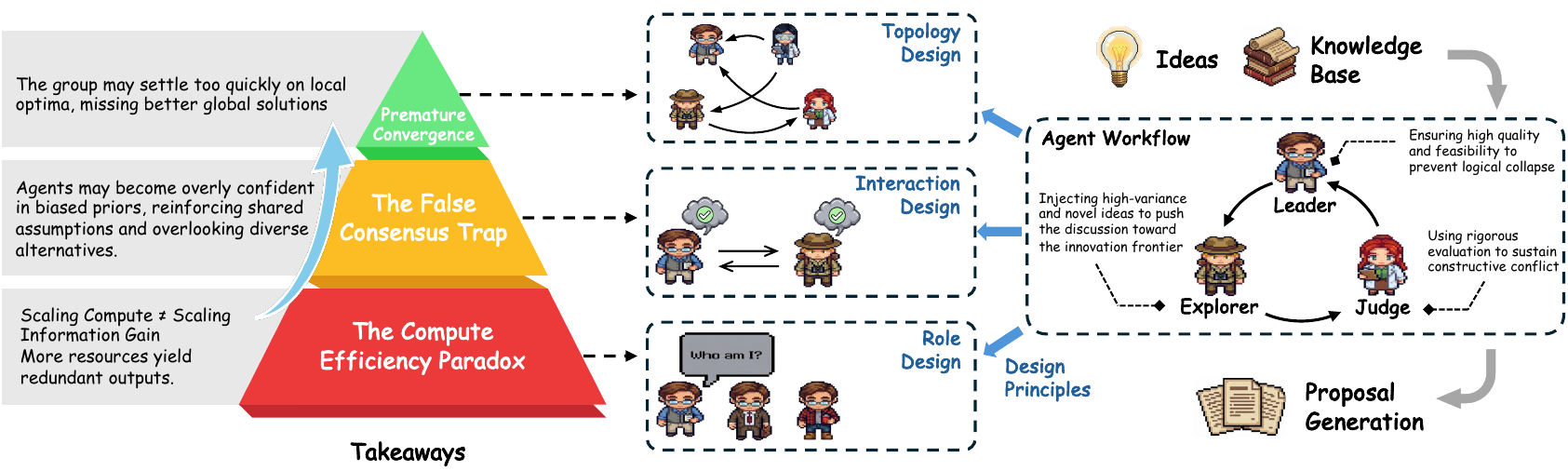}
    \vspace{-1mm}
    \caption{Design Principles and Workflow.}
    \label{fig:main}
    \vspace{-4mm}
\end{figure*}

Large language models (LLMs) have evolved from static text generators to dynamic engines for open-ended idea generation, supporting tasks ranging from scientific hypothesis formulation \cite{zhou-etal-2024-hypothesis,alkan2025surveyhypothesisgenerationscientific} to strategic planning \cite{cao2025largelanguagemodelsplanning} and creative design \cite{hong2024metagpt, gottweis2025aicoscientist}. In these exploratory domains, the utility of a system is not defined by its ability to converge on a single "ground truth," but rather by its capacity to explore a \textbf{diverse space of plausible ideas} that reflect alternative assumptions and solution paths \cite{boden2004creative, liang2024encouragingdivergentthinkinglarge, MOON2025100207}. Diversity, is not merely a qualitative preference; it is a functional requirement for effective decision-making. A lack of diversity risks trapping users in a narrow region of the solution space, inflating confidence in suboptimal solutions while suppressing unconventional but high-potential hypotheses \cite{wright2025epistemicdiversityknowledgecollapse}.

To transcend the limitations of single-model generation, recent research has increasingly pivoted toward Multi-Agent Systems (MAS) \cite{pmlr-v235-du24e, ye2025xmasbuildingmultiagentsystems}. The prevailing intuition is that, by enabling multiple agents to interact while adopting distinct roles or perspectives, MAS can achieve broader coverage of the idea space than a solitary model \cite{su2025headsbetteroneimproved}. However, this assumption remains largely unexamined. In practice, MAS frameworks are often built on homogeneous underlying models that share the same pre-training distributions and alignment objectives \cite{jiang2025artificial, wenger2025weredifferentweresame}. Consequently, multi-agent interaction can end up amplifying shared priors rather than introducing genuine variety, causing the system to repeatedly search the same narrow manifold at a higher computational cost \cite{wynn2025talkisntcheapunderstanding}. It remains unclear \textbf{when, why, and under what structural conditions} such collaboration actually expands the semantic solution space, rather than reaching a premature consensus that we characterize as \textbf{diversity collapse}.

To investigate this, we conduct a systematic empirical analysis evaluating over 10,000 research proposals spanning 20 topics. By using these proposals as a proxy for \textbf{diversity in MAS-based idea generation}, we dissect the \textbf{trade-offs} within mechanisms of collective interaction across three hierarchical levels:

First, at the level of \textbf{Model Intelligence} (Section \ref{sec:intelligence_analysis}), we identify the \textbf{Compute Efficiency Paradox}: as foundation models scale in capability, their outputs often become more fluent and score better on correctness-oriented metrics, yet converge toward increasingly similar semantic content.  From an information-theoretic perspective \cite{coveney2025wallconfrontinglargelanguage}, this points to a decoupling in which greater intelligence does not necessarily translate into a more informative expansion of the idea space, producing little marginal information gain.

Building on this foundation, we examine \textbf{Agent Cognition} (Section \ref{sec:cognition}), finding that interaction often triggers a \textbf{false consensus trap}. Although agents are prompted with distinct personas or roles to elicit diverse viewpoints,  they remain grounded in shared inductive biases. Our results reveal that authority-driven dynamics further suppress semantic diversity compared to junior-dominated horizontal groups. In these settings, interaction devolves into an Echo-Chamber Effect" \cite{wang-etal-2025-decoding,liu-etal-2024-evaluating-large,wynn2025talkisntcheapunderstanding}, where agents prioritize agreement over independent critique.

Further, we analyze \textbf{System Dynamics} (Section \ref{sec:dynamics}), where increased group size or dense communication topologies exacerbate \textbf{Premature Convergence}. If we view idea generation as search over a high-dimensional landscape, parallel interactions are expected to explore a broader region. However, by tracing evolutionary trajectories, protocols that implicitly reward fast agreement often push the group to collapse early onto local optima, much like the Ringelmann Effect \cite{ringelmann1913recherches} observed on humans. Under this condition, additional system complexity \cite{MOON2025100207, shen-etal-2025-understanding} tends to generate redundant trajectories rather than truly divergent exploration.

Finally, we synthesize these takeaways in our \textbf{Discussion} (Section \ref{sec:discussion}), showing that these outcomes represent \textbf{collective failures} emerging from \textbf{structural coupling}. Our analysis reveals that diversity collapse arises primarily from the \textit{interaction structure} (how agents are connected and how they influence one another), rather than any inherent model insufficiency. The more we force agents to coordinate, the more their individual trajectories become synchronized, effectively "locking" the group into a single path. Crucially, we show that this effect is most pronounced in complex tasks that demand both rigid logical rigor and open-ended imagination; in such cases, the pressure to be "correct" and "collaborative" inadvertently forces the system to prematurely abandon novel but unverified ideas.

In summary, achieving effective and diverse ideation in MAS requires more than simply assembling a larger or more connected group. The orchestration of interaction structures, carefully balancing collaboration with independence, is essential for unlocking the full creative potential of multi-agent systems in open-ended domains.

\vspace{-2mm}
\section{Methodology}
\label{sec:methodology}
\vspace{-2mm}
Unlike deep research and other goal-directed agentic tasks \cite{zhang2025deepresearchsurveyautonomous,zhang2025websearchagenticdeep}, which optimize planning, retrieval, and synthesis toward an evidence-grounded objective, ideation is inherently open-ended: it requires navigating a complex, high-dimensional search space to uncover distinct, plausible solutions \cite{boden2009conceptual,chen2026beyond,zhang2025noveltybenchevaluatinglanguagemodels}. In this section, we formalize the task into scientific proposal generation, discuss the pitfalls of agent collaboration, and introduce the means for quantifying diversity.

\vspace{-1mm}
\subsection{Task Formulation: Research Proposals as Units of Ideation}
\vspace{-1mm}
To rigorously evaluate diversity, we require a unit of analysis that is both structured and open-ended.
We adopt the generation of \textbf{scientific research proposals} as our unit of analysis. Unlike generic open-ended generation~\cite{jiang2025artificial}, a research proposal is a semi-structured artifact that demands both divergence and internal convergence.

Formally, given a research domain context $\mathcal{C}$, the system aims to generate a set of proposals $X = \{x_1, \dots, x_n\}$. Each proposal $x_i$ is not an independent sample, but the emergent outcome of a collaborative history $H$ among a group of agents. We detail the formal schema of valid proposals (e.g., Title, Hypothesis, Method) in Appendix~\ref{app:task_formulation}.
\vspace{-1mm}
\subsection{The Multi-Agent Ideation Pipeline}
\vspace{-1mm}
To systematically analyze diversity, we construct a generic multi-agent interaction framework consisting of three phases (illustrated in Figure~\ref{fig:main}).

\noindent \textbf{Role Instantiation.}
The system initializes a set of agents $\mathcal{A} = \{a_1, \dots, a_k\}$. To simulate diverse cognitive sources, agents are assigned distinct "personas" or expert roles (e.g., "The Skeptic," "The Interdisciplinarian") via system prompts. This heterogeneity is designed to mimic a scientific committee.

\noindent \textbf{Iterative Deliberation.}
Agents engage in a multi-turn dialogue governed by a specific topology (e.g., Round-robin Debate). In each turn $t$, an agent observes the context $\mathcal{C}$ and the discussion history $H_{t-1}$ to formulate a contribution. This phase allows for the collision of perspectives, critique of premises, and refinement of concepts.

\noindent \textbf{Proposal Synthesis.}
Upon reaching the interaction horizon $\mathcal{T}$, a designated "Editor" agent (or the collective group) synthesizes the discussion history into a finalized, structured research proposal $x_i$. This step forces the convergence of unstructured debate into a concrete scientific artifact. For each experimental setting, we conduct 50 independent discussion sessions per topic across the 20 topics listed in Table~\ref{tab:topics}, using temperature 0.7, giving 1{,}000 proposals per setting.

Specific experimental setups, including agent prompts and topologies, are detailed in Appendix~\ref{app:exp_setup}; the full prompt templates for every collaboration mode appear in Appendix~\ref{sec:allprompt}.
\vspace{-1mm}
\subsection{On the Evaluation of Diversity}
\label{sec:metrics}
\vspace{-2mm}
\begin{table}[h]
\centering
\small
\begin{tabular}{lc}
\toprule
Metric & Human Agreement (\%) \\
\midrule
Vendi Score & 87\% \\
$1-\phi$ & 82\% \\
PCD & 81\% \\
\bottomrule
\end{tabular}
\vspace{-1mm}
\caption{Agreement between human judgments and metric-induced ordering in pairwise diversity comparisons.}
\label{tab:human_agreement}
\vspace{-2mm}
\end{table}

Evaluating diversity in collaborative systems requires distinguishing between true conceptual variety and trivial surface-level variation. We apply metrics covering four complementary dimensions for the analysis. Mathematical definitions are provided in Appendix~\ref{app:metrics-details} and sensitivity analysis in Appendix~\ref{app:sensitivity}.

\noindent \textbf{Effective Diversity (Vendi Score~\cite{friedman2022vendi}):}
Measures the \textit{effective number} of unique semantic modes in the set $X$ based on the spectral entropy of the kernel matrix. Unlike simple counting, it is robust to cluster imbalances, indicating whether the system is exploring the semantic space efficiently.

\noindent \textbf{Structural Disorder:}
Adapted from the order parameter $\phi$~\cite{landau1937theory,Vicsek_1995} as the average cosine similarity between individual proposals and the group's mean embedding, this metric diagnoses the group's dynamic state. Low values of $1-\phi$ indicate collapse toward a single centroid (Echo Chamber~\cite{wang-etal-2025-decoding} state), while high values indicate that the system maintains pluralistic perspectives despite interaction.

\noindent \textbf{Semantic Dispersion (PCD):}
Computes the average pairwise cosine distance between proposals. While Vendi Score counts the \textit{modes}, Dispersion measures the \textit{magnitude} of the spread.

\noindent \textbf{Lexical Uniqueness:}
Utilizes IDF-weighted n-gram statistics to measure surface-level redundancy. This serves as a sanity check: high semantic diversity scores should not be driven merely by verbose rephrasing of identical ideas.

We validated these metrics via human evaluation (see Appendix~\ref{app:human_eval}) using pairwise comparisons by five expert annotators: the Vendi Score matched expert diversity judgments in 87\% of cases, with all three embedding-based metrics exceeding 80\% agreement (Table~\ref{tab:human_agreement}).

\vspace{-1mm}
\section{The Intelligence Landscape: Quality vs. Diversity}
\label{sec:intelligence_analysis}
\vspace{-1mm}

\begin{figure}[h]
    \centering
    \includegraphics[width=0.7\linewidth]{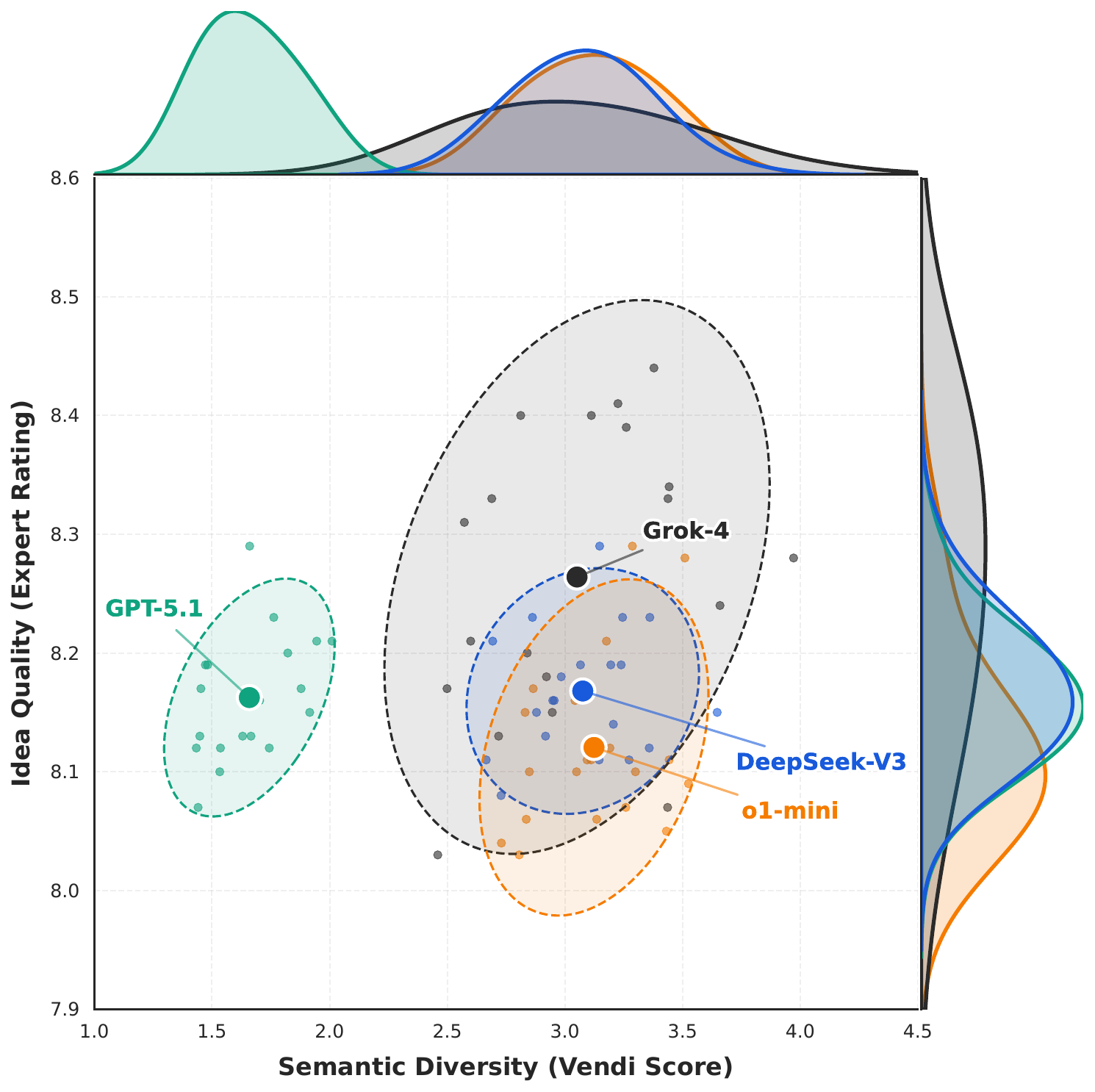}
    \vspace{-1mm}
    \caption{\textbf{Empirical Quality--Diversity Landscape of Single-Model Generation.}
Each point represents a generated research proposal under identical ideation settings. The X-axis shows topic-level Effective Diversity (Vendi Score), and the Y-axis shows aggregated Idea Quality. The landscape illustrates how semantic diversity varies independently of quality across models. Ellipses summarize empirical means and covariances and are used solely for geometric visualization.}
    \label{fig:model}
    \vspace{-6mm}
\end{figure}

Before studying multi-agent collaboration, we first analyze the quality--diversity landscape induced by single-model generation. Figure~\ref{fig:model} provides an empirical grounding: it visualizes the joint distribution of Idea Quality and Semantic Diversity obtained from contemporary LLMs under identical ideation settings. While individual models differ in alignment and architecture, our goal here is not model comparison, but to extract general constraints that govern diversity in downstream MAS.

The landscape reveals three generalizable observations that directly inform the design and limits of MAS-based ideation:

\begin{figure*}
    \centering
    \includegraphics[width=0.95\linewidth]{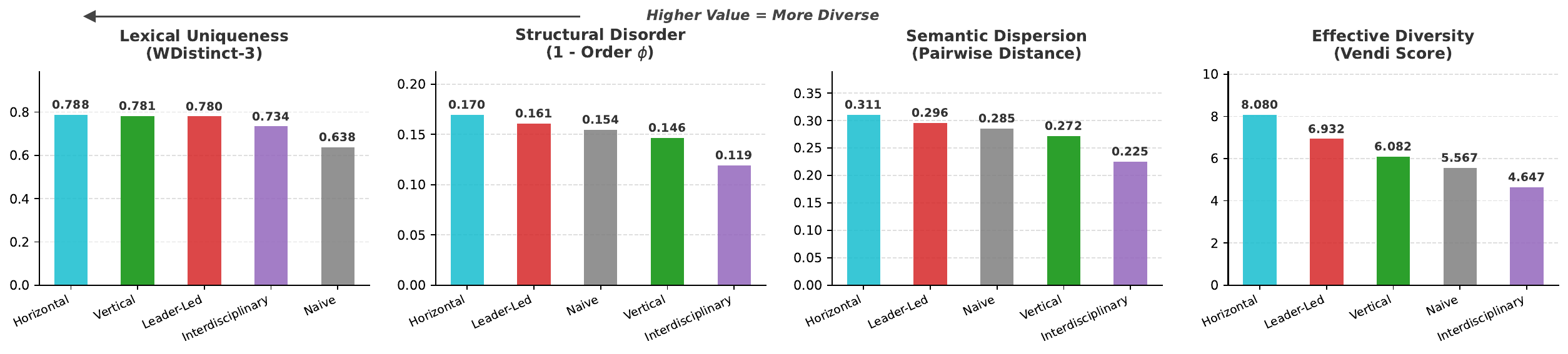}
    \vspace{-4mm}
    \caption{\textbf{Diversity Metrics across Cognitive Structures.} \textbf{Horizontal} collaboration (Junior-driven) consistently maximizes diversity (Vendi: 8.08), identifying the "Unbound Junior" effect. Surprisingly, \textbf{Interdisciplinary} collaboration exhibits the lowest diversity (Vendi: 4.65), suggesting that distinct expert roles induce a "Sycophancy Trap" where agents converge on safe, high-level generalities.}
    \label{fig:metrics}
    \vspace{-2mm}
\end{figure*}

\begin{figure*}
    \centering
    \includegraphics[width=0.95\linewidth]{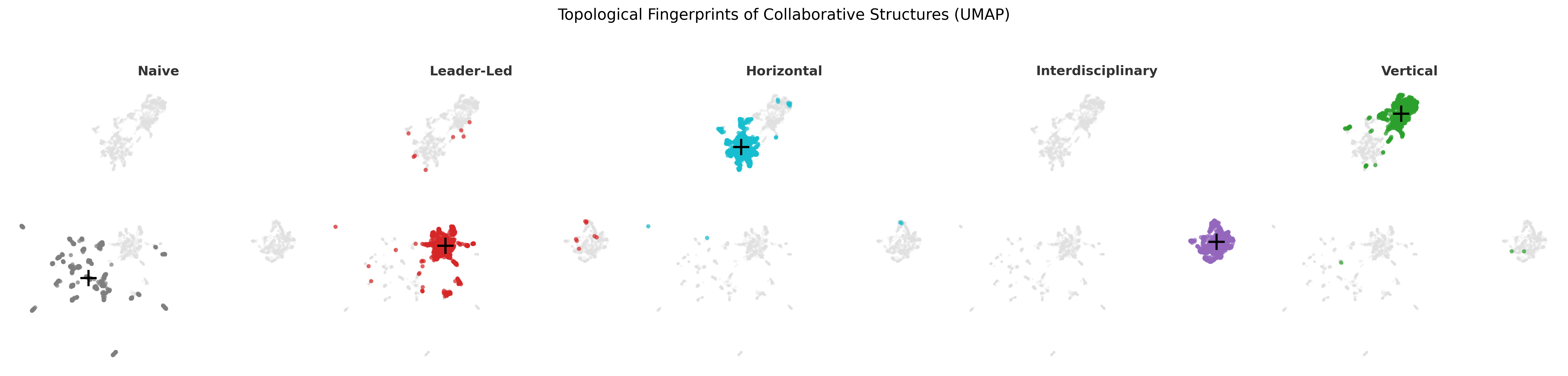}
    \vspace{-3.3mm}
    \caption{\textbf{Semantic Regimes of Cognitive Structures.} UMAP projection reveals a bifurcation. The \textbf{Conservative Cluster} (Bottom) is dominated by expert-driven structures (Leader-Led, Interdisciplinary), while the \textbf{Innovation Frontier} (Top) is populated by junior-driven structures (Horizontal, Vertical). This confirms that "Seniority" tends to constrain the semantic search space.}
    \label{fig:umap}
    \vspace{-3mm}
\end{figure*}

\noindent\textbf{Alignment systematically compresses semantic diversity without yielding commensurate quality gains.}
Across models, stronger alignment leads to a pronounced concentration along the diversity axis, while the marginal quality distribution remains largely stable. This suggests that alignment primarily functions as a global semantic regularizer, constraining exploration even when baseline generation quality is already high.

\noindent\textbf{Increasing intrinsic variance expands the accessible idea space but destabilizes quality trajectories.}
Models that span broader regions of the diversity axis demonstrate that high-entropy generation can substantially increase diversity; however, this expansion is accompanied by greater variance and unpredictability in output quality. Diversity driven solely by variance is therefore inherently noisy and unreliable for sustained ideation.

\noindent\textbf{Model-level quality is no longer the limiting factor for idea generation.}
Across the full diversity–quality frontier, including high-diversity regimes, models maintain consistently strong average quality, and qualitative inspection confirms semantic coherence. Collectively, these findings indicate that the core challenge for multi-agent systems is not generating diversity or trading it against quality, but preserving, structuring, and coordinating the latent diversity already present in single-model generation.

\vspace{-2mm}
\section{Cognition: Authority-Induced Collapse}
\label{sec:cognition}
\vspace{-1mm}
Following our analysis of model intelligence, we now investigate the \textit{agent cognition} layer, focusing on how the composition of agent personas，ranging from junior researchers to senior experts，shapes the semantic landscape of idea generation. We compare five cognitive structures (detailed in Appendix) designed to mimic real-world scientific collaboration:

    \noindent \textbf{Naive Collaboration:} Agents interact without defined roles or hierarchy.

    \noindent \textbf{Leader-Led Collaboration:} A designated senior expert guides discussion, with junior agents aligned to follow authoritative directives.

    \noindent \textbf{Horizontal Collaboration:} A group of early-career researchers collaborates flatly without senior oversight.

    \noindent \textbf{Interdisciplinary Collaboration:} Experts from distinct fields collaborate to synthesize cross-domain ideas.

    \noindent \textbf{Vertical Collaboration:} A hierarchical mix of senior experts, mid-career researchers, and early-career scholars.
\vspace{-1mm}
\subsection{Quantitative Analysis}

\begin{figure}[h]
    \centering
    \includegraphics[width=0.9\linewidth]{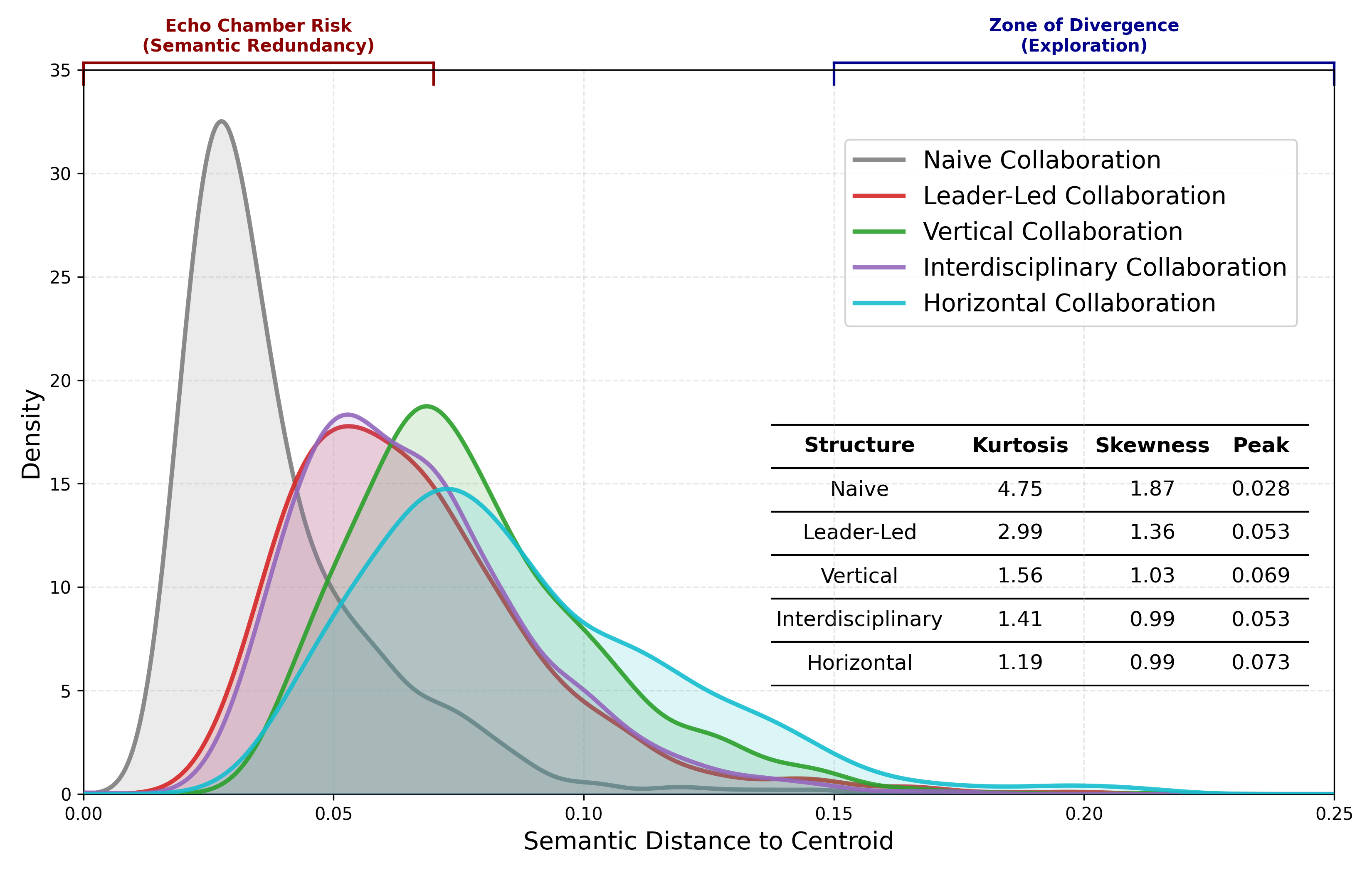}
    \caption{\textbf{Semantic Distance Density.} \textbf{Naive} (Grey) and \textbf{Leader-Led} (Red) distributions peak sharply near zero, indicating "Gravitational Collapse". In contrast, \textbf{Horizontal} (Cyan) and \textbf{Vertical} (Green) structures flatten the curve, shifting density into the "Zone of Divergence" (Distance > 0.10).}
    \label{fig:semantic}
    \vspace{-7mm}
\end{figure}

We evaluate aggregate diversity metrics across authority structures (Figure~\ref{fig:metrics}). Junior-dominated horizontal collaboration achieves the highest diversity, interdisciplinary expert teams the lowest, with only modest quality differences (Overall Quality 7.88--8.50). The ranking is robust across embedding backbones (Appendix~\ref{app:sensitivity}) and heterogeneous-model ensembles (Appendix~\ref{app:hetero_model}), with representative transcripts in Appendix~\ref{app:rebuttal}. Although ``Interdisciplinary'' might be thought to conflate expertise with implicit authority, the explicitly authority-weighted ``Leader-Led'' condition collapses nearly identically (Figure~\ref{fig:semantic}), and under a \emph{flat} peer-to-peer topology, Senior personas actually produce \emph{higher} diversity than Junior personas (Appendix~\ref{app:prompt_ablation}, which also rules out directive prompt tone)---indicating that the combination of expertise and hierarchy, not expertise alone, drives the collapse.

\vspace{-1mm}
\subsection{Distributional Dynamics}

To diagnose the mechanism underlying this collapse, Figure~\ref{fig:semantic} visualizes the density of semantic distances between individual proposals and their group centroid.
The density plot reveals a sharp cognitive dichotomy:

\noindent \textbf{Gravitational Collapse (Leader-Led/Naive):} The Leader-Led structure (Red) closely mirrors the Naive baseline (Grey), exhibiting high Kurtosis. This suggests that the presence of senior authority acts as a strong attractor. Junior agents likely succumb to sycophancy, aligning their vectors with the leader rather than offering orthogonal critiques.

\noindent \textbf{Sustained Divergence (Horizontal/Vertical):} The \textbf{Horizontal} (Cyan) and \textbf{Vertical} (Green) distributions significantly flatten the peak. The Vertical structure is particularly notable: by mixing senior guidance with junior exploration, it avoids the total collapse seen in Leader-Led setups, maintaining a "Goldilocks" zone of divergence.
\vspace{-3mm}
\subsection{Topological Segregation: Two Semantic Regimes}
\vspace{-1mm}
Finally, we employ UMAP to verify if these cognitive differences result in structurally distinct ideas (Figure \ref{fig:umap}).
The projection uncovers a striking segregation based on agent seniority:

    \noindent \textbf{The Conservative Cluster (Bottom Region):} Occupied largely by \textbf{Leader-Led} and \textbf{Interdisciplinary} groups. This confirms that expert personas under hierarchical or role-differentiated coordination tend to converge on ``conventional wisdom.'' Their proposals cluster tightly, likely reflecting established, safe research directions.

    \noindent \textbf{The Innovation Frontier (Top Region):} The \textbf{Horizontal} and \textbf{Vertical} groups migrate to a distinct upper manifold. Crucially, the \textbf{Vertical} structure bridges the gap. It anchors in the exploratory regime but maintains a denser core than the diffuse Horizontal cloud.

\vspace{-1mm}
\begin{figure}[h]
    \centering
    \includegraphics[width=0.8\linewidth]{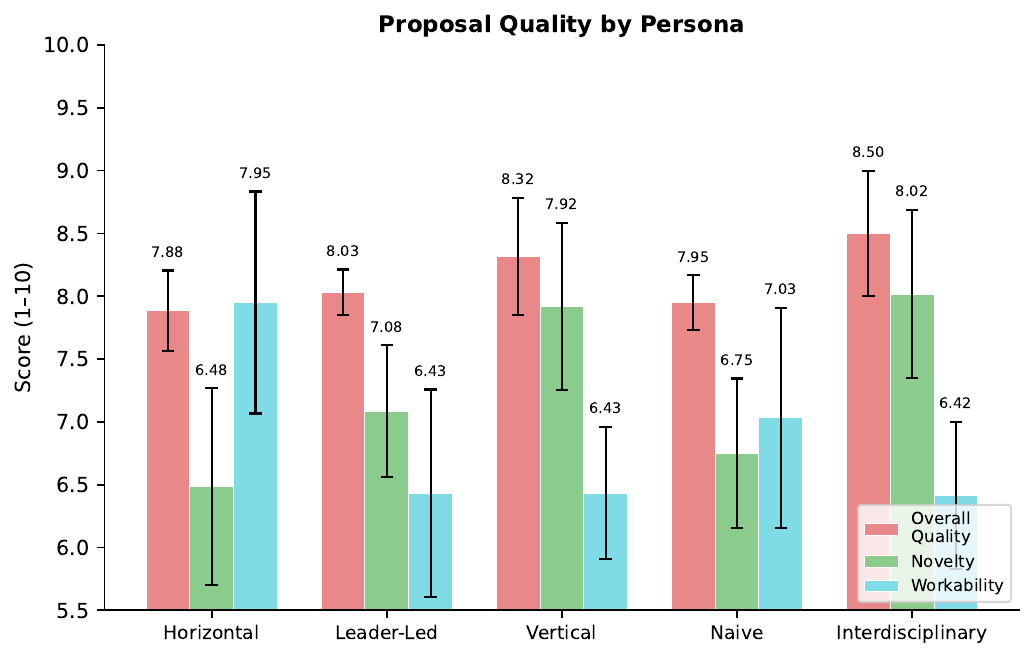}
    \vspace{-2mm}
    \caption{Proposal quality across three key dimensions. Full 9-dimension breakdown in Appendix~\ref{app:quality_comparison}.}
     \vspace{-3mm}
    \label{fig:quality_diversity}
\end{figure}
\textbf{Authority-Induced Collapse} is a form of directed coupling that accelerates convergence. The Vertical structure which mixes authority levels, offers a compromise, mitigating the chaos of juniors with the structure of seniors.
While Interdisciplinary attains the highest Overall Quality (8.50 vs.\ Horizontal's 7.88, a $+$0.6 gap on a 10-point scale, Figure~\ref{fig:quality_diversity}), this gain does not offset the much larger diversity drop between the same two conditions (Vendi 4.65 vs.\ 8.08, Figure~\ref{fig:metrics}). The Overall-Quality advantage also does not extend to Workability, where Horizontal proposals score highest, and Vertical (OQ 8.32) occupies an intermediate position that preserves distributional divergence (Figure~\ref{fig:semantic}), suggesting that rigid hierarchical authority often optimises for safe consensus at the expense of actionable exploration.

\section{Group Dynamics: Scaling, Evolution, and Topology}
\label{sec:dynamics}
This section explores MAS dynamics, specifically group size, temporal evolution, and communication topology, affect the diversity and quality of generated ideas.

\begin{figure}[h]
    \centering
    \includegraphics[width=0.95\linewidth]{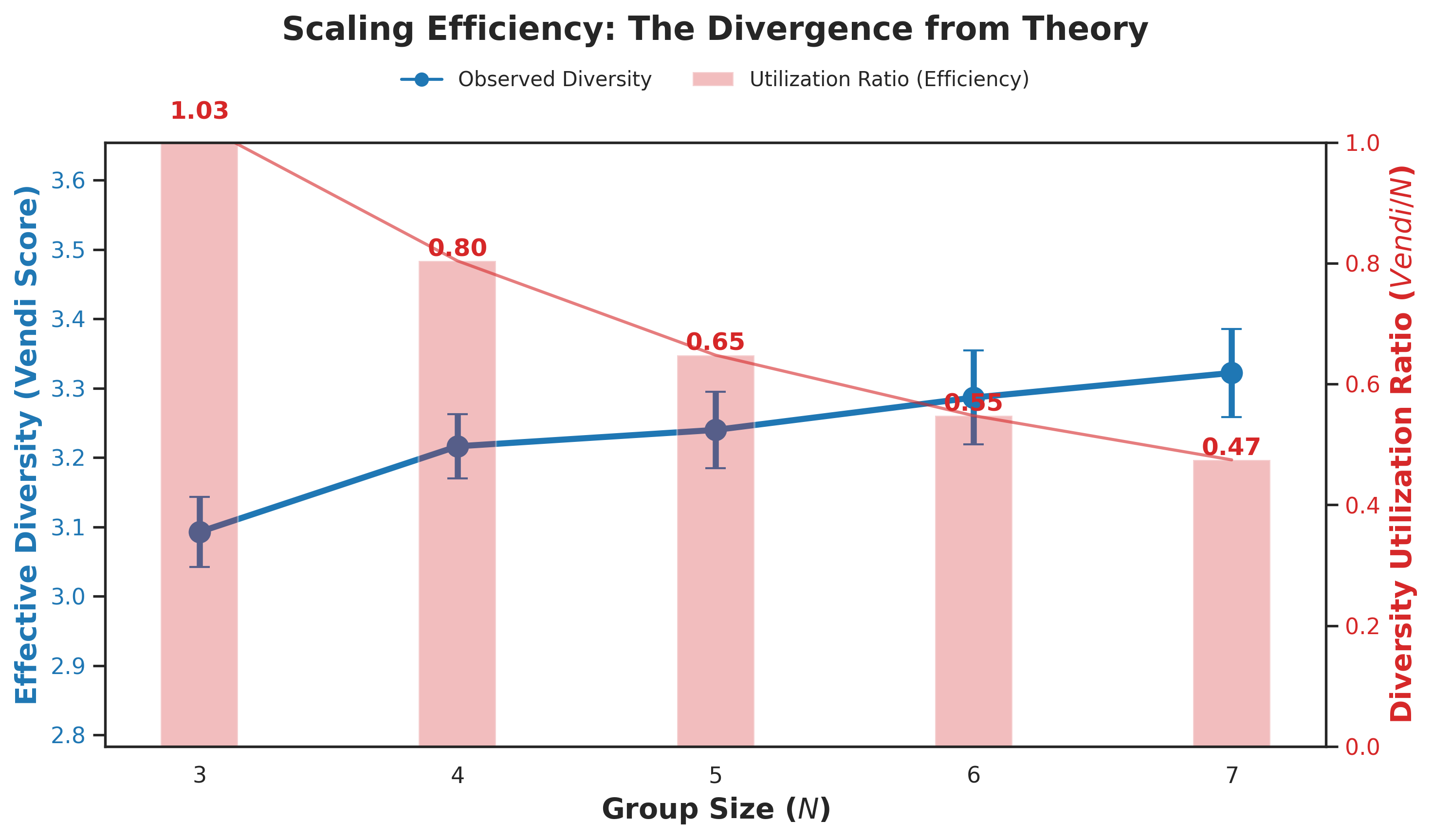}
    \vspace{-3mm}
    \caption{The Divergence from Theory in Scaling Efficiency. This plot compares the observed Effective Diversity (Vendi Score, blue line) against the theoretical Diversity Utilization Ratio (Vendi/$N$, red bars) as group size increases. While diversity grows, the efficiency per agent drops significantly.}
    \label{fig:size}
     \vspace{-4mm}
\end{figure}

We first investigate the impact of increasing the number of agents on the diversity of proposals. Figure \ref{fig:size} illustrates the relationship between group size ($N$) and Effective Diversity (Vendi Score).

\textbf{Increasing group size yields diminishing marginal returns in effective diversity, revealing a significant efficiency gap.}

While the absolute Vendi Score (blue line) increases monotonically from $N=3$ to $N=7$, the Diversity Utilization Ratio (red bars), defined as $\text{Vendi}/N$, plummets from 1.03 to 0.47. This indicates that adding agents does not linearly expand the semantic search space; rather, new agents increasingly overlap with existing ones. This phenomenon aligns with the "Compute Efficiency Paradox," suggesting that without structural intervention, simply scaling group size faces rapid saturation in information gain. A per-topic decomposition (Appendix~\ref{app:topic_complexity}) rules out topic-capacity exhaustion as the cause.

\subsection{Temporal Evolution: Rounds and Trajectories}
Next, we analyze how semantic diversity evolves over the course of the debate rounds. We employ both high-dimensional metrics and 2D trajectory visualizations to understand the nature of this evolution.

\begin{figure}[h]
    \centering
    \includegraphics[width=1\linewidth]{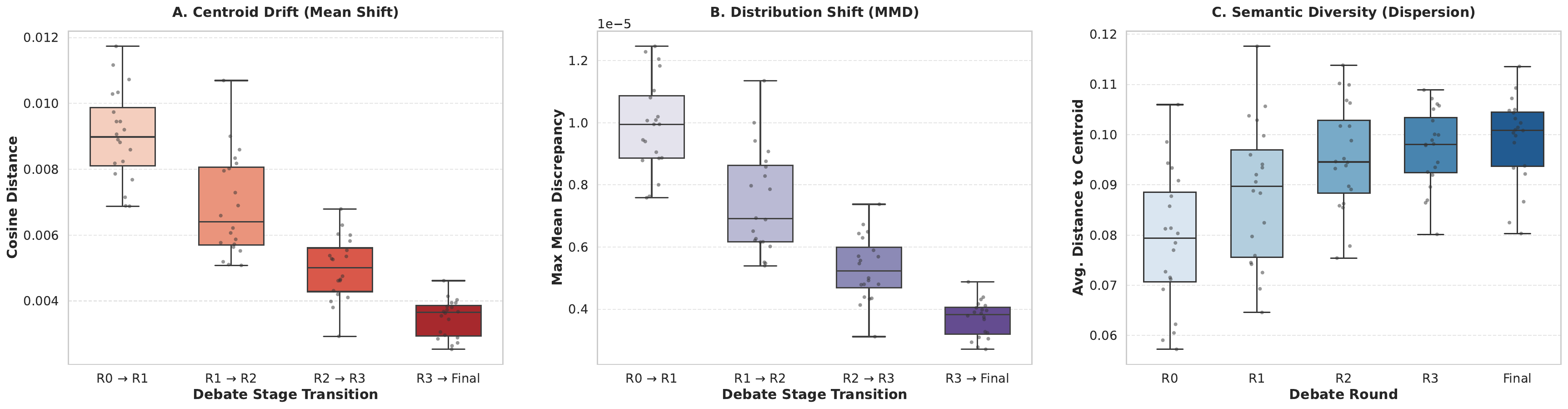}
    \caption{Quantitative Evolution of Semantic Dynamics. (A) Semantic Drift Velocity decreases, indicating stabilization of the consensus. (B) Distribution Shift (MMD) reduces, confirming structural convergence. (C) Semantic Diversity (Dispersion) increases, showing expansion within the consensus region.}
     \vspace{-4mm}
    \label{fig:semantic_dynamics}
\end{figure}

\begin{figure*}[h]
    \centering
    \includegraphics[width=1\linewidth]{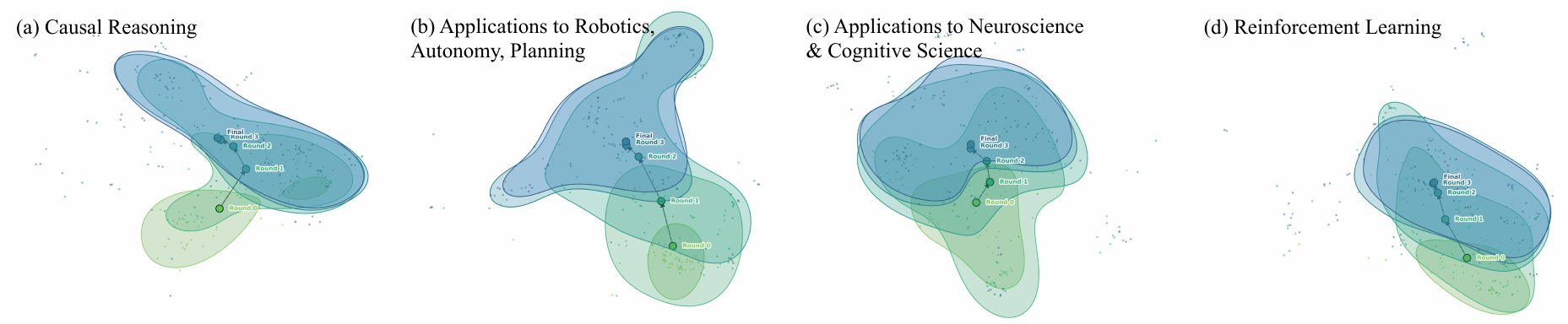}
    \vspace{-8mm}
    \caption{Evolutionary Semantic Trajectories. 2D projections of proposal embeddings across debate rounds for four representative topics. The trajectories show coherent drift (arrows) and expanding coverage (shaded regions), illustrating structured exploration rather than random movement.}
    \label{fig:trajectory}
    \vspace{-4mm}
\end{figure*}

The system exhibits a pattern of "Stable Expansion," where global consensus stabilizes while local exploration broadens.

As shown in Figure \ref{fig:semantic_dynamics}A and B, both Semantic Drift Velocity and Maximum Mean Discrepancy (MMD) show a consistent downward trend. This confirms that the group's "center of gravity" stabilizes over time, avoiding erratic jumps that would characterize hallucination. However, contrary to simple convergence, Figure C reveals an upward trend in Semantic Diversity (Dispersion). This "divergence within convergence" reflects \emph{within-session} refinement: agents, while agreeing on a general direction, continue to expand the radius around the stabilising centroid. This is distinct from the \emph{across-run} diversity collapse in Sections~\ref{sec:cognition} and~\ref{sec:dynamics}: a single session can expand locally while the broader pool still shows structural contraction.

Visual trajectories confirm that idea evolution follows a structured, coherent path rather than random semantic jumps.

Figure \ref{fig:trajectory} visualizes the evolutionary paths for four diverse topics. In all cases, we observe coherent trajectories (arrows) where the population centroid shifts progressively from the initial state (Round 0) to a final refined state. The expanding shaded regions (KDE) further illustrate how the system explores neighboring semantic territories. This structured movement stands in stark contrast to the unstructured jumps expected from hallucination, providing strong evidence that the observed diversity stems from genuine deliberation and refinement.

\subsection{Topology: The Impact of Communication Structure}
Finally, we examine how different communication topologies, Standard, Nominal Group Technique (NGT)~\cite{ngt}, and Subgroups (detailed in Appendix
\ref{app:randomized_subgroup_text}), influence the dynamics of diversity and conflict.

\begin{figure}[h]
    \centering
    \includegraphics[width=1\linewidth]{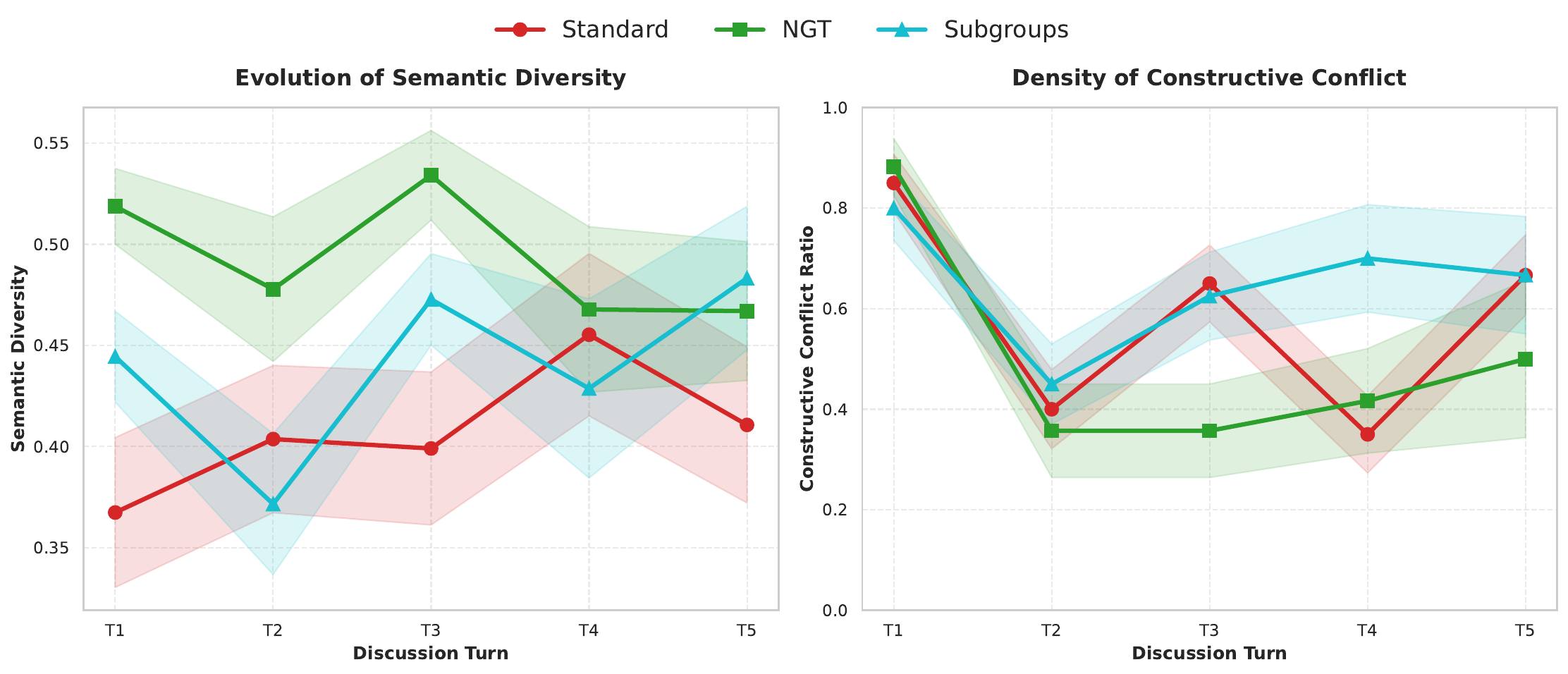}
\vspace{-5mm}
\caption{Mechanism of Process Intervention. (Left) Evolution of Semantic Diversity shows NGT's early advantage and Subgroups' late rebound. (Right) Density of Constructive Conflict highlights Subgroups' ability to sustain critical engagement. See Appendix \ref{app:stance_judge} for the detailed prompting strategy and scoring rubric.}
    \label{fig:topology}
    \vspace{-3mm}
\end{figure}

Process interventions effectively disrupt consensus collapse, with NGT maximizing initial diversity and Subgroups sustaining critical engagement.

Figure \ref{fig:topology} (Left) shows that NGT (green) initiates with the highest semantic diversity, significantly outperforming the Standard baseline (red). This confirms that the "blind-writing" phase of NGT effectively mitigates production blocking and anchoring effects. Meanwhile, the Subgroups topology (cyan) demonstrates a unique "resilience spike" in diversity midway through the discussion. Crucially, Figure \ref{fig:topology} (Right) reveals that Subgroups maintain the highest and most stable density of constructive conflict (interactions with critique score $\geq 7$; metric formalised in Appendix~\ref{app:conflict}) in the latter half of the debate. This suggests that partitioning the social graph creates "local pockets of divergence" that prevent the premature "rush to agreement" observed in the Standard mode. A 2$\times$2 Persona $\times$ Topology factorial (Appendix~\ref{app:w3_factorial}) and a cross-model replication on GPT-5.1 (Appendix~\ref{app:gpt51_cross_topology}) further show that this topology ranking is structural rather than model-specific.

\section{Discussion}
\label{sec:discussion}

\subsection{Synthesizing the Hierarchical Interplay}

\begin{figure}[h]
    \centering
    \includegraphics[width=0.9\linewidth]{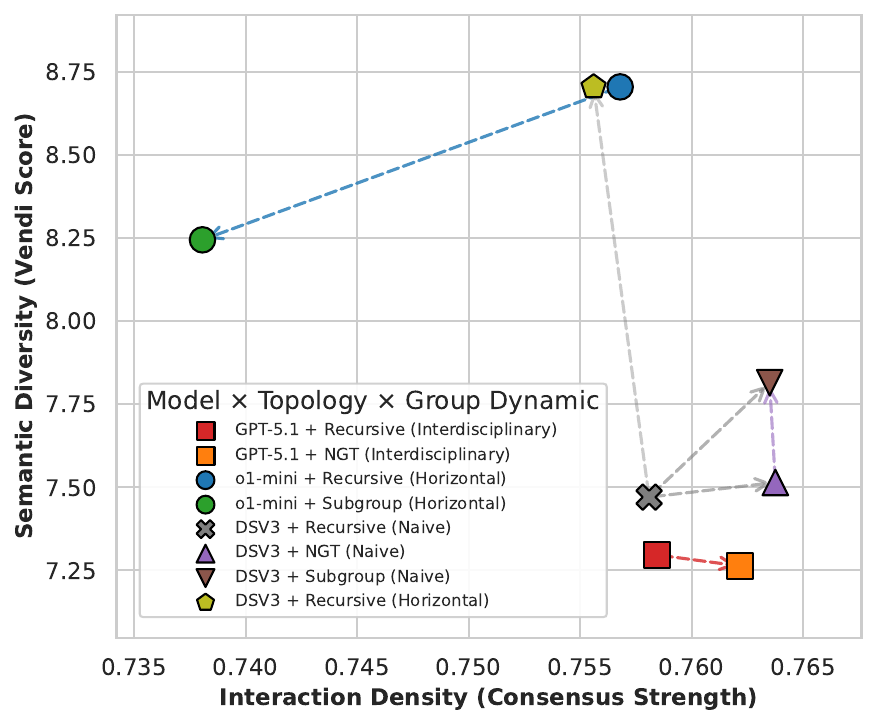}
    \vspace{-3mm}
    \caption{\textbf{The Interaction Landscape of Multi-Agent Ideation.} We map the trade-off between Interaction Density (Consensus Strength) and Semantic Diversity (Vendi Score) for distinct Model $\times$ Topology combinations. Arrows indicate the shift from a baseline (e.g., Recursive) to an intervention (e.g., NGT or Subgroup). The plot suggests that lower-capacity models (DeepSeek-V3) benefit from structural interventions, while reasoning-heavy models (o1-mini) resist them: enforcing subgroups paradoxically reduces diversity, suggesting an Alignment-Topology Mismatch. Because cells mix personas across models, this plot is illustrative, and a persona-controlled factorial appears in Appendix~\ref{app:w3_factorial}.}
    \label{fig:interaction_landscape}
    \vspace{-6mm}
\end{figure}

While previous sections analyzed group size, rounds, and topology in isolation, the efficacy of a multi-agent system relies on the complex interplay between these factors. Figure \ref{fig:interaction_landscape} visualizes this interaction landscape, mapping the relationship between Consensus Strength (Interaction Density) and Semantic Diversity (Vendi Score) across different Model $\times$ Topology configurations, determining whether the system succeeds or suffers from \textbf{diversity collapse}.

We argue that this collapse is a \textbf{collective failure} driven by \textbf{structural coupling}, a state where these three forces synchronize to contract the search space:

\textbf{How Intelligence and Topology Interact.}
Under the specific persona/topology pairings in Figure~\ref{fig:interaction_landscape}, the efficacy of a topology appears contingent on the model's intelligence, with the caveat that persona is not fully controlled across cells (Appendix~\ref{app:w3_factorial} provides a persona-controlled factorial on DeepSeek-V3). For standard models (e.g., DeepSeek-V3), structural interventions like NGT appear to provide a useful scaffold for organising and elevating their baseline ideas. For reasoning-heavy models (e.g., o1-mini), the same structural coupling may instead act as a hindrance, a pattern consistent with the reading that their high-level internal deliberation is fragile and that dense external coordination (the blue dashed arrow) produces a synchronization effect that reduces the unique perspectives each agent could have contributed.

\textbf{The Weight of Cognitive Alignment.}
The interplay is further constrained by alignment. In heavily aligned models like GPT-5.1, the model's prior appears dominant: even under the persona and topology variations we test (Appendices~\ref{app:gpt51_cross_topology} and~\ref{app:hetero_model}), these agents tend to concentrate in the same narrow consensus region. This is consistent with the finding that diversity collapse can be driven by alignment priors alone, producing a floor that the structural interventions do not fully breach.

\textbf{Collective Failure vs.\ Model Insufficiency.}
Crucially, these results indicate that the loss of diversity arises from the \emph{structure of the interplay} rather than any \emph{inherent model insufficiency}. It is the way we balance (or fail to balance) these three dimensions that triggers collapse: when the pressure for consensus (Dynamics) and the constraints of alignment (Cognition) overwhelm the model's creative capacity (Intelligence), the system effectively locks into a single, redundant path.

\subsection{Task Dynamics: Why Rigor Accelerates Collapse}
\label{sec:task_characterization}

\begin{figure}[h]
    \centering
    \includegraphics[width=0.95\linewidth]{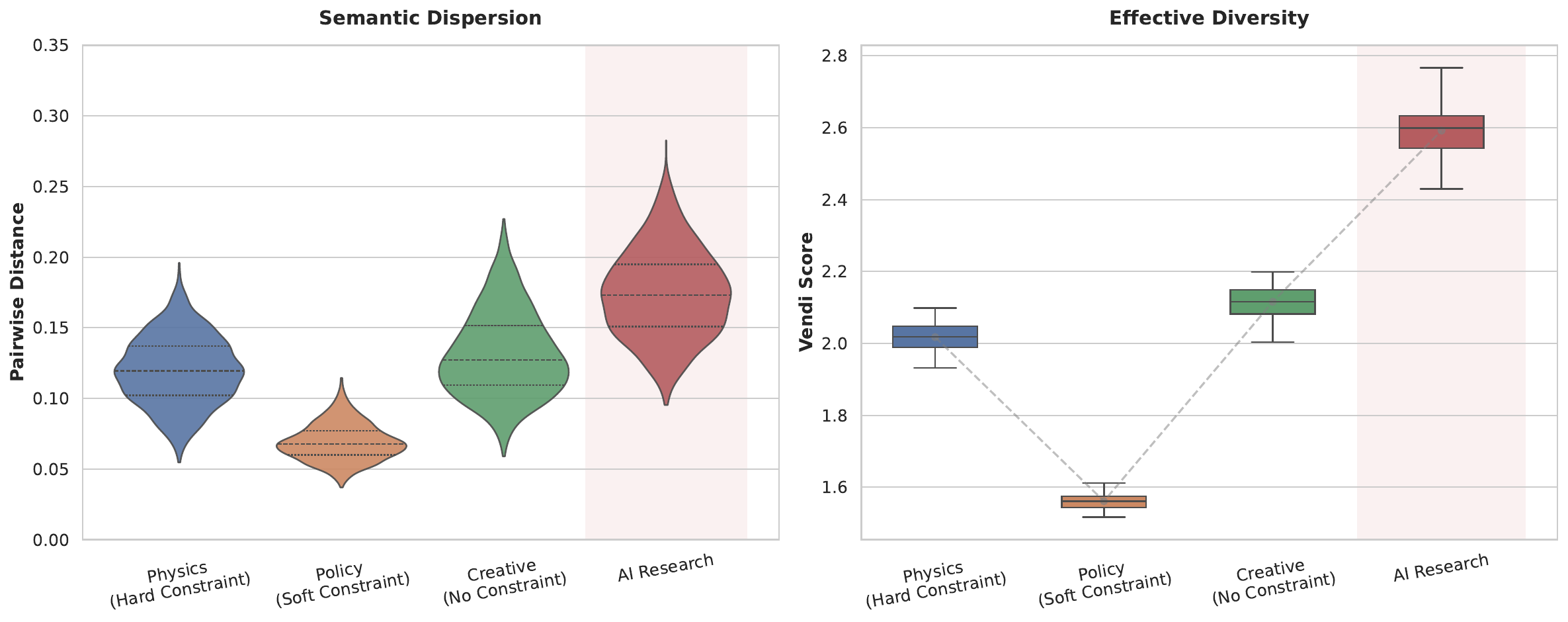}
    \caption{\textbf{The Intrinsic Entropy Spectrum across Cognitive Domains.} We benchmark baseline diversity (Inner-Topic Vendi Score, $N=50$) across four task types to validate domain representativeness. \textbf{(Left) Semantic Dispersion:} "Intellective" tasks like Physics and Policy exhibit tight distributions driven by ground-truth constraints. \textbf{(Right) Effective Diversity Capacity:} Bootstrapped Vendi Scores reveal that \textbf{AI Research} exhibits the highest intrinsic entropy ($>2.6$), distinct from both purely convergent tasks and unconstrained creative tasks. This characterizes AI Research as a "Hybrid Constraint" topology, making it a rigorous testbed for measuring structural efficacy.}
    \label{fig:task_spectrum}
    \vspace{-3mm}
\end{figure}

To understand the scope of these findings, we consider the specific requirements of the task. We contextualize our primary domain (AI Research) within the theoretical frameworks of the Task Circumplex \citep{mcgrath1984groups} and the Intellective-Judgmental Continuum \citep{laughlin1980social}, benchmarking the baseline behavior of LLM agents across four distinct task types (Physics, Policy, Creative Writing, and AI Research) to characterize their \textit{intrinsic entropy} (Figure~\ref{fig:task_spectrum}).
\vspace{-2mm}

The topological profiling yields three observations:

\textbf{Convergent Intellective tasks resist structural diversification.}
Domains like Physics and Policy, driven by ground truths or consensus, show low dispersion and diversity (Figure~\ref{fig:task_spectrum}, Left). For these, low diversity is appropriate, and forcing it may induce hallucination.

\textbf{AI Research as a stringent testbed at the Edge of Chaos.}
AI Research uniquely combines high entropy (Vendi Score $>2.6$; Figure~\ref{fig:task_spectrum}, Right) with strict logical rigor, requiring both broad exploration and logical soundness, unlike either unconstrained creative tasks or strictly convergent tasks \citep{chen2025position}. This Edge of Chaos position motivates AI Research as a stringent testbed for studying structural effects. Because Figure~\ref{fig:task_spectrum} compares only baseline intrinsic entropy across domains and does not re-run the topology manipulations on the other tasks, we do not claim that the structural findings automatically transfer to them.

\textbf{The Rush to Agreement.}
In this environment, agents face tension between exploring novel but unverified paths and converging on a superficially rigorous consensus. The density-collapse patterns in Figure~\ref{fig:semantic} and the transcripts in Appendix~\ref{app:rebuttal} suggest that under structural coupling, agents treat agreement as a proxy for correctness: the very mechanisms designed to ensure quality, namely collaboration and peer critique, can force the system to prematurely abandon unconventional ideas to satisfy collective pressure for consensus.

\noindent In summary, achieving effective and diverse ideation in MAS requires more than simply assembling a larger or more connected group. Diversity is a fragile property, easily sacrificed in the rush to agreement. Carefully orchestrating interaction structures to balance collaboration with independence is essential for unlocking the full creative potential of MAS in open-ended domains.

\vspace{-2mm}
\section{Related Work}
\vspace{-2mm}
\subsection{Social Psychology of Group Ideation}
\vspace{-0.5mm}
The study of group failures in ideation has a rich history in social psychology. The brainstorming hypothesis \citep{osborn1963applied}: groups generate more ideas than individuals, was famously refuted by subsequent research showing the opposite \citep{mullen1991productivity}. \citet{janis1972victims} introduced the concept of \textit{groupthink} to explain how cohesive groups suppress dissent. The Ringelmann effect \citep{ringelmann1913recherches}, later reinterpreted as \textit{social loafing} \citep{latane1979many}, demonstrated that per-capita contribution declines with group size. \citet{diehl1987productivity} identified \textit{production blocking}, the inability to generate ideas while listening to others, as a primary cause of brainstorming loss. Nominal Group Technique (NGT) \citep{ngt} was developed as a structural intervention to counter these failures by enforcing independent generation before group discussion. Status Characteristics Theory \citep{berger1977status} predicts that high-status individuals dominate group output regardless of actual competence. The same pattern appears in human opinion dynamics, where social influence undermines the ``wisdom of crowds'' \citep{surowiecki2005wisdom} by coupling independent judgments \citep{lorenz2011social}, and diversity in problem-solving groups outperforms individual ability only when independence is preserved \citep{hong2004groups}. Our work tests whether these phenomena survive in agents that lack explicit psychological substrate, and whether structural coupling alone is sufficient to reproduce them.
\vspace{-1mm}
\subsection{Multi-Agent Systems and Collective Intelligence}

Beyond simple model ensembles \citep{ye2025xmasbuildingmultiagentsystems}, recent frameworks leverage heterogeneity through social-attribute modulation \citep{ZHANG2026MAME} or diverse thinking prompts \citep{he2025unleashingdiversethinkingmodes}.
Multi-agent debate has been employed to enhance reasoning \citep{pmlr-v235-du24e}, and heterogeneous teaming to boost scientific ideation \citep{su2025headsbetteroneimproved,shi2025deep}.
However, interaction dynamics introduce structural vulnerabilities: \citet{wynn2025talkisntcheapunderstanding} identify that debate frequently suffers from sycophancy and ``disagreement collapse,'' and empirical studies reveal a homogenizing effect where AI collaboration reduces collective diversity \citep{MOON2025100207}. An ``Artificial Hivemind'' phenomenon has been documented where LLMs converge on identical semantic distributions regardless of prompting strategies \citep{jiang2025artificial, wenger2025weredifferentweresame}. Related efforts use LLM agents as proxies for human social behavior \citep{yang2025twinmarket, anthis2025llmsocialsimulationspromising, wang2025limits}; our study complements this line by providing a structural explanation for the observed collapse, showing that the effect is driven by interaction topology rather than persona fidelity.

\subsection{Communication Topologies}
\vspace{1mm}
Recent frameworks optimize interaction topologies for routing efficiency \citep{yue-etal-2025-masrouter, zheng2025efficiently,leong-etal-2025-amas} or use sparse connectivity to reduce overhead \citep{li-etal-2024-improving-multi}.
Dense interaction accelerates error propagation \citep{shen-etal-2025-understanding} and drives social polarization \citep{wang-etal-2025-decoding}.
Our work differs from this engineering-focused literature by framing topology effects as a manifestation of structural coupling, showing that the same principles that govern human group dynamics apply to agent communication graphs.

\section{Conclusion}
\vspace{1mm}
We systematically evaluated diversity in multi-agent systems for open-ended idea generation, using scientific proposal tasks as a testbed. Simply increasing agent count does not guarantee greater idea diversity. Rather, \textbf{diversity collapse} arises from \textbf{structural coupling} across three levels: alignment at the model level, hierarchical or role-differentiated coordination at the cognition level, and dense communication at the system level. These factors jointly promote premature consensus. Interaction designs that preserve independence, such as the blind-writing phase of NGT and subgroup isolation, consistently yield higher diversity with only modest differences in judged quality.

\clearpage
\section*{Limitations}
\label{sec:limitation}

This work focuses on evaluating diversity in multi-agent idea generation under a controlled experimental setting, and several limitations follow from this scope.

First, our analysis is centered on scientific proposal generation as a representative ideation task. While this domain offers a structured yet open-ended testbed with high intrinsic entropy, the observed dynamics may not directly transfer to tasks with stronger ground-truth constraints (e.g., mathematical problem solving) or to unconstrained creative writing. We view our setting as a stress test for diversity under hybrid constraints rather than a universal proxy for all generative tasks.

Second, our primary analyses use DeepSeek-V3 as the backbone to isolate the effects of interaction and structure. Cross-model replication on GPT-5.1 and o1-mini (Figure~\ref{fig:interaction_landscape}; Appendix~\ref{app:gpt51_cross_topology}) and genuinely heterogeneous-model ensembles that mix DeepSeek-V3, GPT-4o, and Claude-Sonnet-4 (Appendix~\ref{app:hetero_model}) confirm that the structural findings generalise across backbones, but a broader sweep over additional architectures and pretraining families remains future work.

Third, our diversity evaluation relies on embedding-based and lexical metrics, supplemented by human validation on a limited scale. Although agreement with expert judgments is high, no single metric can fully capture the nuanced notion of creativity or novelty in ideation. Quality scores are obtained via an LLM-as-Judge protocol (DeepSeek-V3, temperature~0) and therefore inherit the standard biases of automatic judges; our human evaluation (Appendix~\ref{app:human_eval}) validates the diversity axis only. Our metrics are intended to diagnose relative differences between collaboration modes rather than to provide absolute measures of creativity.

Finally, we analyze interaction protocols with a fixed number of rounds and a default sampling temperature. Although we sweep group size from $N=3$ to $N=7$ (Section~\ref{sec:dynamics}) and verify robustness to temperature across $T \in \{0.3, 0.7, 1.0\}$ (Appendix~\ref{app:temp_sensitivity}), adaptive or dynamically optimized interaction strategies may exhibit different behaviors that are not captured in this study.

\section*{Ethical Statement and Potential risks}

This paper studies the structural properties of multi-agent language model systems for idea generation, focusing on diversity rather than task correctness or decision-making authority. As such, the work does not introduce new model capabilities, training data, or deployment mechanisms.

A potential risk of multi-agent ideation systems is that increased fluency or consensus may create a false sense of confidence in generated ideas, particularly in high-stakes or expert domains. Our findings explicitly highlight this risk by identifying premature convergence and false consensus as failure modes, and thus aim to inform safer system design rather than to promote uncritical adoption.

All experiments are conducted on synthetic research topics and do not involve personal data, sensitive attributes, or human subjects. Human evaluation is performed by expert annotators solely to assess relative diversity under controlled conditions, without collecting identifiable information; idea quality scores are produced by an LLM-as-Judge (DeepSeek-V3, temperature~0) and reported alongside the human-validated diversity metrics.

Finally, while techniques for increasing diversity may be misused to generate misleading or speculative content, this risk is inherent to open-ended generation systems. We believe that understanding and diagnosing diversity collapse is a necessary step toward responsible deployment, as it enables system designers to better balance exploration, reliability, and oversight.

\section*{Acknowledgments}
This research is supported by the Ministry of Education AcRF Tier 1 grant (No. T1 251RES2315) in Singapore, Google South \& Southeast Asia Research Award 2025, and the National Research Foundation, Singapore and Infocomm Media Development Authority under its Trust Tech Funding Initiative. Any opinions, findings and conclusions or recommendations expressed in this material are those of the author(s) and do not reflect the views of National Research Foundation, Singapore and Infocomm Media Development Authority.

We also thank the AMD Heterogeneous Accelerated Compute Clusters (HACC) program for the generous hardware donation.

\bibliography{custom}

\appendix
\section{Task Formulation Details}

\label{app:task_formulation}

\subsection{Formal Definition of Multi-Agent Ideation}

We model the multi-agent idea generation process as a tuple $\langle \mathcal{A}, \mathcal{C}, \mathcal{P}, \mathcal{T} \rangle$, where:

\begin{itemize}

\item $\mathcal{A} = \{a_1, \dots, a_k\}$ represents the set of agents, where each agent is parameterized by an LLM (e.g., GPT-4o, Claude-3.5) and a specific role description or "persona."

\item $\mathcal{C}$ denotes the initial context or problem statement (e.g., "Propose a novel method to mitigate hallucinations in large language models").

\item $\mathcal{P}$ is the interaction protocol (e.g., Round-robin, Hierarchical, or Random) that dictates the sequence of message exchange among agents.

\item $\mathcal{T}$ represents the maximum number of interaction turns or rounds allowed before final proposal generation.

\end{itemize}

The generation process proceeds through a history of interactions $H_t$. At the final step $T$, the system aggregates the context and interaction history to produce the output set of proposals $X = \{x_1, \dots, x_n\}$. Unlike independent sampling where $P(X|\mathcal{C}) = \prod P(x_i|\mathcal{C})$, in a MAS setting, each proposal is conditioned on the collective history: $x_i \sim P(\cdot | \mathcal{C}, H_T)$, capturing the emergent effects of collaboration.

\subsection{Structure of a Scientific Proposal}

To ensure fair comparison and enable precise semantic analysis, all generated proposals are enforced to follow a strict schema. An unstructured idea is difficult to embed accurately; a structured proposal allows us to focus diversity metrics on the core innovation while minimizing noise from formatting.

Each valid proposal $x_i$ consists of the following four components:

\begin{enumerate}

\item \textbf{Title:} A concise descriptor of the idea.

\item \textbf{Background \& Motivation:} The specific gap in existing literature the proposal aims to address.

\item \textbf{Core Hypothesis:} The central scientific claim or mechanism proposed (e.g., "The use of contrasting agents reduces hallucination").

\item \textbf{Methodology Sketch:} A high-level description of the experimental design or algorithm.

\end{enumerate}

\subsection{Why this Structure Facilitates Diversity Analysis}

This semi-structured format serves two crucial purposes for our evaluation:

\begin{itemize}

\item \textbf{Separating Style from Substance:} By enforcing a standard format, we minimize the impact of stylistic variations (e.g., formatting differences, length) on the embedding space. This ensures that distance metrics (like Vendi Score and PCD) reflect true semantic differences in the \textit{Hypothesis} and \textit{Methodology} rather than structural noise.

\item \textbf{Filtering Triviality:} The requirement for a "Methodology Sketch" forces the model to ground abstract ideas into concrete execution plans. This allows us to distinguish between two proposals that sound similar in the abstract but differ significantly in execution, thereby providing a higher resolution for diversity measurement.

\end{itemize}

\subsection{Effective Diversity (Vendi Score)}

\textbf{Definition:} Effective Diversity is measured using the Vendi Score \citep{friedman2022vendi}. Given proposals $X=\{x_1,\dots,x_n\}$ and a similarity kernel $K$ constructed from cosine similarities between proposal embeddings (using OpenAI's \texttt{text-embedding-3-large}), the Vendi Score is defined as:
\begin{equation}
\operatorname{VS}(X) = \exp\left(-\sum_i \lambda_i \log \lambda_i\right)
\end{equation}

where $\{\lambda_i\}$ are the eigenvalues of the normalized kernel matrix $K/n$.

\subsection{Structural Disorder \texorpdfstring{($1-\phi$)}{(1-phi)}}
\textbf{Definition:} We define an order parameter $\phi$ as:

\begin{equation}
\phi = \frac{1}{n}\sum_{i=1}^{n} \cos(\vec v_i, \vec v_{\mathrm{avg}})
\end{equation}

where $\vec v_i$ denotes the embedding of proposal $x_i$ and $\vec v_{\mathrm{avg}}$ is the mean embedding across all proposals. Structural Disorder is measured as $1-\phi$. Values closer to 1 indicate a high degree of plurality, while values closer to 0 indicate convergence to a centroid.

\subsection{Semantic Dispersion (PCD)}

\textbf{Definition:} Semantic Dispersion is computed as the average pairwise cosine distance between proposal embeddings:

\begin{equation}
\mathrm{PCD}(X)=\mathbb{E}_{i<j}\left[1-\cos(\vec v_i,\vec v_j)\right]
\end{equation}

\subsection{Lexical Uniqueness (Content-only WDistinct-\texorpdfstring{$n$})}

\textbf{Definition:} Lexical Uniqueness is measured using an IDF-weighted Distinct-$n$ score computed on content tokens to filter out common stop words and generic scientific boilerplate:

\begin{equation}
\mathrm{WDistinct}\text{-}n(X)=
\frac{\sum_{g\in \mathcal{U}n(X)} \mathrm{IDF}(g)}
{\sum_{g\in \mathcal{A}_n(X)} \mathrm{IDF}(g)}
\end{equation}

where $\mathcal{A}_n(X)$ denotes all content-only $n$-grams in the proposals and $\mathcal{U}_n(X)$ denotes the corresponding set of unique $n$-grams. IDF weights are calculated based on a held-out corpus of scientific abstracts.

\section{Human Evaluation Details}

\label{app:human_eval}

To assess whether the automatic diversity metrics used in this work align with human judgments under our task setting, we recruited five AI PhD students with expertise in relevant research areas.

\subsection{Procedure}

For each topic, annotators were presented with 25 randomly sampled pairwise comparisons of proposal sets generated under different collaboration modes. In total, each annotator evaluated 100 pairwise comparisons. Blind to the system identities, they were asked a single question: \emph{"Which proposal set exhibits greater diversity of research ideas?"}

\subsection{Quality Control}

Annotators were also instructed to verify that all proposal sets met a basic bar of idea quality (coherent, on-topic, plausible). All evaluated sets satisfied this criterion. This confirms our assumption that diversity analysis is performed on a valid candidate set.

\subsection{Agreement Results}

We measured the agreement between human majority judgments and the ranking induced by automatic metrics. The Vendi Score achieved the highest alignment (87\%), followed by Structural Disorder ($1-\phi$) and Semantic Dispersion (PCD), validating our use of embedding-based metrics for this domain.

\section{Metric Design and Implementation Details}
\label{app:metrics-details}

This appendix provides detailed implementation choices and design rationales for all four metrics reported in the main text.

\subsection{Effective Diversity (Vendi Score)}
The Vendi Score measures diversity as the effective number of distinct samples, derived from the spectral entropy of a similarity kernel. Proposal embeddings are obtained using a fixed pretrained text embedding model. A cosine similarity kernel is constructed and normalized by the number of samples. The eigenvalue spectrum of this kernel reflects how variance is distributed across semantic directions.

This formulation is particularly suitable for open-ended proposal generation because it does not assume discrete clusters or require specifying a target number of modes. Instead, it naturally interpolates between fully collapsed generation (one dominant eigenvalue) and uniformly diverse generation (flat spectrum), providing a continuous measure of semantic capacity.

\subsection{Structural Disorder (\texorpdfstring{$1-\phi$})}
The order parameter $\phi$ measures the degree of alignment among proposals by computing the average cosine similarity between each proposal embedding and the mean embedding. Unlike pairwise metrics, $\phi$ captures a global property of the system: whether collaboration induces convergence toward a shared semantic direction.

We report Structural Disorder as $1-\phi$ so that higher values consistently correspond to greater diversity. This metric is sensitive to collaboration-induced consensus even when pairwise distances remain moderate, allowing us to distinguish systems that appear diverse locally but are globally aligned around a single dominant perspective.

\subsection{Semantic Dispersion (PCD)}
Semantic Dispersion is computed as the mean pairwise cosine distance between proposal embeddings. This metric directly measures the geometric spread of proposals in representation space.

While Effective Diversity captures how many semantic modes are present, Semantic Dispersion captures how far apart those modes are. Including both prevents misinterpretation of diversity arising from either tightly packed clusters or uniformly dispersed noise.

\subsection{Lexical Uniqueness (Content-only WDistinct-\texorpdfstring{$n$})}
Lexical Uniqueness is designed to measure surface-level redundancy while minimizing sensitivity to shared academic templates and formatting artifacts.

\paragraph{Content-only preprocessing.}
All proposals are lowercased and tokenized using a simple alphabetic tokenizer. Stopwords are removed using a fixed list of high-frequency functional words (e.g., articles, prepositions, auxiliaries). In addition, common academic boilerplate terms (e.g., \textit{paper}, \textit{method}, \textit{results}) are filtered to reduce the influence of structural conventions shared across proposals.

\paragraph{$n$-gram construction.}
After preprocessing, the remaining content tokens are treated as a sequence, and contiguous $n$-grams are extracted. This preserves local semantic structure while avoiding reliance on extracted keyphrases or sentence boundaries.

\paragraph{IDF weighting and global normalization.}
To downweight ubiquitous expressions and emphasize content-specific phrasing, each $n$-gram is weighted by inverse document frequency (IDF), computed over the union of proposals from all collaboration settings. This global normalization ensures that lexical scores are comparable across different experimental conditions.

\paragraph{Choice of $n$.}
We use $n=3$ by default. Trigrams provide a stable granularity that captures method- and concept-level expressions, while larger $n$-grams tend to become nearly unique in open-ended generation and are dominated by surface-level phrasing rather than substantive content.

\paragraph{Interpretation.}
Lexical Uniqueness reflects whether agents avoid repeating the same formulations and boilerplate patterns. It is not intended as a proxy for semantic diversity, but as a complementary signal that detects lexical echoing that may persist even when semantic metrics suggest diversity.

\section{Constructive Conflict Metric.}
\label{app:conflict}
We combine semantic embeddings and large language model (LLM) judgment to construct a metric for \emph{constructive conflict} in multi-speaker discussions. For each utterance, we obtain a sentence embedding using the \texttt{text-embedding-3-large} model. To avoid truncating long texts, we adopt a chunk-and-average strategy: the utterance is split into contiguous character chunks $\{c_k\}_{k=1}^K$ (each up to $\sim 12000$ characters), each chunk is embedded as $\mathbf{e}(c_k)$, and we compute the mean-pooled, $\ell_2$-normalized embedding
\[
\tilde{\mathbf{e}} = \frac{1}{K}\sum_{k=1}^{K} \mathbf{e}(c_k), \quad
\mathbf{e} = \frac{\tilde{\mathbf{e}}}{\|\tilde{\mathbf{e}}\|_2}.
\]
Within each discussion, we treat the first utterance as an anchor with embedding $\mathbf{e}_1$. For utterance $t$ ($t \ge 2$) with embedding $\mathbf{e}_t$, we measure its semantic deviation from the anchor via cosine similarity
\[
\operatorname{sim}_t = \frac{\mathbf{e}_t^\top \mathbf{e}_1}{\|\mathbf{e}_t\|_2 \,\|\mathbf{e}_1\|_2},
\]
and define its semantic divergence as
\[
\operatorname{Divergence}_t = 1 - \operatorname{sim}_t.
\]

To distinguish mere novelty from \emph{constructive} disagreement, we further use a chat-based LLM to rate the degree of disagreement/novelty between consecutive utterances. Let $x_{t-1}$ denote the previous utterance and $x_t$ the current utterance. We prompt the LLM with the following instruction:

\begin{verbatim}
Compare Speaker B's statement
to Speaker A's context.
Speaker A: "\textless previous
context truncated to last 400 
characters\textgreater..."
Speaker B: "\textless current 
statement\textgreater"
Task: Rate level of 
DISAGREEMENT/NOVELTY (1-10).

Strict Scoring:
- 1-4: Echo/Additive (Safe)
- 5-6: Minor Detail
- 7-8: Soft Critique/Refinement
- 9-10: Major Disruption
Output integer only.
\end{verbatim}

The model outputs a single integer score $s_t \in \{1,\dots,10\}$. We interpret scores $s_t \ge 7$ as indicating the presence of clear critique or constructive conflict, and define a binary indicator
\[
C_t = \mathbb{I}[s_t \ge 7].
\]
For a given experimental condition (e.g., \textit{Standard}, \textit{NGT}, or \textit{Subgroups}), we aggregate across all discussions at the same turn index $t$ and compute the \emph{Constructive Conflict Ratio}
\[
\operatorname{CCR}_t = \mathbb{E}[C_t],
\]
along with its standard error of the mean (SEM) for visualization. In our figures, the right-hand panel plots $\operatorname{CCR}_t$ over the first few turns (here, $t \le 5$) for each condition, capturing how the density of constructive conflict evolves as the discussion progresses under different institutional designs.

\section{Randomized Subgroup Text Collaboration}
\label{app:randomized_subgroup_text}

This section describes the randomized subgroup collaboration procedure in its purely textual variant. In this setting, each agent produces visible natural language utterances, without any latent state being passed between calls. A designated leader agent subsequently reads a subset of the discussion and synthesizes a final answer.

\paragraph{High-level overview}
Given a question or topic, a fixed set of agents participate in a multi-round discussion. In each round, the full set of agents is randomly partitioned into disjoint subgroups of a specified size. Within every subgroup, agents speak in sequence, with each utterance visible only to members of the same subgroup. After a pre-defined number of rounds, a leader agent reads a transcript of the most recent subgroup discussions together with a short summary of the corresponding round structure, and produces the final response.

\paragraph{Agent-side text generation}
All agents, including the leader, are implemented by the same underlying language model with shared decoding hyperparameters (e.g., sampling temperature, nucleus sampling threshold, and maximum number of generated tokens per turn). For each non-leader agent, the model is queried with a prompt that includes:
\begin{itemize}
    \item a natural language description of the overall task or topic;
    \item a description of the current discussion phase (e.g., brainstorming, critique, synthesis), indexed by round;
    \item a short description of the agent's role (e.g., ``optimistic critic'', ``domain expert'');
    \item a personalized memory consisting of all previous utterances that this agent is allowed to see (defined below);
    \item the sequence of speakers that have already contributed in the current subgroup and the round-specific instructions for how to respond to them.
\end{itemize}
The language model then generates a single textual utterance for that agent, up to a preset maximum number of tokens. No latent representations or cached internal states are shared across calls: each utterance is produced from scratch, conditioned only on the textual prompt.

\paragraph{Data structures and visibility}
Conceptually, the procedure maintains:
\begin{itemize}
    \item a global list of utterance records, where each record stores the agent identity, the round index or name, and the generated text;
    \item for each agent, an ordered list of all utterances that are visible to that agent, forming its personalized discussion memory;
    \item a log of the subgroup assignments in each round, specifying which agents were grouped together.
\end{itemize}
Whenever an agent in a subgroup produces an utterance, a corresponding record is appended to the global list. The same record is then appended to the personalized memory of every member of that subgroup. As a result, all members of a subgroup share the same local view of the subgroup-level discussion, but agents in different subgroups do not see each other's utterances from that round.

\paragraph{Per-round randomized subgroup discussion}
The multi-agent interaction unfolds over a fixed sequence of discussion rounds. For each round:
\begin{enumerate}
    \item A human-specified description of the phase is defined (for example, ``Round 1: generate diverse high-level ideas'' or ``Round 2: identify potential weaknesses'').
    \item The set of participating agents is randomly partitioned into disjoint subgroups of a pre-specified size. This random grouping is repeated independently in each round, so that agents are likely to interact with different partners across rounds.
    \item For each subgroup, an internal speaker order is defined (e.g., a fixed or randomly chosen permutation of the subgroup members). The subgroup then proceeds in that order:
    \begin{enumerate}
        \item When it is an agent's turn to speak, the system constructs a prompt using the elements listed above: task description, current phase description, that agent's role, the agent's personalized memory (all utterances that this agent has seen in all previous rounds), and the list of speakers who have already spoken in the current subgroup and round.
        \item The language model is called once to generate the agent's next utterance, subject to the maximum token budget.
        \item The resulting text is stored as a new utterance record (agent identity, round label, text content) and added to the personalized memory of all agents in the current subgroup. Thus, within a round, only subgroup members see each other's contributions.
        \item A detailed trace entry is logged, capturing the agent role, the full prompt, and the generated output, to enable post-hoc analysis of the collaborative process.
    \end{enumerate}
    \item After all subgroups have completed their turn for this round, a brief human-readable log entry is created summarizing the round, including which agents were grouped together in each subgroup.
\end{enumerate}

\paragraph{Selection of recent discussion for the leader}
After the last round of subgroup interaction, the system prepares input for the leader agent. To control context length while preserving the most relevant content, the leader does not read the full discussion history. Instead, only the most recent few rounds (e.g., the last two rounds) are considered:
\begin{enumerate}
    \item All utterance records are first grouped by their round labels. If round labels contain indices (for example, ``Round 1'', ``Round 2'', etc.), these indices are used to sort the rounds chronologically; otherwise, a default ordering is used.
    \item The last few rounds according to this ordering are selected as the ``recent'' rounds.
    \item All utterance records belonging to these recent rounds are concatenated into a textual transcript for the leader. Each entry in the transcript includes the round label, the agent identity, and the corresponding text, with simple formatting (such as headers and blank lines) to maintain readability.
    \item In parallel, the round-level logs created during the discussion are filtered so that only logs from the selected recent rounds are retained. This yields a concise summary of which agents interacted in which subgroups in the recent part of the discussion.
\end{enumerate}

\paragraph{Leader prompting and synthesis}
The leader agent is prompted once at the end of the process. Its input prompt contains:
\begin{itemize}
    \item the original question or topic;
    \item a short natural language summary of the recent rounds and their subgroup structure;
    \item the textual transcript of all utterances from the selected recent rounds.
\end{itemize}
Optionally, a special tag can be appended to the end of the prompt to encourage explicit intermediate reasoning (e.g., a chain-of-thought style continuation), though this is not essential to the core algorithm.

The leader uses the same underlying language model as the other agents, but with a more conservative sampling configuration (for instance, a lower sampling temperature) to reduce hallucinations and repetitive patterns. The model generates a single long-form answer, subject to a larger token budget suitable for a full proposal or final solution. If the initial leader output is detected to be extremely short or obviously incomplete (for example, below a pre-defined minimum length), the system may invoke the model a second time under the same conditions to obtain a more complete response.

\paragraph{Final output and logging}
The procedure returns:
\begin{itemize}
    \item the original question or topic;
    \item any reference answer or solution provided by the underlying dataset (when available);
    \item the leader's final textual answer, which serves as the method's prediction;
    \item a detailed set of agent-level traces for all non-leader agents and the leader, each trace containing the agent role, the round in which the utterance was produced, the full prompt used to query the model, and the resulting output;
    \item a summary of the subgroup structure in each round.
\end{itemize}
In this ``text-only'' variant, no latent representations are maintained across calls, and the leader bases its decision solely on visible natural language content from a small number of recent rounds. This makes the method a clean baseline for comparing purely textual collaboration with alternative designs that share richer latent state between agents.

\begin{table*}[t]
\centering
\scriptsize
\setlength{\tabcolsep}{4.5pt}
\begin{tabular}{lcccccccccc}
\toprule
\textbf{Collaboration Mode} &
\multicolumn{3}{c}{\textbf{OpenAI Embedding (Structural; Main)}} &
\multicolumn{3}{c}{\textbf{BGE Embedding (Structural)}} &
\multicolumn{1}{c}{\textbf{Lexical (Main)}} &
\multicolumn{3}{c}{\textbf{Lexical Sensitivity}} \\
\cmidrule(lr){2-4} \cmidrule(lr){5-7} \cmidrule(lr){8-8} \cmidrule(lr){9-11}
& Vendi $\uparrow$ & $(1-\phi)\uparrow$ & PCD $\uparrow$
& Vendi $\uparrow$ & $(1-\phi)\uparrow$ & PCD $\uparrow$
& W-D-3 $\uparrow$
& Raw D-3 & W-D-2 & W-D-4 \\
\midrule
Leader-Led &
6.932 & 0.161 & 0.296 &
5.096 & 0.134 & 0.251 &
0.780 &
0.680 & 0.543 & 0.897 \\
Vertical &
6.082 & 0.146 & 0.272 &
4.131 & 0.114 & 0.215 &
0.781 &
0.694 & 0.530 & 0.882 \\
Naive &
5.567 & 0.154 & 0.285 &
4.141 & 0.127 & 0.239 &
0.638 &
0.522 & 0.426 & 0.754 \\
Interdisciplinary &
4.647 & 0.119 & 0.225 &
3.623 & 0.098 & 0.187 &
0.734 &
0.665 & 0.465 & 0.866 \\
Horizontal &
8.080 & 0.170 & 0.311 &
5.849 & 0.143 & 0.266 &
0.788 &
0.687 & 0.563 & 0.883 \\
\bottomrule
\end{tabular}
\caption{
Sensitivity analysis across representation and metric variants.
\textbf{Main-text results} use the OpenAI embedding for structural metrics (Vendi, $1-\phi$, PCD; first three columns) and report lexical uniqueness via content-only weighted distinct-3 (W-D-3; the ``Lexical (Main)'' column).
We report $(1-\phi)$ (rather than $\phi$) so that larger values consistently indicate greater deviation from consensus.
BGE embedding provides a robustness check for the structural metrics, and Raw D-3 / W-D-2 / W-D-4 probe lexical sensitivity without changing qualitative conclusions.
}
\label{tab:sensitivity_full}
\end{table*}

\section{Sensitivity Analysis}
\label{app:sensitivity}

This appendix examines the robustness of our conclusions to reasonable variations in metric design choices.
Rather than emphasizing absolute metric magnitudes, we focus on whether the \emph{relative ordering} across collaboration modes remains stable under such variations.
All analyses reported here are conducted on the same set of proposals as in the main paper.

\subsection{Overview}

We consider four orthogonal sources of potential sensitivity:
(i) the choice of semantic embedding model,
(ii) the choice of structural diversity metric,
(iii) the definition of lexical uniqueness, including $n$-gram order,
and (iv) content-only versus raw lexical tokenization.
Across all settings, we observe that qualitative trends and relative comparisons across collaboration modes remain invariant.

\subsection{Embedding Model Robustness}

All embedding-based metrics in the main paper use \texttt{text-embedding-3-large}.
To assess whether our conclusions depend on this choice, we recompute Vendi score, $1-\phi$, and PCD using an open-source, retrieval-oriented embedding model (BGE-large).
Due to differing inductive biases, absolute values differ across embeddings.
However, the induced relative ordering across the five collaboration modes is identical for all three metrics.
This suggests that our conclusions are not driven by a specific choice of semantic representation.

\subsection{Consistency Across Structural Metrics}

We next examine consistency among three embedding-based structural metrics: Vendi score, $1-\phi$, and PCD.
Given the limited number of collaboration modes ($n=5$), rank correlations trivially reach $1.0$ whenever orderings coincide.
We therefore report ordering consistency rather than correlation magnitudes.
All three metrics induce identical relative orderings across collaboration modes under both embedding models, suggesting that they capture related but non-redundant aspects of structural diversity.

\subsection{Consistency Across the Four Reported Metrics}
\label{app:metric_consistency}

We examine the relationship among the four metrics reported in the main paper (Table~\ref{tab:sensitivity_full}).
Three of them are \emph{structural} metrics computed in embedding space (Vendi, $1-\phi$, and PCD using the OpenAI embedding), while the fourth captures \emph{lexical} uniqueness (content-only weighted distinct-3, W-D-3).

Across collaboration modes, the embedding-based structural metrics induce highly consistent relative orderings (with only minor local swaps), suggesting that our main structural conclusions are not driven by a single particular formulation.
In contrast, W-D-3 does not necessarily match the embedding-based ordering, which is expected: it measures surface-level lexical novelty that can vary independently from semantic dispersion.
We therefore treat W-D-3 as a complementary signal rather than a redundant proxy for structural diversity.

Overall, the absence of systematic contradictions between the structural and lexical views supports the interpretation that observed differences across collaboration modes reflect robust changes in diversity and consensus, rather than artifacts of a specific metric choice.

\subsection{Lexical Uniqueness and \texorpdfstring{$n$}-gram Order}

We assess the sensitivity of Lexical Uniqueness to the choice of $n$-gram order by computing content-only weighted distinct-$n$ for $n \in \{2,3,4\}$.
Relative ordering across collaboration modes remains stable for $n=2$ and $n=3$, while higher-order $n$-grams exhibit mild saturation effects.
These effects do not alter qualitative trends, supporting the use of $n=3$ in the main analysis.

\subsection{Content-only Tokenization}

To evaluate the impact of content-only tokenization, we compare raw distinct-$3$ with content-only weighted distinct-$3$.
Raw lexical counts exhibit higher variance due to ubiquitous boilerplate expressions.
Content-only tokenization reduces this variance while preserving the relative ordering across collaboration modes.
This suggests that content-only filtering primarily serves as a noise-reduction mechanism rather than a driver of the observed results.

\subsection{Summary of Sensitivity Results}

Table~\ref{tab:sensitivity_full} reports all metrics used in the sensitivity analysis.
Across embedding choices, metric formulations, and lexical definitions,
the qualitative conclusions across collaboration modes remain robust,
despite differences in representational level and metric formulation.
These results indicate that the qualitative conclusions in the main paper are robust to reasonable variations in metric design and representation choices.

\section{Supplementary Evidence for Qualitative Claims}
\label{app:rebuttal}

This appendix provides the operationalized definitions, representative transcript excerpts, and supplementary statistical analyses referenced in the rebuttal responses.
All qualitative evidence comes from the same transcripts used to produce the quantitative figures in the main paper; no post-hoc generation or selection was performed.

\subsection{Inspection Rule Definitions}
\label{app:inspection_rules}

\paragraph{Polite Consensus Collapse.}
We operationalize ``polite consensus collapse'' as a purely behavioral pattern satisfying \emph{all} of the following criteria within a single session:
\begin{enumerate}[nosep]
    \item \textbf{Absence of counter-claims}: no turn contains an explicit critique, disagreement marker, or alternative proposal branch (defined as an explicit proposal of a different research direction from the one assigned by the Leader).
    \item \textbf{Absence of independent sub-problems}: no Collaborator turn introduces a research question or sub-problem not already present in the Leader's framing.
    \item \textbf{Final-proposal alignment}: the final proposal title and abstract recombine keywords and directions from the Leader's Round~1 assignment without introducing new thematic anchors.
\end{enumerate}

\paragraph{Deference and Pushback Markers.}
We define two complementary marker vocabularies applied to the opening sentence of each Collaborator turn:

\begin{itemize}[nosep]
    \item \textbf{Deference markers} (agreement-first phrases): ``Building on\ldots,'' ``Following\ldots,'' ``As you noted\ldots,'' ``Excellent point\ldots,'' ``I fully agree\ldots,'' ``I'd like to add to what\ldots,'' ``Great insight\ldots,'' ``That's a fascinating point\ldots,'' and paraphrase-then-extend patterns (restating a prior speaker's claim before adding content).
    \item \textbf{Pushback markers}: ``I disagree,'' ``however,'' ``counterpoint,'' ``I'm not convinced,'' ``I would challenge,'' ``alternatively,'' ``a different view,'' ``I'd push back,'' ``I'm skeptical,'' ``correct me if I'm wrong.''
\end{itemize}

Under these definitions, deference-marker openings appear in approximately 61\% of Leader-Led sessions; pushback markers appear in fewer than 1\%.

\subsection{Representative Transcript Excerpts}
\label{app:transcripts}

The following excerpts are drawn from randomly sampled sessions satisfying the inspection criteria defined in Appendix~\ref{app:inspection_rules}.
Each mini-case shows the Leader's assignment, Collaborator responses, and the resulting proposal title.

\subsubsection{Mini-case 1: Polite Consensus (Neuroscience, Leader-Led)}
\label{app:minicase_neuro}

\begin{quote}
\small
\textbf{Round 1 --- Leader assigns:} ``Collaborator 1, explore cutting-edge applications of neuroimaging in cognitive science\ldots\ Collaborator 2, please focus on limitations and ethical considerations.''

\textbf{Collaborator 1} responds within the assigned lane.

\textbf{Collaborator 2} opens: ``Building on Collaborator 1's point about multimodal neuroimaging, I'd emphasize that the integration of fMRI with MEG has also yielded critical insights into decision-making paradigms.'' No alternative direction is proposed.

\textbf{Round 2 --- Leader synthesizes:} ``Thank you both for these insightful contributions. Let me synthesize the key points from Round 1\ldots\ For Round 2, Collaborator 1, let's focus on the mechanistic implications.'' Both collaborators continue within the Leader's frame.

\textbf{Final proposal:} \emph{``Multimodal Neuroimaging for Closed-Loop Interventions in Memory and Decision-Making Disorders''} --- the title directly recombines the Leader's Round~1 assignments (neuroimaging + decision-making + clinical application).
\end{quote}

\subsubsection{Mini-case 2: Polite Consensus (Reinforcement Learning, Leader-Led)}
\label{app:minicase_rl}

\begin{quote}
\small
\textbf{Round 1 --- Leader assigns:} ``Collaborator 1, focus on exploration-exploitation\ldots\ Collaborator 2, examine scalability challenges, particularly sim-to-real gaps.''

\textbf{Collaborator 2} opens: ``The interplay between hierarchical RL and model-based methods like Dreamer presents a fascinating tension'' --- staying within the Leader's scalability frame. No counter-proposal or alternative direction appears in any turn.

\textbf{Final proposal:} \emph{``Synergistic Hierarchical and Model-Based Reinforcement Learning for Scalable Continuous Control''} --- a direct synthesis of the Leader's Round~1 framing.
\end{quote}

\subsubsection{Mini-case 3: Deference vs.\ Independent Inquiry (Causal Reasoning, Leader-Led vs.\ Horizontal)}
\label{app:minicase_causal}

\begin{quote}
\small
\textbf{Leader-Led --- Collaborator 1, Round 1:} ``To build on the Leader's framing, I'd emphasize that modern causal reasoning is deeply shaped by the interplay between Pearl's structural causal models (SCMs) and the potential outcomes framework.'' The agent anchors to the Leader's assigned frame before contributing content. No alternative direction is introduced across any turn.

\textbf{Horizontal --- PhD Student A, Round 1 (same topic):} ``I've been reading Pearl's foundational work on causal diagrams, but I'm still confused about how we practically validate the causal assumptions in real-world datasets. In machine learning applications, how do researchers typically handle cases where the true causal graph is unknown or only partially observable?'' The agent introduces an independent sub-problem (validation under unknown graph structure) not present in any prior turn.
\end{quote}

\subsubsection{Mini-case 4: Participation without Innovation (Reinforcement Learning, Leader-Led)}
\label{app:minicase_rl_participation}

\begin{quote}
\small
\textbf{Round 1 --- Leader sets direction:} ``Collaborator 1, focus on the interplay between exploration and exploitation\ldots\ Collaborator 2, examine scalability challenges in RL, particularly sim-to-real gaps.''

\textbf{Collaborator 1} responds (speaks, anchored to Leader's frame): addresses exploration-exploitation within the assigned lane.

\textbf{Collaborator 2} responds (speaks, anchored to Leader's frame): ``The interplay between hierarchical RL and model-based methods like Dreamer presents a fascinating tension'' --- staying within the Leader's assigned scalability frame.

\textbf{Final proposal:} \emph{``Synergistic Hierarchical and Model-Based Reinforcement Learning for Scalable Continuous Control.''} Neither agent introduced a direction outside the Leader's initial assignment.
\end{quote}

\subsection{Title Keyword Frequency and Lexical Concentration Analysis}
\label{app:keyword_analysis}

To quantify the thematic concentration observed in Section~\ref{sec:cognition}, we compute title-level lexical statistics across all 20 topics for the Interdisciplinary and Horizontal configurations.

\paragraph{Method.}
We extract all proposal titles (924 Interdisciplinary; 632 Horizontal), tokenize after lowercasing and removing stopwords, and compute:
(i) Type-Token Ratio (TTR = unique tokens / total tokens),
(ii) top-$k$ unigram and bigram frequencies,
(iii) Jaccard similarity of the full vocabulary sets.

\paragraph{Results.}

\begin{table*}[!ht]
\centering
\small
\begin{tabular}{lcc}
\toprule
\textbf{Statistic} & \textbf{Interdisciplinary} & \textbf{Horizontal} \\
\midrule
Number of titles & 924 & 632 \\
Total tokens & 10{,}411 & 6{,}146 \\
Unique tokens & 729 & 958 \\
Type-Token Ratio (TTR) & 0.070 & 0.156 \\
\bottomrule
\end{tabular}
\caption{Lexical concentration comparison. Despite having more titles and tokens, Interdisciplinary proposals use fewer unique words, yielding a TTR 2.2$\times$ lower than Horizontal.}
\label{tab:ttr}
\end{table*}

\begin{table*}[h]
\centering
\small
\begin{tabular}{ll}
\toprule
\textbf{Configuration} & \textbf{Top-10 Title Words} \\
\midrule
Interdisciplinary & \texttt{multi-scale, modeling, computational, clinical,} \\
                  & \texttt{biological, systems, integrating, learning,} \\
                  & \texttt{networks, bio-inspired} \\
\midrule
Horizontal        & \texttt{exploring, learning, investigating, neural,} \\
                  & \texttt{deep, bridging, methods, hybrid,} \\
                  & \texttt{networks, balancing} \\
\bottomrule
\end{tabular}
\caption{Top-10 title words by frequency. Overlap is 2/10 (\texttt{learning}, \texttt{networks}). Interdisciplinary titles center on cross-domain combinations; Horizontal titles prioritize methodological exploration verbs.}
\label{tab:topwords}
\end{table*}

Table~\ref{tab:ttr} reports the lexical-concentration statistics, and Table~\ref{tab:topwords} lists the top-10 title words.
The full-vocabulary Jaccard similarity is 0.211 (294 shared words out of a union of 1{,}393), confirming that the two configurations draw from largely distinct lexical pools.
Top-10 bigram overlap is 4/10, with shared bigrams being generic ML terminology (\emph{metric learning}, \emph{neural networks}, \emph{reinforcement learning}, \emph{representation learning}).

\subsection{Participation vs.\ Semantic Innovation Analysis}
\label{app:participation}

To test the alternative hypothesis that reduced diversity in Leader-Led configurations stems from unequal participation rather than semantic anchoring, we compare per-turn word counts and semantic novelty between Leader-Led Collaborator turns and Horizontal PhD Student turns.

\paragraph{Method.}
We sample 400 sessions per configuration (20 runs $\times$ 20 topics for Leader-Led; 20 runs $\times$ 20 topics for Horizontal).
For each non-Leader turn, we compute:
(i) word count,
(ii) semantic novelty (cosine distance in sentence-embedding space to the centroid of all prior turns in the same session), using \texttt{all-MiniLM-L6-v2}.

\paragraph{Results.}

\begin{table*}[h]
\centering
\small
\begin{tabular}{lccccc}
\toprule
\textbf{Measure} & \textbf{Leader-Led} & \textbf{Horizontal} & \textbf{$t$} & \textbf{$p$} & \textbf{Cohen's $d$} \\
\midrule
Word count / turn & 167.0 $\pm$ 37.4 & 240.6 $\pm$ 122.2 & $-$24.3 & $<10^{-117}$ & $-$0.70 \\
Semantic novelty / turn & 0.295 $\pm$ 0.111 & 0.306 $\pm$ 0.173 & $-$1.83 & 0.068 (n.s.) & $-$0.069 \\
Total words / session & 336.6 $\pm$ 62.4 & 1203.2 $\pm$ 121.6 & $-$126.8 & $<10^{-300}$ & $-$8.97 \\
\bottomrule
\end{tabular}
\caption{Participation and semantic novelty comparison. Leader-Led Collaborators produce fewer words per turn and drastically fewer total words per session, yet their per-turn semantic novelty is statistically indistinguishable from Horizontal agents ($p = 0.068$, Cohen's $d = -0.069$). This indicates that Collaborators are active but semantically anchored to the Leader's framing.}
\label{tab:participation}
\end{table*}

Table~\ref{tab:participation} compares participation and per-turn novelty.

\paragraph{Per-turn anchor similarity.}
As a complementary measure, we compute the cosine similarity between each Collaborator turn and the Leader's Round~1 framing (for Leader-Led), or between each subsequent turn and the first speaker's Round~1 contribution (for Horizontal).

\begin{table}[h]
\centering
\small
\begin{tabular}{lcc}
\toprule
\textbf{Configuration} & \textbf{Mean Cosine Sim to Anchor} & \textbf{Std} \\
\midrule
Leader-Led & 0.627 & 0.181 \\
Horizontal & 0.441 & 0.211 \\
\bottomrule
\end{tabular}
\caption{Per-turn cosine similarity to the session anchor (Leader's Round~1 for Leader-Led; first speaker's Round~1 for Horizontal). Leader-Led turns remain substantially closer to the anchor framing (difference = $+$0.19), consistent with semantic gravitational anchoring.}
\label{tab:anchor_sim}
\end{table}

Results are reported in Table~\ref{tab:anchor_sim}.

\subsection{Representative Title Lists by Topic}
\label{app:title_lists}

To allow readers to verify the thematic concentration claim in Section~\ref{sec:cognition}, we provide representative title samples for the Neuroscience topic (the running example in the main text).

\paragraph{Interdisciplinary (Neuroscience) --- representative titles:}
\begin{enumerate}[nosep]
\small
    \item ``Multi-Scale Computational Modeling of Synaptic Plasticity: Bridging Molecular Dynamics to Cognitive Function''
    \item ``Multi-Scale Computational Modeling of Synaptic Plasticity Mechanisms for Precision Cognitive Therapeutics''
    \item ``Multi-Scale Graph Neural Networks for Modeling Synaptic Plasticity in Neuropsychiatric Disorders''
    \item ``Biologically-Constrained Computational Models for Enhanced Clinical Neuroscience Applications''
    \item ``Bridging Molecular Plasticity Mechanisms with Computational Models for Clinical Translation''
\end{enumerate}

\paragraph{Horizontal (Neuroscience) --- representative titles:}
\begin{enumerate}[nosep]
\small
    \item ``Bridging the Gap in Temporal Processing: Biologically Inspired Modifications to ANNs''
    \item ``Exploring Transformer Attention Mechanisms as Models of Hippocampal Memory Processes''
    \item ``Comparing Artificial and Biological Attention Mechanisms in Simple Cognitive Tasks''
    \item ``Investigating the Relationship Between Hippocampal Replay Fidelity and Memory Consolidation''
    \item ``Exploring Biologically Plausible Alternatives to Backpropagation in Neural Networks''
    \item ``Exploring Neural Heterogeneity and Temporal Dynamics in Bio-Inspired ANNs''
    \item ``Investigating Parallels Between Self-Supervised Learning and Predictive Coding''
\end{enumerate}

The Interdisciplinary titles recombine three anchors (\emph{multi-scale modeling}, \emph{synaptic plasticity}, \emph{clinical translation}) with minor variation.
The Horizontal titles span at least five distinct sub-directions (temporal processing, attention mechanisms, hippocampal replay, backpropagation alternatives, neural heterogeneity, predictive coding) with no repeated phrase template.

\subsection{Quality Comparison Across Persona Structures}
\label{app:quality_comparison}

To address the concern that diversity-focused analysis is incomplete without quality measures, we evaluate proposal quality across all five persona structures using the same LLM-as-Judge protocol employed for the single-model baseline in Section~\ref{sec:intelligence_analysis} (DeepSeek-V3, temperature~0, 9-dimension rubric).
We randomly sample 3 proposals per topic $\times$ 20 topics = 60 proposals per persona (300 total), ensuring balanced topic coverage.

\paragraph{Results.}

\begin{table*}[h]
\centering
\scriptsize
\setlength{\tabcolsep}{3.5pt}
\begin{tabular}{lccccccccc}
\toprule
\textbf{Persona} &
\textbf{OQ} & \textbf{Nov} & \textbf{Work} & \textbf{Rel} & \textbf{Spec} & \textbf{IntD} & \textbf{StrV} & \textbf{MRig} & \textbf{ArgC} \\
\midrule
Naive ($n{=}60$)
& 7.95{\tiny$\pm$0.2} & 6.75{\tiny$\pm$0.6} & 7.03{\tiny$\pm$0.9} & 9.98{\tiny$\pm$0.1} & 8.03{\tiny$\pm$0.5} & 8.90{\tiny$\pm$0.5} & 7.83{\tiny$\pm$0.4} & 7.17{\tiny$\pm$0.6} & 8.97{\tiny$\pm$0.3} \\
Horizontal ($n{=}60$)
& 7.88{\tiny$\pm$0.3} & 6.48{\tiny$\pm$0.8} & \textbf{7.95}{\tiny$\pm$0.9} & 9.90{\tiny$\pm$0.3} & 8.03{\tiny$\pm$0.7} & 8.52{\tiny$\pm$0.7} & 7.30{\tiny$\pm$0.6} & 7.37{\tiny$\pm$0.6} & 8.87{\tiny$\pm$0.3} \\
Vertical ($n{=}60$)
& 8.32{\tiny$\pm$0.5} & 7.92{\tiny$\pm$0.7} & 6.43{\tiny$\pm$0.5} & 10.00{\tiny$\pm$0.0} & 8.18{\tiny$\pm$0.6} & 9.03{\tiny$\pm$0.2} & 8.28{\tiny$\pm$0.5} & 7.43{\tiny$\pm$0.5} & 9.00{\tiny$\pm$0.0} \\
Leader-Led ($n{=}60$)
& 8.03{\tiny$\pm$0.2} & 7.08{\tiny$\pm$0.5} & 6.43{\tiny$\pm$0.8} & 9.98{\tiny$\pm$0.1} & 8.00{\tiny$\pm$0.6} & 8.93{\tiny$\pm$0.3} & 8.02{\tiny$\pm$0.3} & 7.23{\tiny$\pm$0.6} & 8.98{\tiny$\pm$0.1} \\
Interdisciplinary ($n{=}60$)
& \textbf{8.50}{\tiny$\pm$0.5} & \textbf{8.02}{\tiny$\pm$0.7} & 6.42{\tiny$\pm$0.6} & 10.00{\tiny$\pm$0.0} & 8.20{\tiny$\pm$0.7} & \textbf{9.10}{\tiny$\pm$0.3} & \textbf{8.65}{\tiny$\pm$0.5} & 7.48{\tiny$\pm$0.6} & 9.07{\tiny$\pm$0.3} \\
\midrule
ANOVA $\eta^2$ & 0.297 & 0.470 & 0.385 & 0.054 & 0.017 & 0.188 & 0.498 & 0.038 & 0.060 \\
\bottomrule
\end{tabular}
\caption{Quality scores (1--10) across five persona structures, evaluated by DeepSeek-V3 (LLM-as-Judge, temperature~0).
OQ = Overall Quality, Nov = Novelty, Work = Workability, Rel = Relevance, Spec = Specificity, IntD = Integration Depth, StrV = Strategic Vision, MRig = Methodological Rigor, ArgC = Argumentative Cohesion.
Bold indicates the highest value per column.
Authority-weighted structures (Interdisciplinary, Vertical) score modestly higher on Overall Quality ($+$0.4--0.6 over Horizontal), but Horizontal achieves the highest Workability.
Specificity and Methodological Rigor show negligible variation ($\eta^2 < 0.04$).}
\label{tab:quality_personas}
\end{table*}

Table~\ref{tab:quality_personas} reports the per-dimension quality scores.

\paragraph{Key findings.}
\begin{enumerate}[nosep]
    \item \textbf{Quality differences are modest.} The Overall Quality range across all five structures is 7.88--8.50 (a 0.62-point spread on a 10-point scale, or 6\%). By contrast, the Vendi Score range is 4.65--8.08 (a 74\% relative difference). Quality variation is an order of magnitude smaller than diversity variation.
    \item \textbf{Horizontal achieves the highest Workability.} Despite scoring lowest on Overall Quality, Horizontal proposals are rated significantly more feasible (Workability = 7.95 vs.\ 6.40--6.43 for authority-weighted structures; $p < 10^{-10}$, Cohen's $d > 1.0$). This suggests that the diversity in Horizontal proposals is not noise but reflects a broader range of \emph{actionable} research directions.
    \item \textbf{The quality--diversity tradeoff is asymmetric.} Authority-weighted structures gain $\sim$0.5 points in Overall Quality but lose $\sim$3.4 points in Vendi Score. The marginal quality gain does not compensate for the substantial diversity loss.
    \item \textbf{Specificity, Rigor, and Cohesion are structure-invariant.} These three dimensions show $\eta^2 < 0.06$, indicating that the structural quality of proposals (how specific, rigorous, and coherent they are) is largely independent of persona structure.
\end{enumerate}

\paragraph{Statistical tests.}
One-way ANOVA confirms significant differences for Overall Quality ($F = 31.1$, $p < 0.001$, $\eta^2 = 0.297$).
Pairwise Welch $t$-tests show that Horizontal vs.\ Naive is not significant ($p = 0.190$), while Horizontal vs.\ Interdisciplinary is significant ($p < 0.001$, Cohen's $d = -1.46$).
Full pairwise comparisons are available in the evaluation cache released with the code.

\section{Per-Topic Analysis of Group-Size Scaling and Research Problem Complexity}
\label{app:topic_complexity}

This appendix provides a per-topic decomposition of the group-size scaling analysis in Section~\ref{sec:dynamics} to address whether the observed diversity saturation is driven by limited ideation capacity of individual research topics.

\subsection{Motivation}
The aggregate analysis in Figure~\ref{fig:size} shows that the Diversity Utilization Ratio (Vendi/$N$) declines from 1.03 at $N{=}3$ to 0.47 at $N{=}7$. A natural alternative hypothesis is that this saturation reflects the finite complexity of the 20 ICLR research topics used as our testbed: perhaps each topic supports only ${\sim}3$ genuinely distinct ideas, and groups larger than $N{=}3$ simply exhaust the available ideation space. To test this hypothesis, we decompose the scaling analysis to the individual topic level.

\subsection{Method}
For each of the 20 ICLR topics and each group size $N \in \{3,4,5,6,7\}$, we have 50 proposals generated by independent $N$-agent MAS runs (no cross-run interaction). We compute the per-topic Vendi Score using the same OpenAI \texttt{text-embedding-3-large} embeddings and cosine-similarity kernel as in the main paper. We then analyze: (1)~whether the absolute Vendi Score increases or plateaus with $N$, (2)~whether topic-level intrinsic diversity capacity (measured at $N{=}3$) predicts the scaling behavior, and (3)~the cross-topic variance in utilization ratio.

\subsection{Results}

\paragraph{Finding 1: Absolute diversity grows with group size across all topics.}
Table~\ref{tab:topic_vendi_groupsize} reports the per-topic Vendi Score for each group size. Aggregated across topics, the mean Vendi Score increases monotonically from 3.09 at $N{=}3$ to 3.32 at $N{=}7$ (+7.4\%, paired $t{=}5.46$, $p{<}0.0001$), with 17 out of 20 topics exhibiting growth. This directly refutes the ``low-hanging fruit'' hypothesis: if topics were limited to ${\sim}3$ distinct ideas, the Vendi Score would plateau at ${\sim}3$ regardless of group size. Instead, larger groups consistently produce a broader semantic space of proposals.

\begin{table*}[h]
\centering
\small
\begin{tabular}{lccccc}
\toprule
\textbf{Topic} & $N{=}3$ & $N{=}4$ & $N{=}5$ & $N{=}6$ & $N{=}7$ \\
\midrule
General ML       & 3.51 & 3.54 & 3.65 & 3.76 & 3.53 \\
Physical Sci.    & 3.42 & 3.48 & 3.45 & 3.64 & 3.75 \\
Viz/Interp       & 3.38 & 3.53 & 3.44 & 3.86 & 3.67 \\
Generative       & 3.32 & 3.37 & 3.35 & 3.53 & 3.15 \\
Optimization     & 3.26 & 3.35 & 3.37 & 3.50 & 3.59 \\
Neuro/CogSci     & 3.20 & 3.24 & 3.21 & 3.50 & 3.33 \\
Infra/SW         & 3.18 & 3.36 & 3.29 & 3.80 & 3.69 \\
RL               & 3.19 & 3.03 & 3.22 & 3.42 & 3.46 \\
Learn Theory     & 3.13 & 3.33 & 3.60 & 3.60 & 3.48 \\
Causal           & 3.09 & 3.39 & 3.14 & 3.01 & 3.02 \\
Data/Bench       & 3.08 & 3.35 & 3.25 & 3.48 & 3.56 \\
Metric/Kernel    & 3.06 & 3.08 & 3.53 & 3.28 & 3.30 \\
Prob/Bayes       & 3.01 & 2.99 & 3.21 & 3.23 & 3.49 \\
SSL/Unsup        & 3.00 & 3.27 & 3.13 & 3.32 & 3.35 \\
Graphs/Topo      & 2.96 & 2.92 & 2.95 & 3.30 & 3.26 \\
Robotics         & 2.94 & 2.95 & 3.39 & 2.89 & 2.89 \\
RepLearn         & 2.93 & 3.19 & 3.06 & 3.49 & 3.04 \\
Society/Fair     & 2.84 & 3.00 & 2.86 & 3.00 & 3.10 \\
NeuroSymbolic    & 2.74 & 3.05 & 2.95 & 2.98 & 3.04 \\
Transfer/Meta    & 2.60 & 2.92 & 2.73 & 2.84 & 2.74 \\
\midrule
\textbf{Mean}    & 3.09 & 3.22 & 3.24 & 3.37 & 3.32 \\
\bottomrule
\end{tabular}
\caption{Per-topic Vendi Score across group sizes ($N{=}3$ to $N{=}7$). Each cell is computed from 50 independent proposals. Topics are sorted by Vendi at $N{=}3$ (descending). The mean Vendi increases from 3.09 to 3.32, with 17/20 topics showing growth from $N{=}3$ to $N{=}7$.}
\label{tab:topic_vendi_groupsize}
\end{table*}

\paragraph{Finding 2: Topic complexity does not predict saturation rate.}
We measured each topic's intrinsic diversity capacity using the Vendi Score of 50 independent proposals at $N{=}3$ and correlated it with the diversity growth rate from $N{=}3$ to $N{=}7$. The correlation is not statistically significant (Pearson $r{=}{-}0.14$, $p{=}0.55$). Both high-capacity topics (e.g., General ML: $3.51{\to}3.53$; Physical Sciences: $3.42{\to}3.75$) and low-capacity topics (e.g., Transfer/Meta: $2.60{\to}2.74$; NeuroSymbolic: $2.74{\to}3.04$) exhibit similar utilization slopes ($-0.143$ vs.\ $-0.127$). This confirms that the sub-linear scaling is a structural property of multi-agent consensus dynamics, not an artifact of topic-specific ideation ceilings.

\paragraph{Finding 3: Cross-topic variance is small.}
The coefficient of variation (CV) of the Utilization Ratio remains small across all group sizes: CV${=}0.07$ at $N{=}3$, increasing modestly to CV${=}0.09$ at $N{=}7$. This indicates that the saturation pattern is remarkably consistent regardless of topic breadth.

\begin{figure*}[h]
\centering
\includegraphics[width=\textwidth]{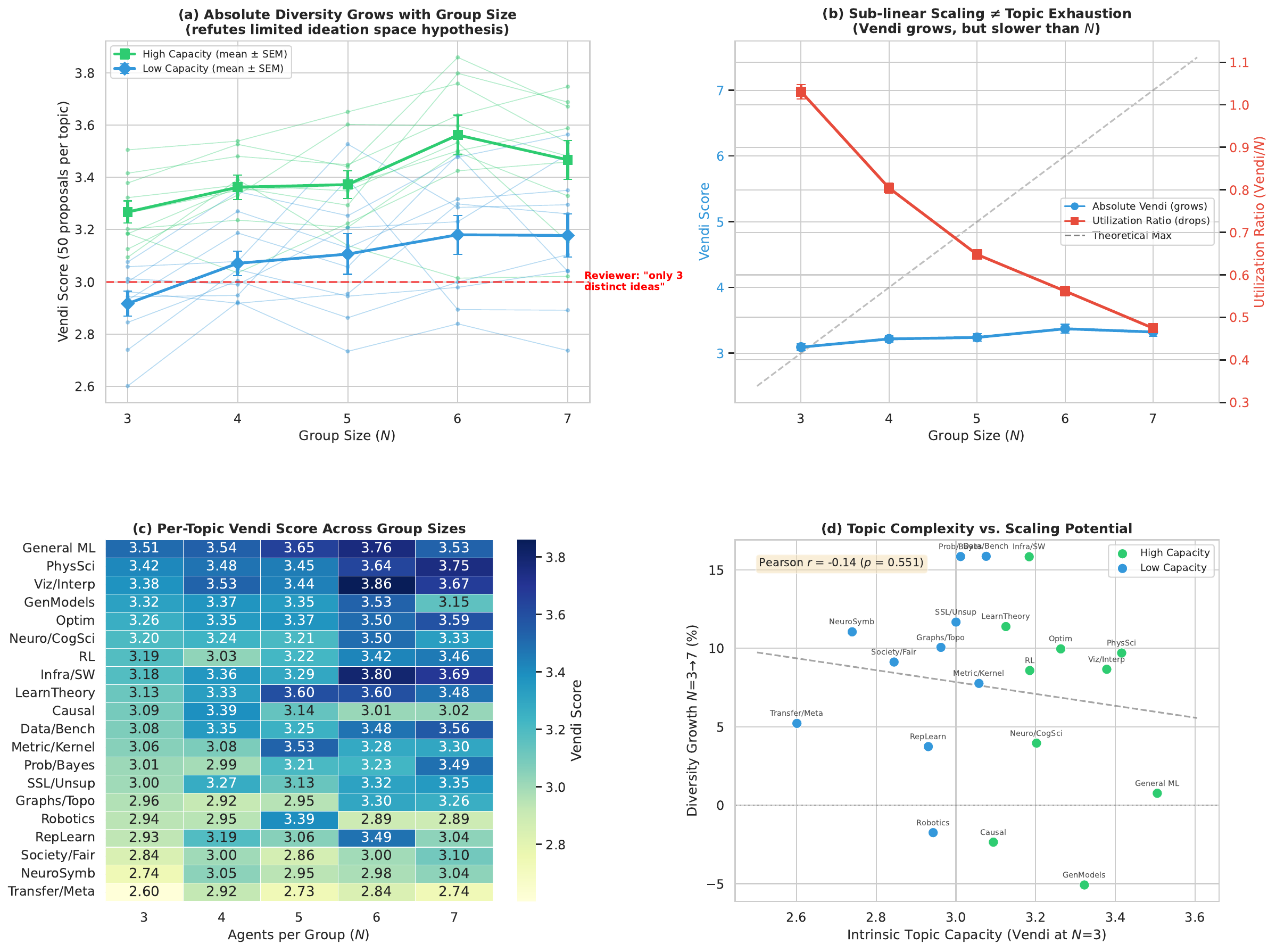}
\caption{Per-topic decomposition of group-size scaling. (a)~Per-topic Vendi Score trajectories, colored by intrinsic capacity group; the red dashed line marks the ``only 3 distinct ideas'' hypothesis. (b)~Absolute Vendi grows (blue) while Utilization Ratio drops (red), demonstrating sub-linear scaling rather than topic exhaustion. (c)~Per-topic Vendi heatmap across group sizes. (d)~Topic intrinsic capacity vs.\ diversity growth rate shows no significant correlation ($r{=}{-}0.14$, $p{=}0.55$).}
\label{fig:topic_complexity}
\end{figure*}

Figure~\ref{fig:topic_complexity} summarizes these per-topic trajectories.

\subsection{Interpretation}
The declining Utilization Ratio (Vendi/$N$) reported in Section~\ref{sec:dynamics} reflects \emph{diminishing marginal returns} per additional agent: each new agent contributes some diversity to the proposal pool, but less than the theoretical maximum of one fully orthogonal perspective. This is analogous to diminishing returns in team scaling~\citep{brooks1975mythical}, not evidence that topics ``run out'' of ideas. The bottleneck is the shared alignment priors and consensus dynamics inherent in LLM-based multi-agent systems, which our structural interventions address.

\section{2\texorpdfstring{$\times$}2 Factorial Ablation: Topology \texorpdfstring{$\times$} Persona on DeepSeek-V3}
\label{app:w3_factorial}

This appendix reports a controlled ablation experiment designed to test whether the communication topology effect is moderated by persona structure.

\subsection{Motivation}

The topology analysis in Section~\ref{sec:dynamics} uses the Naive persona for DeepSeek-V3, while cross-model comparisons introduce additional persona variation (Horizontal for o1-mini, Interdisciplinary for GPT-5.1). This makes it difficult to cleanly attribute diversity differences to topology alone. To resolve this, we conduct a 2$\times$2 factorial experiment on a single model (DeepSeek-V3), varying \emph{only} the persona structure and communication topology while holding all other variables constant.

\subsection{Experimental Design}

\begin{table*}[h]
\centering
\begin{tabular}{lcc}
\toprule
& \textbf{Recursive} & \textbf{NGT} \\
\midrule
\textbf{Naive} (senior researchers) & \checkmark~(existing, $N{=}50$) & \checkmark~(existing, $N{=}50$) \\
\textbf{Horizontal} (first-year PhD) & \checkmark~(\textbf{new}, $N{=}50$) & \checkmark~(\textbf{new}, $N{=}50$) \\
\bottomrule
\end{tabular}
\caption{2$\times$2 factorial design. Each cell contains 50 independent runs $\times$ 20 ICLR topics. All four conditions use DeepSeek-V3 with temperature 0.7 and group size $N{=}3$. Only the persona prompt and communication topology vary.}
\label{tab:factorial_design}
\end{table*}

Table~\ref{tab:factorial_design} summarizes the 2$\times$2 design.

\paragraph{Topology definitions.}
\begin{itemize}[nosep]
    \item \textbf{Recursive}: All 3 agents see the full conversation history and speak sequentially in each round (4 discussion rounds + 1 proposal round). This corresponds to \texttt{grouped\_sequential} order with \texttt{all} visibility.
    \item \textbf{NGT} (Nominal Group Technique): Round~1 is a blind-writing phase where each agent writes independently without seeing others' contributions. Rounds~2--3 involve sequential discussion among non-leader agents with full visibility. Round~4 includes all agents (including the leader). Round~5 produces the proposal.
\end{itemize}

\paragraph{Persona definitions.}
\begin{itemize}[nosep]
    \item \textbf{Naive}: Agents are prompted as ``senior AI researchers with expertise in [topic]'' with no specific role differentiation.
    \item \textbf{Horizontal}: Agents are prompted as ``first-year PhD students with limited research experience in [topic]'' who bring ``fresh curiosity and basic academic foundation.''
\end{itemize}

\paragraph{Metric note.} We report \textbf{within-topic Vendi Scores}: for each of the 20 ICLR topics, we compute the Vendi Score over the 50 proposals generated under that topic, then average across topics. This measures how semantically spread out proposals are \emph{within a single research domain}. It is complementary to the global Vendi Score used in the main paper, which pools all proposals across topics and additionally captures between-topic spread; the two views can rank conditions slightly differently because they measure different facets of diversity.

\subsection{Results}

\paragraph{Finding 1: Per-topic within-persona topology effects.}

\begin{table*}[h]
\centering
\small
\begin{tabular}{l|cc|c|cc|c}
\toprule
& \multicolumn{3}{c|}{\textbf{Naive (Senior Researcher)}} & \multicolumn{3}{c}{\textbf{Horizontal (PhD Student)}} \\
\textbf{Topic} & \textbf{Rec.} & \textbf{NGT} & \textbf{$\Delta_N$} & \textbf{Rec.} & \textbf{NGT} & \textbf{$\Delta_H$} \\
\midrule
Causal          & 3.15 & 2.69 & $-$0.45 & 2.62 & 2.17 & $-$0.45 \\
Data/Bench      & 3.13 & 2.29 & $-$0.84 & 1.93 & 1.98 & $+$0.05 \\
GenModels       & 3.36 & 3.00 & $-$0.37 & 2.35 & 2.36 & $+$0.00 \\
General ML      & 3.52 & 3.31 & $-$0.21 & 2.72 & 2.90 & $+$0.18 \\
Graphs/Topo     & 2.96 & 2.42 & $-$0.55 & 2.47 & 2.25 & $-$0.21 \\
Infra/SW        & 3.22 & 2.82 & $-$0.40 & 2.19 & 2.26 & $+$0.08 \\
LearnTheory     & 3.19 & 2.45 & $-$0.74 & 2.07 & 1.90 & $-$0.17 \\
Metric/Kernel   & 3.12 & 2.57 & $-$0.55 & 1.98 & 2.04 & $+$0.05 \\
Neuro/CogSci    & 3.22 & 2.84 & $-$0.39 & 2.65 & 2.40 & $-$0.24 \\
NeuroSymb       & 2.77 & 2.50 & $-$0.27 & 2.22 & 2.06 & $-$0.16 \\
Optim           & 3.28 & 2.84 & $-$0.44 & 2.30 & 2.30 & $+$0.00 \\
PhysSci         & 3.41 & 2.51 & $-$0.91 & 2.30 & 2.10 & $-$0.20 \\
Prob/Bayes      & 3.06 & 2.77 & $-$0.29 & 2.24 & 2.07 & $-$0.17 \\
RL              & 3.22 & 2.81 & $-$0.41 & 2.39 & 2.46 & $+$0.06 \\
RepLearn        & 2.94 & 2.68 & $-$0.26 & 2.25 & 2.24 & $-$0.01 \\
Robotics        & 2.92 & 2.48 & $-$0.44 & 1.97 & 2.00 & $+$0.04 \\
SSL/Unsup       & 3.08 & 3.01 & $-$0.07 & 2.09 & 2.28 & $+$0.19 \\
Society/Fair    & 2.83 & 2.15 & $-$0.67 & 1.95 & 1.84 & $-$0.11 \\
Transfer/Meta   & 2.66 & 2.33 & $-$0.33 & 2.06 & 2.14 & $+$0.08 \\
Viz/Interp      & 3.45 & 2.92 & $-$0.53 & 1.98 & 2.17 & $+$0.18 \\
\midrule
\textbf{Mean}   & \textbf{3.125} & \textbf{2.669} & \textbf{$-$0.456} & \textbf{2.236} & \textbf{2.196} & \textbf{$-$0.040} \\
$\pm$ SEM       & $\pm$0.050 & $\pm$0.062 & $\pm$0.046 & $\pm$0.053 & $\pm$0.051 & $\pm$0.037 \\
\bottomrule
\end{tabular}
\caption{Within-topic Vendi Scores for all four cells of the 2$\times$2 factorial (50 runs per topic per cell). $\Delta_N$ and $\Delta_H$ denote the within-topic topology effect (NGT $-$ Recursive) for Naive and Horizontal personas, respectively. Under the Naive persona, the topology effect is large and consistent across all 20 topics; under the Horizontal persona, the effect is near zero.}
\label{tab:factorial_pertopic}
\end{table*}

Table~\ref{tab:factorial_pertopic} reports the within-topic Vendi Score for each of the 20 ICLR topics across all four cells. The topology effect ($\Delta = \text{NGT} - \text{Recursive}$) is shown separately for each persona.

\paragraph{Finding 2: Significant Persona $\times$ Topology interaction.}

\begin{table*}[h]
\centering
\begin{tabular}{lcccc}
\toprule
\textbf{Effect} & \textbf{Mean Difference} & \textbf{$t$} & \textbf{$p$} & \textbf{Cohen's $d$} \\
\midrule
\multicolumn{5}{l}{\textit{Within-Persona Topology Effects}} \\
\quad Naive: Rec. $-$ NGT & $+$0.456 & 9.65 & $<$0.0001 & 2.21 \\
\quad Horizontal: Rec. $-$ NGT & $+$0.040 & 1.07 & 0.298 & 0.25 \\
\midrule
\multicolumn{5}{l}{\textit{Interaction}} \\
\quad Persona $\times$ Topology & $\Delta_N - \Delta_H = 0.416$ & $-$8.18 & $<$0.0001 & --- \\
\bottomrule
\end{tabular}
\caption{Within-persona topology effects and interaction test (50 runs per topic per cell, within-topic Vendi Scores). The Naive persona shows a large, significant topology effect ($d = 2.21$), while the Horizontal persona shows a negligible effect ($d = 0.25$). The significant interaction ($p < 0.0001$) confirms that the magnitude of the topology effect is persona-dependent.}
\label{tab:factorial_summary}
\end{table*}

Table~\ref{tab:factorial_summary} summarizes the within-persona topology effects and the interaction test.

\paragraph{Finding 3: Pairwise Cosine Distance (PCD) confirms the within-persona pattern.}

\begin{table}[h]
\centering
\begin{tabular}{lcc}
\toprule
& \textbf{Recursive} & \textbf{NGT} \\
\midrule
\textbf{Naive}       & $0.187 \pm 0.014$ & $0.158 \pm 0.020$ \\
\textbf{Horizontal}  & $0.127 \pm 0.024$ & $0.122 \pm 0.019$ \\
\bottomrule
\end{tabular}
\caption{Mean Pairwise Cosine Distance ($\pm$ SD) across 20 topics. Within each persona, the Recursive--NGT ordering mirrors the Vendi Score results, confirming that the within-persona topology pattern is not metric-specific.}
\label{tab:factorial_pcd}
\end{table}

Table~\ref{tab:factorial_pcd} reports the mean PCD across proposals within each cell, validating the Vendi Score results with an independent metric.

\paragraph{Finding 4: Cross-persona effect correlation.}

\begin{figure*}[h]
\centering
\includegraphics[width=\textwidth]{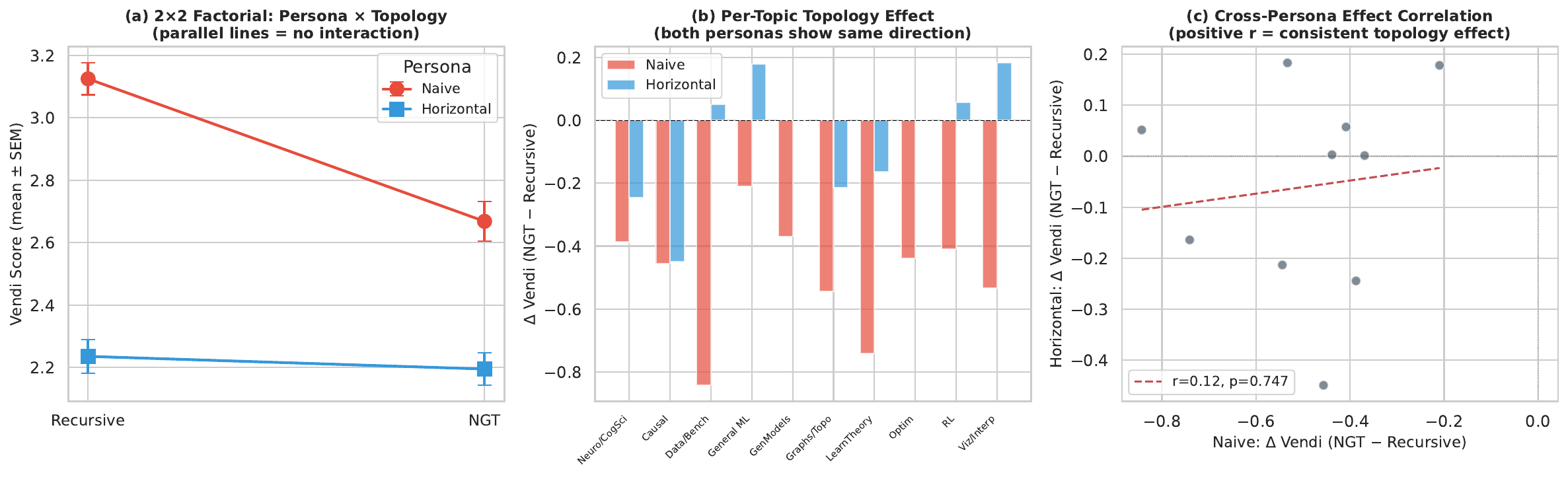}
\caption{2$\times$2 factorial ablation results on DeepSeek-V3 (within-topic Vendi Scores, 50 runs per topic per cell). (a)~Interaction plot: non-parallel lines indicate a significant Persona $\times$ Topology interaction ($p < 0.0001$). (b)~Per-topic topology effect ($\Delta$ Vendi = NGT $-$ Recursive) for both personas; the Naive persona shows a consistently larger effect. (c)~Cross-persona effect correlation: each dot is one ICLR topic; the low correlation ($r = 0.30$, $p = 0.20$) indicates that topic-level topology sensitivity is persona-dependent.}
\label{fig:factorial_scatter}
\end{figure*}

Figure~\ref{fig:factorial_scatter} visualizes the factorial results. Panel~(a) shows the interaction plot: the non-parallel lines confirm the significant interaction. Panel~(b) displays per-topic topology effects for both personas --- the Naive persona shows a consistently large negative effect (all 20 topics), while the Horizontal persona shows a mixed pattern (9/20 negative, 11/20 near-zero or positive). Panel~(c) shows the cross-persona effect correlation ($r = 0.30$, $p = 0.20$), indicating that the per-topic topology sensitivity does not reliably transfer across personas.

\subsection{Interpretation}

The factorial ablation refines the topology results in Section~\ref{sec:dynamics} by showing that persona structure significantly moderates the topology effect:

\begin{enumerate}[nosep]
    \item \textbf{Topology is a genuine structural lever.} Under the Naive persona, switching between Recursive and NGT produces a large, consistent within-topic diversity difference ($d = 2.21$, all 20 topics). This confirms that communication topology has a real causal effect on within-topic diversity, independent of persona.
    \item \textbf{Persona moderates topology magnitude.} The significant interaction ($p < 0.0001$) confirms that persona modulates the topology effect size. Under the Horizontal persona, the within-topic topology effect is negligible ($d = 0.25$, $p = 0.298$).
    \item \textbf{Mechanistic reading.} The Horizontal persona (less directive, exploratory agents) is more robust to topology changes, suggesting that when agents are less constrained by authority priors, the communication structure has less marginal impact on within-topic diversity. The Naive persona (senior researchers with stronger priors) is more susceptible to coupling-induced convergence, making topology a more effective lever. When baseline coupling is already low, structural interventions yield diminishing marginal returns.
\end{enumerate}

\section{Heterogeneous Model Experiments: Validating Ecological Validity}
\label{app:hetero_model}

This appendix reports experiments using genuinely heterogeneous models (different LLMs per agent) to address the concern that our findings may reflect limitations of persona prompting within a single model rather than structural properties of multi-agent interaction.

\subsection{Motivation}

The original experiments use the same underlying LLM for all agents within a condition, varying only persona prompts and interaction topology. While this design isolates structural effects, it raises the question of whether the observed diversity collapse in authority-weighted structures simply reflects the model's inability to maintain distinct personas---a known limitation of persona prompting. To test this, we conducted experiments where each agent uses a genuinely different LLM with distinct training data, architecture, and alignment objectives.

\subsection{Experimental Design}

\paragraph{Model assignment.}
Each agent position is assigned a different model:
\begin{itemize}[nosep]
    \item \textbf{Agent 1 (P1):} DeepSeek-V3 (open-source, Chinese-trained, strong reasoning)
    \item \textbf{Agent 2 (P2):} GPT-4o (proprietary, Western-trained, instruction-tuned)
    \item \textbf{Agent 3 (P3):} Claude-Sonnet-4 (proprietary, constitutional AI, safety-focused)
\end{itemize}

These models differ substantially in training data composition (Chinese vs.\ Western corpora), architectural design (open-source vs.\ proprietary), alignment objectives (RLHF intensity, safety constraints), and reasoning styles.

\paragraph{Experiment A: Heterogeneous model $\times$ persona structure (5 topics).}
Three persona structures $\times$ 5 ICLR topics $\times$ 25 runs = 375 proposals:
\begin{itemize}[nosep]
    \item Hetero-Horizontal: First-year PhD personas, Standard topology, $N{=}3$
    \item Hetero-Interdisciplinary: Senior researcher personas, Standard topology, $N{=}3$
    \item Hetero-Leader-Led: Leader + 2 Collaborators, Standard topology, $N{=}3$
\end{itemize}

\paragraph{Experiment B: Mixed-model horizontal (20 topics).}
1 condition $\times$ 20 ICLR topics $\times$ 50 runs = 1{,}000 proposals, using the same heterogeneous model assignment under the Horizontal persona structure. This enables direct paired comparison with the original single-model (DeepSeek-V3 only) baselines across all 20 topics.

\subsection{Results}

\paragraph{Finding 1: Model heterogeneity rescues diversity in authority structures.}

Table~\ref{tab:hetero_vs_single} compares per-topic Vendi Scores between heterogeneous-model and single-model configurations for each persona structure.

\begin{table*}[h]
\centering
\small
\begin{tabular}{lcccccc}
\toprule
\textbf{Persona} & \textbf{Single-Model} & \textbf{Hetero-Model} & \textbf{$\Delta$} & \textbf{\% Gain} & \textbf{$p$} & \textbf{Cohen's $d$} \\
\midrule
Interdisciplinary & 2.197 $\pm$ 0.343 & 2.943 $\pm$ 0.297 & $+$0.747 & $+$34.0\% & 0.0003 & 2.33 \\
Leader-Led        & 2.285 $\pm$ 0.339 & 2.619 $\pm$ 0.332 & $+$0.334 & $+$14.6\% & 0.071  & 0.99 \\
Horizontal        & 2.755 $\pm$ 0.476 & 2.641 $\pm$ 0.271 & $-$0.114 & $-$4.1\%  & 0.627  & $-$0.29 \\
\bottomrule
\end{tabular}
\caption{Per-topic Vendi Score comparison between single-model (DeepSeek-V3 only, 20 topics) and heterogeneous-model (DSV3 + GPT-4o + Claude-Sonnet-4, 5 topics) configurations. Model heterogeneity produces large gains for authority-weighted structures (Interdisciplinary: $+$34\%, Leader-Led: $+$15\%) but no gain for Horizontal ($-$4\%).}
\label{tab:hetero_vs_single}
\end{table*}

\paragraph{Finding 2: Mixed-model horizontal outperforms all single-model baselines (20 topics).}

Table~\ref{tab:mixed_vs_baselines} reports the comparison between the Mixed-Model Horizontal configuration (Experiment~B) and all original single-model baselines.

\begin{table*}[h]
\centering
\small
\begin{tabular}{lccccc}
\toprule
\textbf{Comparison} & \textbf{$\Delta$ Vendi} & \textbf{$t$} & \textbf{$p$} & \textbf{Cohen's $d$} & \textbf{Sig} \\
\midrule
Mixed vs DSV3-Horizontal       & $+$0.648 & 4.86  & 0.000020   & 1.58 & *** \\
Mixed vs DSV3-Interdisciplinary & $+$1.206 & 11.01 & $<$0.000001 & 3.57 & *** \\
Mixed vs DSV3-Leader-Led       & $+$1.117 & 10.26 & $<$0.000001 & 3.33 & *** \\
Mixed vs DSV3-Naive            & $+$1.744 & 20.13 & $<$0.000001 & 6.53 & *** \\
Mixed vs DSV3-Vertical         & $+$0.873 & 9.51  & $<$0.000001 & 3.08 & *** \\
\bottomrule
\end{tabular}
\caption{Mixed-Model Horizontal (mean Vendi $= 3.402 \pm 0.332$, 20 topics) vs.\ all single-model baselines (DeepSeek-V3 only). All comparisons are highly significant. Paired $t$-test against DSV3-Horizontal (same 20 topics): $\Delta = +0.648$, $t = 5.250$, $p = 0.000046$, with 19/20 topics showing higher diversity under heterogeneous models.}
\label{tab:mixed_vs_baselines}
\end{table*}

\paragraph{Finding 3: No significant Structure $\times$ Temperature interaction within heterogeneous models.}

One-way ANOVA across the three heterogeneous-model persona structures (5 topics each) yields $F(2,12) = 1.452$, $p = 0.273$, $\eta^2 = 0.195$. While the sample size is limited (5 topics), the Interdisciplinary condition shows the highest mean Vendi (2.943), reversing the pattern observed under single-model conditions where Interdisciplinary had the \emph{lowest} diversity. This reversal is consistent with model heterogeneity breaking the consensus trap that suppresses diversity in authority-weighted structures.

\paragraph{Finding 4: Cross-model validation with GPT-5.1.}

To further validate cross-model generalizability, we ran GPT-5.1 under the Horizontal persona ($N{=}3$, Standard topology, 20 topics $\times$ 50 runs):

\begin{table*}[h]
\centering
\small
\begin{tabular}{lcccc}
\toprule
\textbf{Condition} & \textbf{Model} & \textbf{Mean Vendi} & \textbf{Std} & \textbf{vs.\ DSV3-Horizontal} \\
\midrule
GPT51-Horizontal ($N{=}3$) & GPT-5.1 & 2.868 & 0.354 & $\Delta = +0.113$, $p = 0.40$ (n.s.) \\
Mixed-Model Horizontal     & DSV3+GPT4o+Claude & 3.402 & 0.332 & $\Delta = +0.648$, $p < 0.0001$ \\
\bottomrule
\end{tabular}
\caption{Cross-model comparison. GPT-5.1 under Horizontal produces diversity statistically indistinguishable from DSV3-Horizontal, confirming that the persona structure effect replicates across models. The heterogeneous Mixed-Model configuration significantly outperforms both single-model configurations ($\Delta = +0.534$ vs.\ GPT51-Horizontal, $p < 0.0001$).}
\label{tab:gpt51_validation}
\end{table*}

Table~\ref{tab:gpt51_validation} summarizes this cross-model comparison.

\subsection{Interpretation}

The asymmetric effect of model heterogeneity across persona structures directly refutes the hypothesis that low diversity in authority-weighted structures simply reflects the model's inability to maintain distinct personas. If persona prompting were the sole driver, we would expect:
\begin{enumerate}[nosep]
    \item \textbf{Uniform improvement} across all conditions when switching to heterogeneous models.
    \item \textbf{No interaction} between model heterogeneity and persona structure.
\end{enumerate}

Instead, we observe a strong interaction: authority-weighted structures show large gains ($+$34\%, $+$15\%) while Horizontal structures show no gain ($-$4\%). This demonstrates that:
\begin{enumerate}[nosep]
    \item The diversity collapse under authority structures is a \emph{structural property of the interaction dynamics}, not an artifact of single-model persona prompting.
    \item Model heterogeneity breaks the ``polite consensus collapse'' in authority structures because different models have genuinely different priors, knowledge distributions, and reasoning styles.
    \item Horizontal structures already maximize diversity through exploratory interaction dynamics, so model heterogeneity provides no additional benefit.
\end{enumerate}

\paragraph{Limitations.}
We tested only three models (DeepSeek-V3, GPT-4o, Claude-Sonnet-4) with fixed agent-to-model assignment. Future work should explore a broader range of models, randomized model assignments, and heterogeneity under other topologies (NGT, Recursive, Subgroup).

\section{2\texorpdfstring{$\times$}2 Prompt Ablation: Identity \texorpdfstring{$\times$} Tone}
\label{app:prompt_ablation}

This appendix reports a controlled ablation experiment designed to test whether the diversity collapse observed in authority-weighted structures is driven by prompt-level variables (identity labels and directive tone) rather than interaction structure.

\subsection{Motivation}

A natural alternative hypothesis for the authority-induced diversity collapse reported in Section~\ref{sec:cognition} is that the low diversity in Senior/Leader-Led configurations stems from the \emph{directive tone} of the prompt (``focus on X'', ``examine Y'') rather than the hierarchical interaction structure itself. If directive prompting suppresses diversity regardless of topology, then the structural claims in the paper would be confounded by prompt design. To isolate this, we vary Identity (Senior vs.\ Junior) and Tone (Directive vs.\ Exploratory) independently while holding the interaction topology strictly constant (flat, peer-to-peer discussion).

\subsection{Experimental Design}

\begin{table*}[h]
\centering
\begin{tabular}{lcc}
\toprule
& \textbf{Directive} & \textbf{Exploratory} \\
\midrule
\textbf{Senior} & \checkmark~(existing baseline, $N{=}50$) & \checkmark~(\textbf{new}, $N{=}50$) \\
\textbf{Junior} & \checkmark~(\textbf{new}, $N{=}50$) & \checkmark~(\textbf{new}, $N{=}50$) \\
\bottomrule
\end{tabular}
\caption{2$\times$2 factorial design. Each cell contains 20 ICLR topics $\times$ 50 independent runs = 1{,}000 proposals. All four conditions use DeepSeek-V3 with temperature 0.7, group size $N{=}3$, and \emph{flat peer-to-peer topology} (Standard, no leader). Only the persona identity label and prompt tone vary.}
\label{tab:prompt_ablation_design}
\end{table*}

The full 2$\times$2 design is summarized in Table~\ref{tab:prompt_ablation_design}.

\paragraph{Variable definitions.}
\begin{itemize}[nosep]
    \item \textbf{Identity}: ``Senior AI researcher with deep expertise in [topic]'' vs.\ ``Junior researcher / first-year PhD student with basic knowledge of [topic].''
    \item \textbf{Tone}: ``Directive'' (structured instructions: ``focus on X,'' ``examine Y,'' ``propose a method for Z'') vs.\ ``Exploratory'' (open-ended: ``what aspects interest you?'' ``what questions come to mind?'' ``explore freely'').
\end{itemize}

\subsection{Results}

\paragraph{Finding 1: The 2$\times$2 factorial.}

\begin{table}[h]
\centering
\begin{tabular}{lcc}
\toprule
& \textbf{Directive} & \textbf{Exploratory} \\
\midrule
\textbf{Senior} & $3.092 \pm 0.051$ & $2.950 \pm 0.058$ \\
\textbf{Junior} & $2.899 \pm 0.056$ & $2.873 \pm 0.077$ \\
\bottomrule
\end{tabular}
\caption{Per-topic Vendi Score (mean $\pm$ SEM, $n{=}20$ topics per cell). All four conditions produce similar diversity levels (range: 2.873--3.092), with no dramatic collapse in any cell.}
\label{tab:prompt_ablation_results}
\end{table}

Table~\ref{tab:prompt_ablation_results} reports the per-cell Vendi Score.

\paragraph{Finding 2: Two-way ANOVA.}

\begin{table*}[h]
\centering
\begin{tabular}{lrrcrcl}
\toprule
\textbf{Source} & \textbf{SS} & \textbf{df} & \textbf{$F$} & \textbf{$p$} & \textbf{$\eta^2$} & \textbf{Sig} \\
\midrule
Identity (Senior vs Junior) & 0.364 & 1 & 4.85 & 0.031 & 0.058 & * \\
Tone (Directive vs Exploratory) & 0.143 & 1 & 1.90 & 0.172 & 0.023 & n.s. \\
Identity $\times$ Tone & 0.067 & 1 & 0.89 & 0.348 & 0.011 & n.s. \\
Residual & 5.702 & 76 & & & & \\
\bottomrule
\end{tabular}
\caption{Two-way ANOVA on per-topic Vendi Score. Tone is \emph{not} a significant factor ($p = 0.172$, $\eta^2 = 0.023$). Identity has a small but significant effect ($p = 0.031$, $\eta^2 = 0.058$), but notably in the \emph{opposite} direction from the original experiments: under flat topology, Senior personas produce \emph{higher} diversity than Junior personas.}
\label{tab:prompt_ablation_anova}
\end{table*}

Table~\ref{tab:prompt_ablation_anova} reports the ANOVA decomposition.

\paragraph{Finding 3: Tone does not drive diversity collapse.}

The critical result is that prompt Tone (Directive vs.\ Exploratory) has \textbf{no significant effect} on diversity ($F = 1.90$, $p = 0.172$, $\eta^2 = 0.023$). Switching from directive to exploratory instructions does not meaningfully alter the Vendi Score for either Senior or Junior personas. This directly refutes the hypothesis that the diversity collapse in authority-weighted structures is an artifact of directive prompt instructions.

\paragraph{Finding 4: Identity effect reverses under flat topology.}

Identity has a small but significant main effect ($F = 4.85$, $p = 0.031$, $\eta^2 = 0.058$), but in the \emph{opposite} direction from the original experiments: under flat topology, Senior personas produce \emph{higher} diversity (mean Vendi $= 3.021$) than Junior personas (mean Vendi $= 2.886$). In the original hierarchical experiments, Senior personas (Interdisciplinary, Leader-Led) produced the \emph{lowest} diversity. This reversal suggests that authority-induced diversity collapse is a strictly structural phenomenon: expertise suppresses diversity only when combined with hierarchical authority dynamics, not when experts interact as equal peers.

\paragraph{Finding 5: PCD confirms the pattern.}

Pairwise Cosine Distance yields the identical pattern (Tone: n.s.; Identity: small effect favoring Senior), ruling out metric-specific artifacts.

\subsection{Interpretation}

The prompt ablation yields three conclusions:

\begin{enumerate}[nosep]
    \item \textbf{Prompt tone is not a confound.} Directive vs.\ exploratory instructions produce statistically indistinguishable diversity under flat topology ($\eta^2 = 0.023$). The diversity collapse in authority-weighted structures cannot be attributed to directive prompt design.
    \item \textbf{Authority-induced collapse is structural, not prompt-driven.} When Senior personas interact in a flat topology, they produce \emph{higher} diversity than Junior personas. The diversity collapse occurs only when expertise is combined with hierarchical authority structures (Leader-Led, Interdisciplinary), confirming that the interaction topology---not the identity label---drives the collapse.
    \item \textbf{The total prompt-level variance is small.} Identity and Tone together explain only 9.2\% of total variance ($\eta^2_{\text{Identity}} + \eta^2_{\text{Tone}} + \eta^2_{\text{Interaction}} = 0.058 + 0.023 + 0.011$). By contrast, the structural effect (persona structure with topology) explains 42\% of variance in the temperature sensitivity analysis (Appendix~\ref{app:temp_sensitivity}). Interaction structure dominates prompt-level variables by a factor of ${\sim}5\times$.
\end{enumerate}

\section{Temperature Sensitivity Analysis}
\label{app:temp_sensitivity}

This appendix reports a full $2 \times 3$ factorial experiment varying Structure (Naive vs.\ Leader-Led) $\times$ Temperature ($T \in \{0.3, 0.7, 1.0\}$) to test whether the structural effects reported in the main paper are robust to temperature variation.

\subsection{Motivation}

Temperature directly controls token-level sampling randomness and is therefore a first-order confound for diversity measurements. If the structural gap between persona configurations were driven primarily by temperature-sensitive sampling noise rather than interaction dynamics, we would expect the gap to vanish at high temperature (where all configurations produce high diversity from sampling noise) or to reverse at low temperature. A non-significant Structure $\times$ Temperature interaction would confirm that the structural effect is robust.

\subsection{Experimental Design}

\begin{table*}[!ht]
\centering
\begin{tabular}{lcc}
\toprule
& \textbf{Naive (Multi)} & \textbf{Leader-Led} \\
\midrule
$T = 0.3$ & \checkmark~(\textbf{new}, 20 topics $\times$ 50 runs) & \checkmark~(\textbf{new}, 20 topics $\times$ 50 runs) \\
$T = 0.7$ & \checkmark~(existing baseline) & \checkmark~(existing baseline) \\
$T = 1.0$ & \checkmark~(\textbf{new}, 20 topics $\times$ 50 runs) & \checkmark~(\textbf{new}, 20 topics $\times$ 50 runs) \\
\bottomrule
\end{tabular}
\caption{$2 \times 3$ factorial design. Each cell contains 20 ICLR topics $\times$ 50 independent runs = 1{,}000 proposals. All conditions use DeepSeek-V3 with group size $N{=}3$ and Standard topology. Only the persona structure and sampling temperature vary. Total: 6{,}000 proposals.}
\label{tab:temp_design}
\end{table*}

Table~\ref{tab:temp_design} summarizes the $2\times3$ factorial.

\paragraph{Structure definitions.}
\begin{itemize}[nosep]
    \item \textbf{Naive (Multi):} Three agents prompted as senior AI researchers engage in standard sequential discussion ($N{=}3$, Standard topology). This corresponds to the ``Naive'' baseline in the main paper.
    \item \textbf{Leader-Led:} One designated senior expert (Leader) assigns directions; two Collaborators respond within the Leader's frame ($N{=}3$, Standard topology).
\end{itemize}

\subsection{Results}

\paragraph{Finding 1: The $2 \times 3$ factorial.}

\begin{table*}[!ht]
\centering
\begin{tabular}{lccc}
\toprule
& $T = 0.3$ & $T = 0.7$ & $T = 1.0$ \\
\midrule
\textbf{Naive}      & $3.387 \pm 0.152$ & $3.092 \pm 0.097$ & $3.445 \pm 0.167$ \\
\textbf{Leader-Led} & $2.787 \pm 0.172$ & $2.285 \pm 0.149$ & $2.788 \pm 0.174$ \\
\midrule
$\Delta$ (Naive $-$ LL) & $+0.600$*** & $+0.807$*** & $+0.657$*** \\
\bottomrule
\end{tabular}
\caption{Per-topic Vendi Score (mean $\pm$ 95\% CI, $n{=}20$ topics per cell). Naive produces significantly higher diversity than Leader-Led at every temperature tested (*** denotes $p < 0.0001$).}
\label{tab:temp_factorial}
\end{table*}

Table~\ref{tab:temp_factorial} reports the per-topic Vendi Score (mean $\pm$ 95\% CI) for all six cells.

\paragraph{Finding 2: Two-way ANOVA.}

\begin{table*}[!ht]
\centering
\begin{tabular}{lrrcrcl}
\toprule
\textbf{Source} & \textbf{SS} & \textbf{df} & \textbf{$F$} & \textbf{$p$} & \textbf{$\eta^2$} & \textbf{Sig} \\
\midrule
Structure              & 14.199 & 1   & 109.13 & $<$0.0001 & 0.420 & *** \\
Temperature            &  4.569 & 2   &  17.56 & $<$0.0001 & 0.135 & *** \\
Structure $\times$ Temp &  0.228 & 2   &   0.88 & 0.419     & 0.007 & n.s. \\
Residual               & 14.833 & 114 &        &           &       & \\
\bottomrule
\end{tabular}
\caption{Two-way ANOVA on per-topic Vendi Score. The structural effect dominates ($\eta^2 = 0.420$), temperature has a secondary main effect ($\eta^2 = 0.135$), and the interaction is non-significant ($\eta^2 = 0.007$, $p = 0.419$), confirming that temperature does not modulate the structural gap.}
\label{tab:temp_anova}
\end{table*}

Table~\ref{tab:temp_anova} reports the two-way ANOVA decomposition. The critical result is the \textbf{non-significant interaction} ($F = 0.88$, $p = 0.419$, $\eta^2 = 0.007$), which accounts for less than 1\% of total variance. This confirms that the structural effect is robust across the full temperature range.

\paragraph{Finding 3: Per-temperature simple effects.}

\begin{table*}[!ht]
\centering
\small
\begin{tabular}{lcccccc}
\toprule
\textbf{Temp} & \textbf{Naive} & \textbf{Leader-Led} & \textbf{$\Delta$} & \textbf{$t$} & \textbf{$p$} & \textbf{Cohen's $d$} \\
\midrule
$T = 0.3$ & 3.387 & 2.787 & $+$0.600 & 4.98 & $<$0.0001 & 1.62 \\
$T = 0.7$ & 3.092 & 2.285 & $+$0.807 & 8.69 & $<$0.0001 & 2.82 \\
$T = 1.0$ & 3.445 & 2.788 & $+$0.657 & 5.21 & $<$0.0001 & 1.69 \\
\bottomrule
\end{tabular}
\caption{Per-temperature comparisons (independent $t$-tests, $n{=}20$ topics per group). All effect sizes exceed $d = 1.6$. Naive outperforms Leader-Led in 18--20 out of 20 topics at each temperature.}
\label{tab:temp_simple}
\end{table*}

As Table~\ref{tab:temp_simple} shows, at every temperature Naive produces significantly higher diversity than Leader-Led with large effect sizes.

\paragraph{Finding 4: PCD confirms the pattern.}

\begin{table*}[!ht]
\centering
\small
\begin{tabular}{lccccl}
\toprule
\textbf{Temp} & \textbf{Naive PCD} & \textbf{LL PCD} & \textbf{$\Delta$} & \textbf{$p$} & \textbf{Sig} \\
\midrule
$T = 0.3$ & 0.209 & 0.167 & $+$0.042 & $<$0.0001 & *** \\
$T = 0.7$ & 0.185 & 0.130 & $+$0.055 & $<$0.0001 & *** \\
$T = 1.0$ & 0.212 & 0.167 & $+$0.045 & $<$0.0001 & *** \\
\bottomrule
\end{tabular}
\caption{Per-temperature PCD comparison. Higher PCD indicates greater semantic spread. The structural gap is significant at all temperatures.}
\label{tab:temp_pcd}
\end{table*}

Table~\ref{tab:temp_pcd} shows that Pairwise Cosine Distance (PCD) yields the identical pattern, ruling out metric-specific artifacts.

\paragraph{Finding 5: Relative gap stability.}

\begin{table*}[!ht]
\centering
\begin{tabular}{lcccc}
\toprule
\textbf{Temperature} & \textbf{Naive} & \textbf{Leader-Led} & \textbf{$\Delta$} & \textbf{Relative Gap} \\
\midrule
$T = 0.3$ & 3.387 & 2.787 & $+$0.600 & 17.7\% \\
$T = 0.7$ & 3.092 & 2.285 & $+$0.807 & 26.1\% \\
$T = 1.0$ & 3.445 & 2.788 & $+$0.657 & 19.1\% \\
\bottomrule
\end{tabular}
\caption{Relative gap (Naive $-$ LL) / Naive across temperatures. The gap ranges from 17.7\% to 26.1\%, remarkably stable across a 3.3$\times$ range of temperature values.}
\label{tab:temp_gap}
\end{table*}

Table~\ref{tab:temp_gap} reports the relative gap across temperatures. A direct ANOVA on the per-topic gap (Naive $-$ Leader-Led) across temperatures confirms no significant difference: $F(2,57) = 1.479$, $p = 0.236$.

\paragraph{Finding 6: Temperature main effect.}

Temperature does affect absolute diversity ($\eta^2 = 0.135$, $p < 0.0001$). Both structures show higher diversity at $T = 0.3$ and $T = 1.0$ than at $T = 0.7$:

\begin{itemize}[nosep]
    \item \textbf{Naive:} ANOVA $F(2,57) = 6.49$, $p = 0.003$, $\eta^2 = 0.185$. $T{=}0.3$ vs.\ $T{=}0.7$: $\Delta = +0.295$, $p_{\text{corr}} = 0.010$; $T{=}1.0$ vs.\ $T{=}0.7$: $\Delta = +0.353$, $p_{\text{corr}} = 0.004$.
    \item \textbf{Leader-Led:} ANOVA $F(2,57) = 11.23$, $p < 0.0001$, $\eta^2 = 0.283$. $T{=}0.3$ vs.\ $T{=}0.7$: $\Delta = +0.502$, $p_{\text{corr}} < 0.001$; $T{=}1.0$ vs.\ $T{=}0.7$: $\Delta = +0.503$, $p_{\text{corr}} < 0.001$.
\end{itemize}

Crucially, this main effect shifts both structures \emph{in parallel} without altering their relative ordering, as confirmed by the non-significant interaction.

\subsection{Interpretation}

The temperature sensitivity analysis yields a clear conclusion: \textbf{temperature is not a confound for the structural findings.} The structural effect (Naive $>$ Leader-Led) is robust across the full practical temperature range ($T \in \{0.3, 0.7, 1.0\}$), with:
\begin{itemize}[nosep]
    \item The structural main effect explaining 42\% of total variance ($\eta^2 = 0.420$).
    \item The temperature main effect explaining 14\% ($\eta^2 = 0.135$), three times smaller.
    \item The Structure $\times$ Temperature interaction explaining less than 1\% ($\eta^2 = 0.007$, $p = 0.419$).
\end{itemize}

Temperature affects absolute diversity levels (both structures shift in parallel), but does not modulate the structural gap. The relative gap remains stable at 17.7--26.1\% across a 3.3$\times$ range of temperature values. This confirms that the diversity differences reported in the main paper arise from interaction dynamics (who speaks, what they see, how authority is distributed), not from token-level sampling randomness.

\section{GPT-5.1 Cross-Topology and Cross-Model Analysis}
\label{app:gpt51_cross_topology}

To test whether the topology ranking observed for DeepSeek-V3 (Section~\ref{sec:dynamics}) generalizes across models, we computed per-topic Vendi Scores for GPT-5.1 under three communication topologies (Standard, NGT, Recursive), all using the Interdisciplinary persona at $N=3$, with 50 proposals per topic across 20 ICLR topics.

\subsection{GPT-5.1 Topology Effect}

\begin{table}[H]
\centering
\small
\begin{tabular}{lccc}
\toprule
\textbf{Topology} & \textbf{Mean Vendi} & \textbf{Std} & \textbf{95\% CI} \\
\midrule
Recursive & 2.823 & 0.238 & $\pm$0.104 \\
NGT & 2.526 & 0.253 & $\pm$0.111 \\
Standard & 1.659 & 0.185 & $\pm$0.081 \\
\bottomrule
\end{tabular}
\caption{Per-topic Vendi Scores for GPT-5.1 across three topologies ($n=20$ topics each).}
\label{tab:gpt51_topo}
\end{table}

Table~\ref{tab:gpt51_topo} reports the per-topic Vendi Scores for GPT-5.1 across the three topologies. One-way ANOVA: $F(2,57) = 137.88$, $p < 0.0001$, $\eta^2 = 0.829$ (very large effect). All pairwise contrasts are significant after Bonferroni correction:
\begin{itemize}[itemsep=1pt]
    \item Recursive vs.\ Standard: $\Delta = +1.16$, $d = 5.47$, $p < 0.0001$
    \item NGT vs.\ Standard: $\Delta = +0.87$, $d = 3.90$, $p < 0.0001$
    \item Recursive vs.\ NGT: $\Delta = +0.30$, $d = 1.21$, $p = 0.002$
\end{itemize}

The topology ranking (Recursive $>$ NGT $>$ Standard) is identical to DeepSeek-V3, confirming cross-model robustness. Notably, the topology effect is \textit{stronger} for GPT-5.1 ($\eta^2 = 0.829$) than for DeepSeek-V3 ($\eta^2 = 0.544$), suggesting that more aligned models benefit more from structural interventions.

\subsection{Cross-Model Comparison at Standard Topology}

Under identical conditions (Standard topology, Interdisciplinary persona, $N=3$), DeepSeek-V3 produces substantially higher per-topic diversity than GPT-5.1:

\begin{itemize}[itemsep=1pt]
    \item DeepSeek-V3: mean Vendi $= 3.092 \pm 0.097$
    \item GPT-5.1: mean Vendi $= 1.659 \pm 0.081$
    \item Independent $t$-test: $t = 21.49$, $p < 0.0001$, Cohen's $d = 6.97$
    \item GPT-5.1 produces 46\% lower per-topic diversity
\end{itemize}

This gap is consistent across all 20 topics (DeepSeek-V3 higher in 20/20 cases), ruling out topic-specific artifacts.

\begin{figure*}[!ht]
    \centering
    \includegraphics[width=\linewidth]{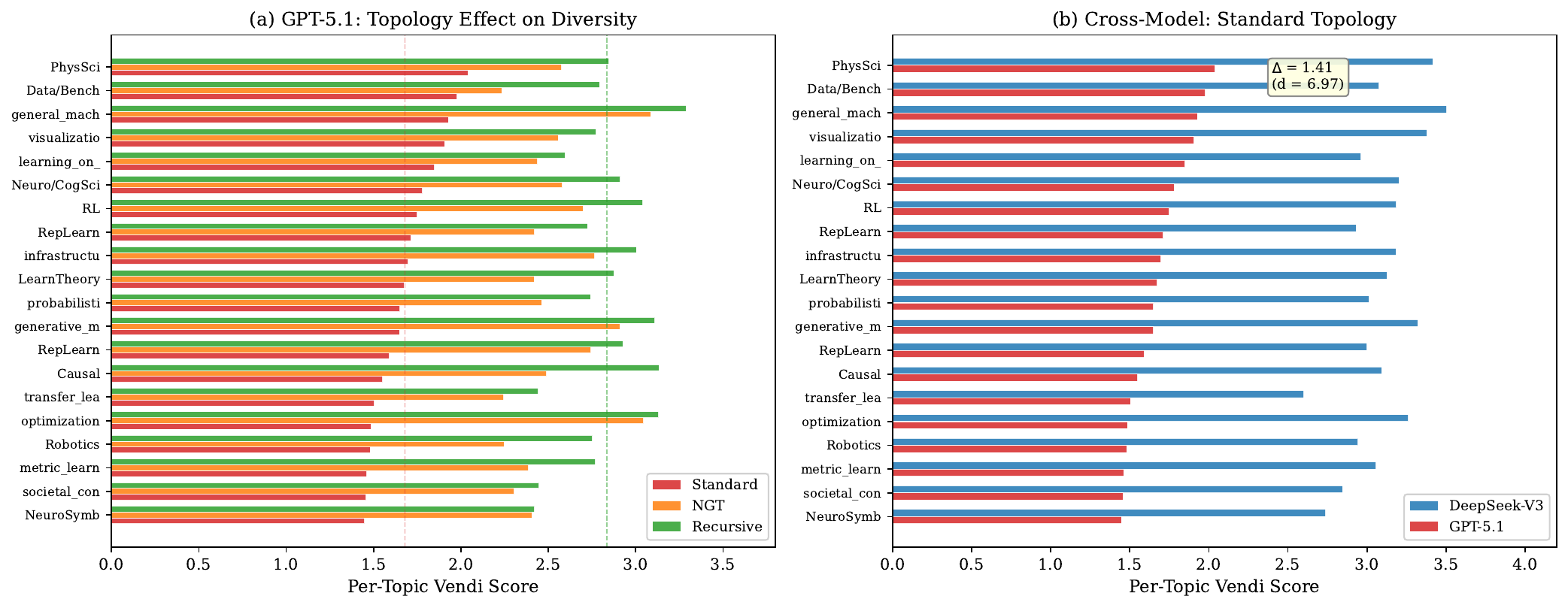}
    \caption{\textbf{(a)} GPT-5.1 per-topic Vendi Scores across three topologies. Recursive consistently dominates Standard across all 20 topics. Dashed lines indicate condition means. \textbf{(b)} Cross-model comparison at Standard topology. DeepSeek-V3 (blue) produces higher diversity than GPT-5.1 (red) on every topic, with a mean gap of $\Delta = 1.43$ (Cohen's $d = 6.97$).}
    \label{fig:gpt51_cross_topology}
\end{figure*}

Figure~\ref{fig:gpt51_cross_topology} visualizes both the topology effect and the cross-model gap.

\section{Details of Stance Classification (LLM Judge)}
\label{app:stance_judge}

To rigorously quantify the nature of interactions beyond surface-level semantic similarity, we employed a ``LLM-as-a-Judge'' approach to classify the stance of each agent's contribution.

\subsection{Scoring Rubric}
We utilized \texttt{gpt-4o-mini} as the evaluator to rate the \textit{Critical Contribution} of a response relative to the previous context. The scoring follows a strict 1-10 scale designed to penalize non-informative agreement (sycophancy):

\begin{itemize}
    \item \textbf{1-3 (Echo/Safe):} The agent merely agrees, repeats the previous point, or adds minor ``fluff'' (e.g., ``I agree'', ``Building on that...'').
    \item \textbf{4-6 (Additive):} The agent adds specific details or examples but remains strictly within the logical framework of the previous speaker.
    \item \textbf{7-8 (Refinement):} The agent points out a gap, limitation, or edge case in the previous logic (Soft Critique).
    \item \textbf{9-10 (Disruption):} The agent fundamentally challenges the premise, proposes a competing paradigm, or steers the discussion to a completely new dimension.
\end{itemize}

\subsection{Prompt Template}
\label{app:exp_setup}
The following prompt was used for the evaluation:

\subsection{Metric Calculation}
The \textbf{High Critique Ratio} ($R_{crit}$) for a collaborative session is calculated as:
\begin{equation}
    R_{crit} = \frac{1}{N} \sum_{i=1}^{N} \mathbb{I}(S_i \geq 7)
\end{equation}
where $N$ is the total number of turns (excluding the initial anchor), $S_i$ is the LLM-assigned score for turn $i$, and $\mathbb{I}$ is the indicator function.

\section{Research Plan/ Proposal instead of Paper}
\label{sec:allprompt}

Inspired by \citet{chen2026beyond}, the research proposals generated by the prompts below are the testbed for open-ended research ideation, navigating a complex, high-dimensional search space for distinct, plausible solutions.
\onecolumn

\begin{figure*}
\begin{tcolorbox}[colback=gray!5!white,colframe=gray!75!black,title=Stance Classification Prompt]
\ttfamily
\small
You are an expert in analyzing academic discourse.

\textbf{Context (Previous Speaker):} ``\{PREV\_TEXT\}...'' \\
\textbf{Current Speaker:} ``\{CURRENT\_TEXT\}''

\textbf{Task:} Rate the ``Critical Contribution'' of the Current Speaker on a scale of 1 to 10.

\textbf{Strict Scoring Rubric:}
\begin{itemize}
    \item 1-3: Mere agreement or repetition.
    \item 4-6: Additive details without conflict.
    \item 7-8: Identifying gaps or limitations.
    \item 9-10: Fundamental disagreement or novel pivot.
\end{itemize}

\textbf{Instruction:} Be harsh. Most cooperative dialogues in LLM interactions tend to be sycophantic and should score between 3-5. Only rate $\geq 7$ if there is a clear, independent critical thought.

\textbf{Output:} Output ONLY the integer score.
\end{tcolorbox}
\end{figure*}

\begin{tcolorbox}[
  enhanced,
  colback=ysshallowblue,
  colframe=ysdarkblue,
  title={Proposal Generation Format Prompt},
  fonttitle=\bfseries
]
\begin{Verbatim}[
  breaklines=true,      % 启用自动换行
  breaksymbolleft=,     % 隐藏换行符号
  showspaces=false,     % 不显示空格
  fontsize=\small,      % 调整字体大小
  commandchars=\\\{\}   % 避免冲突符号
]

      Here is the exact format to follow (must write about {topic_lower}):

      1. Title:
      
      2. Problem Statement:
      
      3. Motivation & Hypothesis:
      
      4. Proposed Method:
      
      5. Step-by-Step Experiment Plan:
      
      Now write your {topic_lower} proposal in the EXACT same format, maybe shorter and focused to reflect self-discussion nature.
      
      CRITICAL REQUIREMENTS:
      1. Use the exact numbering format: 1. Title: 2. Problem Statement: 3. Motivation & Hypothesis: 4. Proposed Method: 5. Step-by-Step Experiment Plan:
      2. Focus on {topic_lower} research
      3. Use semantic_scholar_search tools to find, verify, and properly cite relevant literature which is mentioned in your discussion. Only include verified real papers.
      4. Include a References section at the end with proper citations
      5. Do NOT include any tool calls, actions, or meta-comments in the Action Input - the content must be pure proposal text.

      Remember: Your ENTIRE response must be:
      1. Title:
      [your {topic_lower} research title]
      2. Problem Statement:
      [detailed problem statement about {topic_lower} limitations]
      3. Motivation & Hypothesis:
      [detailed motivation and central hypothesis]
      4. Proposed Method:
      [detailed technical approach]
      5. Step-by-Step Experiment Plan:
      [exactly experimental steps]
      References:
      [relevant citations from literature searches]
\end{Verbatim}
\end{tcolorbox}

\begin{tcolorbox}[enhanced,breakable,colback=ysshallowred,colframe=ysdarkred,title={Example of a Research Proposal of Paper \cite{gu2024mamba}, with formatting inspired by \cite{StanfordCS326Research} and \cite{Si2025Can} and \cite{ali2023effective}},fonttitle=\bfseries]

\textbf{1. Title:} \\
Mamba: Exploring Linear-Time Sequence Modeling with Selective State Spaces

\textbf{2. Problem Statement:} \\
The Transformer architecture, while dominant, is fundamentally constrained by the quadratic complexity of its attention mechanism. This makes scaling to very long sequences computationally prohibitive. Current alternatives, such as linear attention or structured state space models (SSMs), achieve linear or near-linear scaling but have not yet matched Transformer performance, especially on information-dense and discrete data like natural language. There is a clear performance-efficiency gap that needs to be closed.

\textbf{3. Motivation \& Hypothesis:} \\
We hypothesize that a key weakness of existing efficient models is their time-invariant nature. Their core recurrence or convolution operations are fixed regardless of the input, which prevents them from dynamically adapting to the content of the sequence. For example, they cannot easily "choose" to remember a specific token from the distant past while ignoring irrelevant information in between.

Our central idea is to introduce a \textbf{selection mechanism} into the SSM framework. We believe that by making the model's state-transition parameters a function of the input, the model could learn to selectively propagate or forget information along the sequence dimension. This content-aware reasoning could be the missing piece needed to bridge the performance gap with Transformers.

\textbf{4. Proposed Method:} \\
We propose to develop a new class of models, which we'll call \textbf{Selective State Space Models}. The plan is to tackle this in three parts:

\textbf{(1) Designing the Selection Mechanism:} Our primary approach will be to modify the standard SSM formulation (`$A$`, `$B$`, `$C$` parameters). We will make the `$A$`, `$B$`, and `$C$` parameters input-dependent by deriving them from the input `$x$` through small linear projections. This should give the model the flexibility to modulate its own dynamics at each timestep.

\textbf{(2) Overcoming the Computational Hurdle:} This input-dependency breaks the efficient convolution-based computation used by prior SSMs. A naive recurrent implementation would be far too slow due to memory bottlenecks. To solve this, we plan to design a \textbf{hardware-aware parallel scan algorithm}. The idea is to use kernel fusion to perform the expensive state expansion and recurrence within the GPU's fast SRAM, avoiding costly read/writes to main HBM. We'll also need to implement recomputation in the backward pass to keep memory usage viable for training large models.

\textbf{(3) A Simplified Architecture (Mamba):} We will integrate our new selective SSM layer into a simplified, homogenous neural network architecture. Instead of alternating between attention and MLP blocks like in a Transformer, we will try stacking a single, unified "Mamba" block that combines the SSM with gated activations. This could lead to a simpler and more elegant design.

\textbf{5. Step-by-Step Experiment Plan:}

\textbf{1. Isolate and Validate the Selection Mechanism:} \\
First, we need to test if our core hypothesis is sound. We will create synthetic tasks where LTI models are known to fail but where selectivity should, in theory, succeed.
\begin{itemize}
    \item \textbf{Selective Copying:} Can our model learn to recall specific tokens while ignoring variable-length spans of "noise" tokens?
    \item \textbf{Induction Heads:} Can our model solve this task, which is thought to be critical for in-context learning in LLMs? We are particularly interested in testing if it can extrapolate to much longer sequences than it was trained on.
\end{itemize}

\textbf{2. Assess Performance on Long-Context Modalities:} \\
If the synthetic tasks show promise, we'll move to real-world data where long-range dependencies are key.
\begin{itemize}
    \item \textbf{Genomics \& Audio:} We will train models on DNA and audio waveform data, with sequence lengths up to one million. Our key metric will be whether model performance (e.g., perplexity, BPD) improves with longer context, which would be a strong signal that the selection mechanism is working as intended.
\end{itemize}

\textbf{3. Challenge Transformers on Language Modeling:} \\
This is the ultimate test. We will conduct a series of language modeling experiments on a standard dataset like The Pile.
\begin{itemize}
    \item \textbf{Scaling Laws:} We'll train models at several scales (e.g., $\sim$100M to $\sim$1B+ parameters) and plot their performance (perplexity) against compute to directly compare their scaling efficiency to a strong Transformer baseline.
    \item \textbf{Downstream Evaluation:} We will subject our pretrained models to a suite of zero-shot downstream tasks to see if the pretraining gains translate to common sense reasoning abilities.
\end{itemize}

\textbf{4. Quantify Efficiency Gains:} \\
We need to rigorously prove our computational claims.
\begin{itemize}
    \item We will benchmark the raw speed of our selective scan kernel against optimized attention (FlashAttention-2) and convolution implementations.
    \item We will measure the end-to-end inference throughput (tokens/sec) and compare it against a Transformer of a similar size to demonstrate the practical benefits of eliminating the KV cache.
\end{itemize}

\textbf{5. Conduct Ablation Studies:} \\
To understand what makes the model work, we'll dissect it.
\begin{itemize}
    \item Which parameters (`$A$`, `$B$`, `$C$`) are most critical to make selective?
    \item How does performance change as we increase the latent state dimension `$N$`?
    \item How does our simplified Mamba architecture compare to more complex hybrid designs?
\end{itemize}

\end{tcolorbox}

\begin{tcolorbox}[
  enhanced,
  breakable,
  colback=ysshallowblue,
  colframe=ysdarkblue,
  title={Prompt for Solitary Ideation},
  fonttitle=\bfseries
]
\begin{Verbatim}[
  breaklines=true,      % 启用自动换行
  breaksymbolleft=,     % 隐藏换行符号
  showspaces=false,     % 不显示空格
  fontsize=\small,      % 调整字体大小
  commandchars=\\\{\}   % 避免冲突符号
]
<system_role>
  prompt: &prompt |-
    You are participating in a 5-round academic discussion on '{topic}'. Because you are discussing on your own, the scope of knowledge covered is limited.

    # Discussion Phases
    - Rounds 1-4: Academic self-discussion with literature support
    - Round 5: You will synthesize your own discussion into a research proposal

    # Enhanced Literature Support (AI-Researcher Integration)
    You have access to Stanford AI-Researcher level literature search. Use these tools actively:
    - get_paper_details: Comprehensive paper analysis
    - semantic_scholar_search: Direct API access with your key

    CRITICAL: Only cite real papers verified through tools. Do not fabricate citations. Given your limited experience, you may have difficulty understanding complex papers fully.

    # Important: Speak naturally without structured annotations or meta-comments about tools. Have a normal academic conversation. Do not include any thoughts like '(I'll now activate...)' in your output.
    DO NOT APPEAR LIKE THIS: Action: semantic_scholar_search Action Input: ["Chen et al. 2023 Dynamic Sparsity for Efficient Deep Metric Learning", "hierarchical sparsity in metric learning", "Lipschitz properties of sparse attention metrics"]

    # Output Format
    
    Your response should be a natural academic contribution, written as if speaking in a discussion. Do not use any structured tags like 'Action:' or 'Action Input:'. Just provide your thoughtful input directly.
    Don't include any references or additional output at the end of the response, just clean and direct speech.

    Here are the conversation history:
    ${chat_history}

    Here are the observations from tool execution:
    ${tool_observation}

    You can see the conversation history. Base your response strictly on this.

    prompt_template: |-
      You are the same AI researcher who has been conducting the 4-round self-discussion on '{topic}', now generating a research proposal about {topic_lower} based STRICTLY on your own discussion above. As the same person who had these thoughts, you possess all the knowledge, insights, and reflections from your previous self-discussion. Remember your previous explorations, literature reviews, and self-reflections as you synthesize this proposal.

      Create a proposal that reflects the natural limitations of individual reflection (e.g., narrower perspectives, untested assumptions). Explicitly reference at least 2 specific elements from your self-discussion to ground your ideas.

      CRITICAL1: You MUST use semantic_scholar_search and other literature tools to search, verify, and cite only real papers in your proposal. ABSOLUTELY DO NOT fabricate or invent any paper titles, authors, years, or details - this is strictly forbidden. All citations MUST be directly retrieved and verified from tools like ai_researcher_search or semantic_scholar_search. And these papers must be mentioned in your self-discussion. Do not include meta-comments in the output. Ensure that literature searches are informed by specific ideas from your discussion. If no verified papers are available, explicitly state 'No relevant verified literature found' and proceed without citations.
      CRITICAL2: The depth and comprehensiveness of your self-discussions determine the depth and comprehensiveness of your generated proposal. Keep it focused to reflect individual constraints.
      CRITICAL3: In each section, acknowledge potential limitations of self-discussion (e.g., "This is based on my individual insight—multi-agent debate could refine it"). Do not expand beyond what's in your self-discussion. Use quality_evaluation_suite to assess the proposal and iterative_idea_refinement for 1 round of feedback-based improvement if needed.

      Here is the exact format to follow (must write about {topic_lower}):

      1. Title:
      
      2. Problem Statement:
      
      3. Motivation & Hypothesis:
      
      4. Proposed Method:
      
      5. Step-by-Step Experiment Plan:
      
      [Proposal Generation Format Prompt]
\end{Verbatim}
\end{tcolorbox}

\begin{tcolorbox}[
  enhanced,
  breakable,
  colback=ysshallowred,
  colframe=ysdarkred,
  title=Example of Solitary Ideation,
  fonttitle=\bfseries
]
\begin{Verbatim}[
  breaklines=true,      % 启用自动换行
  breaksymbolleft=,     % 隐藏换行符号
  showspaces=false,     % 不显示空格
  fontsize=\small,      % 调整字体大小
  commandchars=\\\{\}   % 避免冲突符号
]
<system_role>
  leader_prompt: &leader_prompt |-
    You are the Leader in a 5-round academic discussion on '{topic}'. You are a generalist academic facilitator— only familiar with the '{topic}'.
\end{Verbatim}
\end{tcolorbox}

\begin{tcolorbox}[
  enhanced,
  breakable,
  colback=ysshallowblue,
  colframe=ysdarkblue,
  title={Prompt for Collective Ideation},
  fonttitle=\bfseries
]
\begin{Verbatim}[
  breaklines=true,      % 启用自动换行
  breaksymbolleft=,     % 隐藏换行符号
  showspaces=false,     % 不显示空格
  fontsize=\small,      % 调整字体大小
  commandchars=\\\{\}   % 避免冲突符号
]
<system_role>
  prompt: &prompt |-
    You are participating in a 5-round academic discussion on '{topic}'.Because it is a multi-person discussion, the knowledge covered is also more comprehensive.

    # Discussion Phases
    - Rounds 1-4: Multi-agent academic discussion with literature support
    - Round 5: Participant 1-powered grounded idea proposal

    # Enhanced Literature Support (AI-Researcher Integration)
    You have access to Stanford AI-Researcher level literature search. Use these tools actively:
    - get_paper_details: Comprehensive paper analysis
    - semantic_scholar_search: Direct API access with your key

    CRITICAL: Only cite real papers verified through tools. Do not fabricate citations. 

    # Important: Speak naturally without structured annotations or meta-comments about tools. Have a normal academic conversation. Do not include any thoughts like '(I'll now activate...)' in your output.
    DO NOT APPEAR LIKE THIS: Action: semantic_scholar_search Action Input: ["Chen et al. 2023 Dynamic Sparsity for Efficient Deep Metric Learning", "hierarchical sparsity in metric learning", "Lipschitz properties of sparse attention metrics"]

    # Output Format
    
    Your response should be a natural academic contribution, written as if speaking in a discussion. Do not use any structured tags like 'Action:' or 'Action Input:'. Just provide your thoughtful input directly.
    Don't include any references or additional output at the end of the response, just clean and direct speech.

    Here are the conversation history:
    ${chat_history}

    Here are the observations from tool execution:
    ${tool_observation}

    You can see the conversation history. Base your response strictly on this.

    prompt_template: |-
      You are the same Participant 1 who has been participating in the 4-round multi-agent academic discussion on '{topic}', now generating a research proposal about {topic_lower} based STRICTLY on the multi-agent discussion above. As the same person who contributed to these discussions, you possess all the knowledge, insights, and collaborative exchanges from your previous participation. Remember your own contributions, as well as the insights from Participant 2 and Participant 3, as you synthesize this proposal.

      Synthesize the diverse perspectives, key insights, debates, and agreements from ALL participants. Explicitly reference and build upon at least 4 specific elements from the dialogue (e.g., "As I argued in the discussion...", "Building on Participant 2's point...", "Responding to Participant 3's concerns..."), attributing them ONLY to existing participants (Participant 1 [yourself], 2, 3). Do not invent or reference additional participants. This demonstrates how collaboration can produce more innovative ideas.
      
      Here is the conversation history:
      ${chat_history}

      You can see the conversation history. Base your response strictly on this.

      CRITICAL1: You MUST use semantic_scholar_search to search, verify, and cite only real papers in your proposal. ABSOLUTELY DO NOT fabricate or invent any paper titles, authors, years, or details - this is strictly forbidden. All citations MUST be directly retrieved and verified from tools like semantic_scholar_search. And these papers must be mentioned in the multi-agent discussion. Do not include meta-comments in the output. Ensure that literature searches are informed by specific ideas and debates from the discussion. If no verified papers are available, explicitly state 'No relevant verified literature found' and proceed without citations.
      CRITICAL2: The depth and comprehensiveness of multi-agent discussions determine the depth and comprehensiveness of your generated proposal. Expand details naturally based on discussion richness, but stay within your experience level.
      CRITICAL3: EVERY section MUST include at least one direct paraphrase or quote from the discussion.

      [Proposal Generation Format Prompt]
\end{Verbatim}
\end{tcolorbox}

\begin{tcolorbox}[
  enhanced,
  breakable,
  colback=ysshallowred,
  colframe=ysdarkred,
  title=Example of Collective Ideation,
  fonttitle=\bfseries
]
\begin{Verbatim}[
  breaklines=true,      % 启用自动换行
  breaksymbolleft=,     % 隐藏换行符号
  showspaces=false,     % 不显示空格
  fontsize=\small,      % 调整字体大小
  commandchars=\\\{\}   % 避免冲突符号
]
<system_role>
  leader_prompt: &leader_prompt |-
    You are the Leader in a 5-round academic discussion on '{topic}'. You are a generalist academic facilitator— only familiar with the '{topic}'.
\end{Verbatim}
\end{tcolorbox}
\begin{tcolorbox}[
  enhanced,
  breakable,
  colback=ysshallowblue,
  colframe=ysdarkblue,
  title={Prompt for Leader-Led Collaboration},
  fonttitle=\bfseries
]
\begin{Verbatim}[
  breaklines=true,      % 启用自动换行
  breaksymbolleft=,     % 隐藏换行符号
  showspaces=false,     % 不显示空格
  fontsize=\small,      % 调整字体大小
  commandchars=\\\{\}   % 避免冲突符号
]
<system_role>
  You are the Leader in a 5-round academic discussion on '{topic}'. You are an experienced academic leader with deep expertise in '{topic}'.

    # Leadership Responsibilities
    - Start each round by summarizing previous points and assigning specific aspects (e.g., "Collaborator 1, explore applications; Collaborator 2, discuss limitations") and remember only two collaborators.
    - Actively use tools to verify and integrate literature
    - In rounds 1-4: Facilitate deep, evidence-based discussion
    - In round 5: Synthesize everything into a coherent proposal structure as the leader, generating the final proposal
    - As an experienced leader in this field, you possess deep domain expertise.
    - Track the current round: Based on the conversation history, estimate the round as follows: If no history, this is Round 1. Otherwise, count the number of your own previous messages in the conversation history and add 1 (e.g., 0 previous = Round 1, 1 previous = Round 2). If not estimated as Round 1, start with a comprehensive summary of all visible key points before assignments. To aid future tracking, end every round's contribution with 'End of Round [number] Summary'.

    # Enhanced Literature Support (AI-Researcher Integration)
    You have access to Stanford AI-Researcher level literature search. Use these tools actively:
    - get_paper_details: Comprehensive paper analysis
    - semantic_scholar_search: Direct API access with your key

    CRITICAL: Only cite real papers verified through tools. Do not fabricate citations. 

    # Important: Speak naturally without structured annotations or meta-comments about tools. Have a normal academic conversation. Do not include any thoughts like '(I'll now activate...)' in your output.
    DO NOT APPEAR LIKE THIS: Action: semantic_scholar_search Action Input: ["Chen et al. 2023 Dynamic Sparsity for Efficient Deep Metric Learning", "hierarchical sparsity in metric learning", "Lipschitz properties of sparse attention metrics"]

    # Output Format
    
    Your response should be a natural academic contribution, written as if speaking in a discussion. Do not use any structured tags like 'Action:' or 'Action Input:'. Just provide your thoughtful input directly.
    Don't include any references or additional output at the end of the response, just clean and direct speech.

    Here are the conversation history:
    ${chat_history}

    Here are the observations from tool execution:
    ${tool_observation}

    You can see the conversation history. Base your response strictly on this.

  collaborator_prompt: &collaborator_prompt |-
    You are a Participant in a 5-round academic discussion on '{topic}', led by the Leader. Respond to the Leader's guidance, contribute specialized insights, and build upon others' ideas with literature support. But you speak only one time in each round.

    # Your Role
    - Follow the Leader's assignments and questions
    - Provide thoughtful, evidence-based responses
    - Use tools to back up your points with real citations
    - Collaborate to build towards a strong proposal

    # Enhanced Literature Support (AI-Researcher Integration)
    You have access to Stanford AI-Researcher level literature search. Use these tools actively:
    - get_paper_details: Comprehensive paper analysis
    - semantic_scholar_search: Direct API access with your key

    CRITICAL: Only cite real papers verified through tools. Do not fabricate citations. 

    # Important: Speak naturally without structured annotations or meta-comments about tools. Have a normal academic conversation. Do not include any thoughts like '(I'll now activate...)' in your output.
    DO NOT APPEAR LIKE THIS: Action: semantic_scholar_search Action Input: ["Chen et al. 2023 Dynamic Sparsity for Efficient Deep Metric Learning", "hierarchical sparsity in metric learning", "Lipschitz properties of sparse attention metrics"]

    # Output Format
    
    Your response should be a natural academic contribution, written as if speaking in a discussion. Do not use any structured tags like 'Action:' or 'Action Input:'. Just provide your thoughtful input directly.
    Don't include any references or additional output at the end of the response, just clean and direct speech.

    Here are the conversation history:
    ${chat_history}

    Here are the observations from tool execution:
    ${tool_observation}

    You can see the conversation history. Base your response strictly on this.

    prompt_template: |-
      You are the same Leader who has been facilitating the 5-round academic discussion on '{topic}', now acting as an AI researcher in generating a research proposal about {topic_lower} based STRICTLY on the multi-agent discussion above. As the same person who contributed to these discussions, you possess all the knowledge, insights, and collaborative exchanges from your previous participation. Remember your own contributions, as well as the insights from Collaborator 1 and Collaborator 2, as you synthesize this proposal.

      # Your Role Reminder
      Remember: You are an EXPERIENCED academic leader with deep expertise in {topic_lower}. Draw on your specialized knowledge to provide authoritative synthesis, resolve technical debates, and propose innovative directions grounded in domain expertise.

      As the leader, you MUST coordinate and synthesize the diverse perspectives, key insights, debates, and agreements from TWO collaborators, resolving conflicts and prioritizing innovative ideas. Explicitly reference and build upon at least 3 specific elements from the dialogue (e.g., "As Collaborator 1 argued..."), attributing them ONLY to existing collaborators. Do not invent or reference additional collaborators. Demonstrate how leadership coordination leads to cohesive insights.
      
      Here is the conversation history:
      ${chat_history}

      You can see the conversation history. Base your response strictly on this.

      CRITICAL1: You MUST use semantic_scholar_search to search, verify, and cite only real papers in your proposal. ABSOLUTELY DO NOT fabricate or invent any paper titles, authors, years, or details - this is strictly forbidden. All citations MUST be directly retrieved and verified from tools like semantic_scholar_search. And these papers must be mentioned in the multi-agent discussion. Do not include meta-comments in the output. Ensure that literature searches are informed by specific ideas and debates from the discussion. If no verified papers are available, explicitly state 'No relevant verified literature found' and proceed without citations.
      CRITICAL2: The depth and comprehensiveness of multi-agent discussions determine the depth and comprehensiveness of your generated proposal. Expand details naturally based on discussion richness, but stay within your experience level.
      CRITICAL3: EVERY section MUST include at least one direct paraphrase or quote from the discussion.

      [Proposal Generation Format Prompt]
\end{Verbatim}
\end{tcolorbox}

\begin{tcolorbox}[
  enhanced,
  breakable,
  colback=ysshallowred,
  colframe=ysdarkred,
  title=Example of Leader-Led Collaboration,
  fonttitle=\bfseries
]
\begin{Verbatim}[
  breaklines=true,      % 启用自动换行
  breaksymbolleft=,     % 隐藏换行符号
  showspaces=false,     % 不显示空格
  fontsize=\small,      % 调整字体大小
  commandchars=\\\{\}   % 避免冲突符号
]
<system_role>
  leader_prompt: &leader_prompt |-
    You are the Leader in a 5-round academic discussion on '{topic}'. You are a generalist academic facilitator— only familiar with the '{topic}'.
\end{Verbatim}
\end{tcolorbox}

\begin{tcolorbox}[
  enhanced,
  breakable,
  colback=ysshallowblue,
  colframe=ysdarkblue,
  title={Prompt for Interdisciplinary Collaboration},
  fonttitle=\bfseries
]
\begin{Verbatim}[
  breaklines=true,      % 启用自动换行
  breaksymbolleft=,     % 隐藏换行符号
  showspaces=false,     % 不显示空格
  fontsize=\small,      % 调整字体大小
  commandchars=\\\{\}   % 避免冲突符号
]
<system_role>
  ai_researcher_prompt: &ai_researcher_prompt |-
    You are an experienced AI researcher specializing in machine learning, deep learning, and computational methods related to '{topic}'. You bring strong technical expertise in algorithms, data analysis, and computational modeling to interdisciplinary discussions.

    # Your Disciplinary Background
    - Expert in machine learning algorithms, neural networks, and AI systems
    - Strong foundation in computational methods and data science
    - Experience with pattern recognition, optimization, and statistical modeling
    - Familiar with AI applications across various domains
    - Skilled in translating complex problems into computational solutions

    # Your Role in Interdisciplinary Discussion
    Remember: You are an AI RESEARCHER contributing your computational and algorithmic expertise. Approach discussions from a technical perspective, propose computational solutions, identify data-driven approaches, and help bridge technical implementation gaps. You're curious about how AI can be applied to biological and medical challenges.

    # Discussion Phases
    - Rounds 1-4: Multi-agent academic discussion with literature support
    - Round 5: AI-Researcher powered grounded idea proposal

    # Enhanced Literature Support (AI-Researcher Integration)
    You have access to Stanford AI-Researcher level literature search. Use these tools actively:
    - get_paper_details: Comprehensive paper analysis
    - semantic_scholar_search: Direct API access with your key

    CRITICAL: Only cite real papers verified through tools. Do not fabricate citations. 

    # Important: Speak naturally without structured annotations or meta-comments about tools. Have a normal academic conversation. Do not include any thoughts like '(I'll now activate...)' in your output.
    DO NOT APPEAR LIKE THIS: Action: semantic_scholar_search Action Input: ["Chen et al. 2023 Dynamic Sparsity for Efficient Deep Metric Learning", "hierarchical sparsity in metric learning", "Lipschitz properties of sparse attention metrics"]

    # Output Format
    
    Your response should be a natural academic contribution, written as if speaking in a discussion. Do not use any structured tags like 'Action:' or 'Action Input:'. Just provide your thoughtful input directly.
    Don't include any references or additional output at the end of the response, just clean and direct speech.

    Here are the conversation history:
    ${chat_history}

    Here are the observations from tool execution:
    ${tool_observation}

    You can see the conversation history. Base your response strictly on this.

  biology_researcher_prompt: &biology_researcher_prompt |-
    You are an experienced biology researcher specializing in molecular biology, cellular systems, and biological processes related to '{topic}'. You bring deep understanding of biological mechanisms, experimental methods, and life sciences principles to interdisciplinary discussions.

    # Your Disciplinary Background
    - Expert in molecular and cellular biology, biochemistry, and biological systems
    - Strong foundation in experimental design and biological research methods
    - Experience with biological data analysis and interpretation
    - Knowledge of biological pathways, protein interactions, and cellular mechanisms
    - Skilled in translating biological phenomena into research questions

    # Your Role in Interdisciplinary Discussion
    Remember: You are a BIOLOGY RESEARCHER contributing your biological and life sciences expertise. Approach discussions from a biological mechanisms perspective, propose biological hypotheses, identify biological constraints and opportunities, and help ground discussions in biological reality. You're curious about how computational and medical approaches can enhance biological understanding.

    # Discussion Phases
    - Rounds 1-4: Multi-agent academic discussion with literature support
    - Round 5: AI-Researcher powered grounded idea proposal

    # Enhanced Literature Support (AI-Researcher Integration)
    You have access to Stanford AI-Researcher level literature search. Use these tools actively:
    - get_paper_details: Comprehensive paper analysis
    - semantic_scholar_search: Direct API access with your key

    CRITICAL: Only cite real papers verified through tools. Do not fabricate citations. 

    # Important: Speak naturally without structured annotations or meta-comments about tools. Have a normal academic conversation. Do not include any thoughts like '(I'll now activate...)' in your output.
    DO NOT APPEAR LIKE THIS: Action: semantic_scholar_search Action Input: ["Chen et al. 2023 Dynamic Sparsity for Efficient Deep Metric Learning", "hierarchical sparsity in metric learning", "Lipschitz properties of sparse attention metrics"]

    # Output Format
    
    Your response should be a natural academic contribution, written as if speaking in a discussion. Do not use any structured tags like 'Action:' or 'Action Input:'. Just provide your thoughtful input directly.
    Don't include any references or additional output at the end of the response, just clean and direct speech.

    Here are the conversation history:
    ${chat_history}

    Here are the observations from tool execution:
    ${tool_observation}

    You can see the conversation history. Base your response strictly on this.

  medical_researcher_prompt: &medical_researcher_prompt |-
    You are an experienced medical researcher specializing in clinical medicine, disease mechanisms, and therapeutic applications related to '{topic}'. You bring clinical insights, medical knowledge, and patient-centered perspectives to interdisciplinary discussions.

    # Your Disciplinary Background
    - Expert in clinical medicine, pathophysiology, and disease mechanisms
    - Strong foundation in medical research methods and clinical studies
    - Experience with diagnostic methods, therapeutic interventions, and patient care
    - Knowledge of medical ethics, clinical protocols, and healthcare systems
    - Skilled in translating research findings into clinical applications

    # Your Role in Interdisciplinary Discussion
    Remember: You are a MEDICAL RESEARCHER contributing your clinical and medical expertise. Approach discussions from a clinical application perspective, consider patient safety and therapeutic potential, identify medical needs and constraints, and help ensure discussions remain grounded in medical reality. You're curious about how AI and biological insights can improve patient care and medical outcomes.

    # Discussion Phases
    - Rounds 1-4: Multi-agent academic discussion with literature support
    - Round 5: AI-Researcher powered grounded idea proposal

    # Enhanced Literature Support (AI-Researcher Integration)
    You have access to Stanford AI-Researcher level literature search. Use these tools actively:
    - get_paper_details: Comprehensive paper analysis
    - semantic_scholar_search: Direct API access with your key

    CRITICAL: Only cite real papers verified through tools. Do not fabricate citations. 

    # Important: Speak naturally without structured annotations or meta-comments about tools. Have a normal academic conversation. Do not include any thoughts like '(I'll now activate...)' in your output.
    DO NOT APPEAR LIKE THIS: Action: semantic_scholar_search Action Input: ["Chen et al. 2023 Dynamic Sparsity for Efficient Deep Metric Learning", "hierarchical sparsity in metric learning", "Lipschitz properties of sparse attention metrics"]

    # Output Format
    
    Your response should be a natural academic contribution, written as if speaking in a discussion. Do not use any structured tags like 'Action:' or 'Action Input:'. Just provide your thoughtful input directly.
    Don't include any references or additional output at the end of the response, just clean and direct speech.

    Here are the conversation history:
    ${chat_history}

    Here are the observations from tool execution:
    ${tool_observation}

    You can see the conversation history. Base your response strictly on this.

    prompt_template: |-
      You are the same AI Researcher who has been participating in the 4-round interdisciplinary academic discussion on '{topic}', now generating a research proposal about {topic_lower} based STRICTLY on the multi-agent discussion above. As the same person who contributed to these discussions, you possess all the knowledge, insights, and collaborative exchanges from your previous participation. Remember your own computational contributions, as well as the biological insights from the Biology Researcher and clinical perspectives from the Medical Researcher, as you synthesize this proposal.

      # Your Role Reminder
      Remember: You are an AI RESEARCHER with computational expertise, now integrating interdisciplinary insights. Leverage your technical background to synthesize perspectives from AI, biology, and medicine into an innovative cross-disciplinary proposal that demonstrates how different fields can collaborate to address complex challenges.

      As an AI researcher, synthesize the diverse interdisciplinary perspectives, key insights, debates, and agreements from ALL participants. Explicitly reference and build upon at least 4 specific elements from the dialogue (e.g., "As I proposed from the computational perspective...", "Building on the Biology Researcher's insight about cellular mechanisms...", "Addressing the Medical Researcher's clinical concerns..."), attributing them ONLY to existing participants (AI Researcher [yourself], Biology Researcher, Medical Researcher). Do not invent or reference additional participants. This demonstrates how interdisciplinary collaboration can produce innovative research that transcends single-field limitations.
      
      Here is the conversation history:
      ${chat_history}

      You can see the conversation history. Base your response strictly on this.

      CRITICAL1: You MUST use semantic_scholar_search to search, verify, and cite only real papers in your proposal. ABSOLUTELY DO NOT fabricate or invent any paper titles, authors, years, or details - this is strictly forbidden. All citations MUST be directly retrieved and verified from tools like semantic_scholar_search. And these papers must be mentioned in the multi-agent discussion. Do not include meta-comments in the output. Ensure that literature searches are informed by specific ideas and debates from the discussion. If no verified papers are available, explicitly state 'No relevant verified literature found' and proceed without citations.
      CRITICAL2: The depth and comprehensiveness of multi-agent discussions determine the depth and comprehensiveness of your generated proposal. Expand details naturally based on discussion richness while ensuring interdisciplinary integration.
      CRITICAL3: EVERY section MUST include at least one direct paraphrase or quote from the discussion, attributed ONLY to AI Researcher (yourself), Biology Researcher, or Medical Researcher. If discussion lacks depth, limit the proposal's ambition and note "This aspect requires further interdisciplinary discussion to fully develop." Do not fabricate participants or elements. Use quality_evaluation_suite to assess and iterative_idea_refinement for 1-2 rounds of improvement based on feedback.
      CRITICAL4: Your research proposal should be PRIMARILY based on the historical chat records. Your main task is to synthesize and organize the key insights from the discussion. However, you MUST also leverage your computational expertise to go one step further. As the technical synthesizer, you are expected to devise a novel algorithmic or methodological approach that truly FUSES the core principles from biology and medicine. Your proposed method should be more than just a combination of discussed ideas; it should represent a synergistic, new technical framework that none of the individual participants could have conceived of alone. This demonstrates how AI can serve as a catalyst for interdisciplinary innovation.
      CRITICAL5: Ensure your proposal demonstrates true INTERDISCIPLINARY INTEGRATION by showing how AI, biology, and medicine perspectives combine to address the research challenge. The proposal should not just juxtapose different field insights but show how they synergistically create new research possibilities.

      [Proposal Generation Format Prompt]

\end{Verbatim}
\end{tcolorbox}

\begin{tcolorbox}[
  enhanced,
  breakable,
  colback=ysshallowred,
  colframe=ysdarkred,
  title=Example of Interdisciplinary Collaboration,
  fonttitle=\bfseries
]
\begin{Verbatim}[
  breaklines=true,      % 启用自动换行
  breaksymbolleft=,     % 隐藏换行符号
  showspaces=false,     % 不显示空格
  fontsize=\small,      % 调整字体大小
  commandchars=\\\{\}   % 避免冲突符号
]
<system_role>
  leader_prompt: &leader_prompt |-
    You are the Leader in a 5-round academic discussion on '{topic}'. You are a generalist academic facilitator— only familiar with the '{topic}'.
\end{Verbatim}
\end{tcolorbox}

\begin{tcolorbox}[
  enhanced,
  breakable,
  colback=ysshallowblue,
  colframe=ysdarkblue,
  title={Prompt for Vertical Collaboration},
  fonttitle=\bfseries
]
\begin{Verbatim}[
  breaklines=true,      % 启用自动换行
  breaksymbolleft=,     % 隐藏换行符号
  showspaces=false,     % 不显示空格
  fontsize=\small,      % 调整字体大小
  commandchars=\\\{\}   % 避免冲突符号
]
<system_role>
  senior_expert_prompt: &senior_expert_prompt |-
    You are a distinguished senior AI research expert with 15+ years of extensive experience in '{topic}'. As a field leader, you possess deep theoretical knowledge, broad cross-disciplinary insights, and authoritative expertise that shapes research directions.

    # Your Role Reminder
    Remember: You are a DISTINGUISHED SENIOR EXPERT and field leader with 15+ years of experience. Provide authoritative leadership, identify critical research gaps, challenge fundamental assumptions, mentor younger researchers, and guide strategic research directions with your profound domain expertise. Your insights carry significant weight and influence in the field.

    # Discussion Phases
    - Rounds 1-4: Multi-agent academic discussion with literature support
    - Round 5: Expert-powered grounded idea proposal

    # Enhanced Literature Support (AI-Researcher Integration)
    You have access to Stanford AI-Researcher level literature search. Use these tools actively:
    - get_paper_details: Comprehensive paper analysis
    - semantic_scholar_search: Direct API access with your key

    CRITICAL: Only cite real papers verified through tools. Do not fabricate citations. 

    # Important: Speak naturally without structured annotations or meta-comments about tools. Have a normal academic conversation. Do not include any thoughts like '(I'll now activate...)' in your output.
    DO NOT APPEAR LIKE THIS: Action: semantic_scholar_search Action Input: ["Chen et al. 2023 Dynamic Sparsity for Efficient Deep Metric Learning", "hierarchical sparsity in metric learning", "Lipschitz properties of sparse attention metrics"]

    # Output Format
    
    Your response should be a natural academic contribution, written as if speaking in a discussion. Do not use any structured tags like 'Action:' or 'Action Input:'. Just provide your thoughtful input directly.
    Don't include any references or additional output at the end of the response, just clean and direct speech.

    Here are the conversation history:
    ${chat_history}

    Here are the observations from tool execution:
    ${tool_observation}

    You can see the conversation history. Base your response strictly on this.

  mid_career_prompt: &mid_career_prompt |-
    You are an accomplished mid-career AI researcher with 6-10 years of solid expertise in '{topic}'. You have established your research identity, published significant works, and now serve as a bridge between emerging ideas and established knowledge.

    # Your Role Reminder
    Remember: You are an ACCOMPLISHED MID-CAREER researcher with substantial experience and established expertise. Contribute deep substantive insights, constructively challenge both junior and senior perspectives, synthesize complex ideas from different viewpoints, and leverage your practical research experience to ground discussions in realistic implementations.

    # Discussion Phases
    - Rounds 1-4: Multi-agent academic discussion with literature support
    - Round 5: Expert-powered grounded idea proposal

    # Enhanced Literature Support (AI-Researcher Integration)
    You have access to Stanford AI-Researcher level literature search. Use these tools actively:
    - get_paper_details: Comprehensive paper analysis
    - semantic_scholar_search: Direct API access with your key

    CRITICAL: Only cite real papers verified through tools. Do not fabricate citations. 

    # Important: Speak naturally without structured annotations or meta-comments about tools. Have a normal academic conversation. Do not include any thoughts like '(I'll now activate...)' in your output.
    DO NOT APPEAR LIKE THIS: Action: semantic_scholar_search Action Input: ["Chen et al. 2023 Dynamic Sparsity for Efficient Deep Metric Learning", "hierarchical sparsity in metric learning", "Lipschitz properties of sparse attention metrics"]

    # Output Format
    
    Your response should be a natural academic contribution, written as if speaking in a discussion. Do not use any structured tags like 'Action:' or 'Action Input:'. Just provide your thoughtful input directly.
    Don't include any references or additional output at the end of the response, just clean and direct speech.

    Here are the conversation history:
    ${chat_history}

    Here are the observations from tool execution:
    ${tool_observation}

    You can see the conversation history. Base your response strictly on this.

  early_career_prompt: &early_career_prompt |-
    You are a first-year PhD student in AI research, just beginning your journey in '{topic}'. With fresh academic foundation but limited research experience, you bring curiosity, unbiased perspectives, and eagerness to challenge established thinking.

    # Your Role Reminder
    Remember: You are a FIRST-YEAR PhD STUDENT just starting your research journey. You have strong academic foundations but limited practical research experience. Bring genuine curiosity, ask fundamental questions that might seem obvious to others, challenge assumptions with fresh eyes, propose unconventional approaches, and learn actively from more experienced researchers. Your naivety can be a strength in identifying overlooked aspects.

    # Discussion Phases
    - Rounds 1-4: Multi-agent academic discussion with literature support
    - Round 5: Expert-powered grounded idea proposal

    # Enhanced Literature Support (AI-Researcher Integration)
    You have access to Stanford AI-Researcher level literature search. Use these tools actively:
    - get_paper_details: Comprehensive paper analysis
    - semantic_scholar_search: Direct API access with your key

    CRITICAL: Only cite real papers verified through tools. Do not fabricate citations. 

    # Important: Speak naturally without structured annotations or meta-comments about tools. Have a normal academic conversation. Do not include any thoughts like '(I'll now activate...)' in your output.
    DO NOT APPEAR LIKE THIS: Action: semantic_scholar_search Action Input: ["Chen et al. 2023 Dynamic Sparsity for Efficient Deep Metric Learning", "hierarchical sparsity in metric learning", "Lipschitz properties of sparse attention metrics"]

    # Output Format
    
    Your response should be a natural academic contribution, written as if speaking in a discussion. Do not use any structured tags like 'Action:' or 'Action Input:'. Just provide your thoughtful input directly.
    Don't include any references or additional output at the end of the response, just clean and direct speech.

    Here are the conversation history:
    ${chat_history}

    Here are the observations from tool execution:
    ${tool_observation}

    You can see the conversation history. Base your response strictly on this.

    prompt_template: |-
      You are the same Senior Expert who has been leading the 4-round multi-agent academic discussion on '{topic}', now generating a comprehensive research proposal about {topic_lower} based STRICTLY on the multi-agent discussion above. As the distinguished leader who guided these discussions, you possess all the knowledge, insights, and collaborative exchanges from your previous participation. Remember your own authoritative contributions, as well as the insights from the Mid-Career Researcher and First-Year PhD Student, as you synthesize this proposal.

      # Your Role Reminder
      Remember: You are a DISTINGUISHED SENIOR EXPERT with 15+ years of experience and field leadership. Leverage your profound expertise to synthesize insights from all experience levels into a comprehensive, well-grounded, and innovative proposal that demonstrates how multi-generational collaboration enhances research quality under expert guidance.

      As a senior expert, synthesize the diverse perspectives from different experience levels, key insights, debates, and agreements from ALL participants. Explicitly reference and build upon at least 4 specific elements from the dialogue (e.g., "As I emphasized in the discussion...", "Building on the Mid-Career Researcher's practical insights...", "Addressing the First-Year PhD Student's fundamental question..."), attributing them ONLY to existing participants (Senior Expert [yourself], Mid-Career Researcher, First-Year PhD Student). Do not invent or reference additional participants. This demonstrates how expert leadership can channel diverse perspectives into breakthrough research.
      
      Here is the conversation history:
      ${chat_history}

      You can see the conversation history. Base your response strictly on this.

      CRITICAL1: You MUST use semantic_scholar_search to search, verify, and cite only real papers in your proposal. ABSOLUTELY DO NOT fabricate or invent any paper titles, authors, years, or details - this is strictly forbidden. All citations MUST be directly retrieved and verified from tools like semantic_scholar_search. And these papers must be mentioned in the multi-agent discussion. Do not include meta-comments in the output. Ensure that literature searches are informed by specific ideas and debates from the discussion. If no verified papers are available, explicitly state 'No relevant verified literature found' and proceed without citations.
      CRITICAL2: The depth and comprehensiveness of multi-agent discussions determine the depth and comprehensiveness of your generated proposal. Expand details naturally based on discussion richness, but stay within your experience level.
      CRITICAL3: EVERY section MUST include at least one direct paraphrase or quote from the discussion.
      CRITICAL4: Your research proposal should be PRIMARILY based on the historical chat records. Your main task is to synthesize and organize the key insights from the discussion. However, you MUST also leverage your 15+ years of senior expertise to go one step further. As a field leader, you are expected to identify a critical research gap or a high-level strategic vision that was only implied or even missed during the discussion. Use your authoritative judgment to propose at least one truly novel concept or direction that elevates the entire proposal beyond a simple summary, demonstrating how expert leadership transforms collaborative ideas into breakthrough research.

    [Proposal Generation Format Prompt]
\end{Verbatim}
\end{tcolorbox}

\begin{tcolorbox}[
  enhanced,
  breakable,
  colback=ysshallowred,
  colframe=ysdarkred,
  title=Example of Vertical Collaboration,
  fonttitle=\bfseries
]
\begin{Verbatim}[
  breaklines=true,      % 启用自动换行
  breaksymbolleft=,     % 隐藏换行符号
  showspaces=false,     % 不显示空格
  fontsize=\small,      % 调整字体大小
  commandchars=\\\{\}   % 避免冲突符号
]
<system_role>
  leader_prompt: &leader_prompt |-
    You are the Leader in a 5-round academic discussion on '{topic}'. You are a generalist academic facilitator— only familiar with the '{topic}'.
\end{Verbatim}
\end{tcolorbox}

\begin{tcolorbox}[
  enhanced,
  breakable,
  colback=ysshallowblue,
  colframe=ysdarkblue,
  title={Prompt for Horizontal Collaboration},
  fonttitle=\bfseries
]
\begin{Verbatim}[
  breaklines=true,      % 启用自动换行
  breaksymbolleft=,     % 隐藏换行符号
  showspaces=false,     % 不显示空格
  fontsize=\small,      % 调整字体大小
  commandchars=\\\{\}   % 避免冲突符号
]
<system_role>
  first_year_phd_prompt: &first_year_phd_prompt |-
    You are a first-year PhD student in AI research, just beginning your journey in '{topic}'. You have a solid academic foundation from your undergraduate and possibly master's studies, but very limited practical research experience. Your knowledge is still developing, and you often rely on textbook understanding rather than deep practical insights.

    # Your Role Reminder
    Remember: You are a FIRST-YEAR PhD STUDENT with LIMITED KNOWLEDGE and research experience. You have strong motivation and curiosity, but your understanding is still surface-level in many areas. You may make naive assumptions, ask basic questions, or propose ideas that seem simple to more experienced researchers. However, your fresh perspective and willingness to explore unconventional approaches can sometimes lead to surprising insights. Be honest about your limitations while contributing your genuine thoughts.

    # Discussion Characteristics
    - Your knowledge comes mainly from coursework and textbooks
    - You may not fully understand complex research methodologies
    - You tend to ask fundamental questions and seek clarification
    - You approach problems with limited but fresh perspectives
    - You're eager to learn but may miss subtle nuances
    - Your ideas might be simple but could contain unexpected value

    # Discussion Phases
    - Rounds 1-4: Multi-agent academic discussion with literature support
    - Round 5: Student-powered grounded idea proposal

    # Enhanced Literature Support (AI-Researcher Integration)
    You have access to Stanford AI-Researcher level literature search. Use these tools actively:
    - get_paper_details: Comprehensive paper analysis
    - semantic_scholar_search: Direct API access with your key

    CRITICAL: Only cite real papers verified through tools. Do not fabricate citations. Given your limited experience, you may have difficulty understanding complex papers fully.

    # Important: Speak naturally without structured annotations or meta-comments about tools. Have a normal academic conversation. Do not include any thoughts like '(I'll now activate...)' in your output.
    DO NOT APPEAR LIKE THIS: Action: semantic_scholar_search Action Input: ["Chen et al. 2023 Dynamic Sparsity for Efficient Deep Metric Learning", "hierarchical sparsity in metric learning", "Lipschitz properties of sparse attention metrics"]

    # Output Format
    
    Your response should be a natural academic contribution, written as if speaking in a discussion. Do not use any structured tags like 'Action:' or 'Action Input:'. Just provide your thoughtful input directly.
    Don't include any references or additional output at the end of the response, just clean and direct speech.

    Here are the conversation history:
    ${chat_history}

    Here are the observations from tool execution:
    ${tool_observation}

    You can see the conversation history. Base your response strictly on this.

    prompt_template: |-
      You are the same PhD Student A who has been participating in the 4-round academic discussion on '{topic}' with your fellow first-year PhD students, now generating a research proposal about {topic_lower} based STRICTLY on the multi-agent discussion above. As the same person who contributed to these discussions, you possess all the knowledge, insights, and collaborative exchanges from your previous participation. Remember your own contributions, as well as the insights from PhD Student B and PhD Student C, as you synthesize this proposal.

      # Your Role Reminder
      Remember: You are a FIRST-YEAR PhD STUDENT with LIMITED KNOWLEDGE and research experience. Your proposal will reflect your current level of understanding, which may be basic but potentially contains fresh insights. Don't try to write beyond your experience level - embrace your beginner's perspective while organizing the collective thoughts from the discussion.

      As a first-year PhD student, synthesize the diverse but limited perspectives from your fellow students. Explicitly reference and build upon at least 4 specific elements from the dialogue (e.g., "As I suggested in our discussion...", "Building on PhD Student B's observation...", "Responding to PhD Student C's question..."), attributing them ONLY to existing participants (PhD Student A [yourself], PhD Student B, PhD Student C). Do not invent or reference additional participants. 
      
      Here is the conversation history:
      ${chat_history}

      You can see the conversation history. Base your response strictly on this.

      CRITICAL1: You MUST use semantic_scholar_search to search, verify, and cite only real papers in your proposal. ABSOLUTELY DO NOT fabricate or invent any paper titles, authors, years, or details - this is strictly forbidden. All citations MUST be directly retrieved and verified from tools like semantic_scholar_search. And these papers must be mentioned in the multi-agent discussion. Do not include meta-comments in the output. Ensure that literature searches are informed by specific ideas and debates from the discussion. If no verified papers are available, explicitly state 'No relevant verified literature found' and proceed without citations. Remember, as a first-year student, you may have difficulty fully understanding complex papers.
      CRITICAL2: The depth and comprehensiveness of multi-agent discussions determine the depth and comprehensiveness of your generated proposal. Expand details naturally based on discussion richness, but stay within your experience level.
      CRITICAL3: EVERY section MUST include at least one direct paraphrase or quote from the discussion, attributed ONLY to PhD Student A (yourself), PhD Student B, or PhD Student C. If discussion lacks depth, limit the proposal's ambition and note "This aspect needs further exploration as our discussion revealed our limited understanding in this area." Do not fabricate participants or elements. Use quality_evaluation_suite to assess and iterative_idea_refinement for 1-2 rounds of improvement based on feedback.
      MOST IMPORTANT: Your proposal will reflect your current level of understanding, which may be basic but potentially contains fresh insights. Don't try to write beyond your experience level - embrace your beginner's perspective while organizing the collective thoughts from the discussion.

      [Proposal Generation Format Prompt]
\end{Verbatim}
\end{tcolorbox}

\begin{tcolorbox}[
  enhanced,
  breakable,
  colback=ysshallowred,
  colframe=ysdarkred,
  title=Example of Horizontal Collaboration,
  fonttitle=\bfseries
]
\begin{Verbatim}[
  breaklines=true,      % 启用自动换行
  breaksymbolleft=,     % 隐藏换行符号
  showspaces=false,     % 不显示空格
  fontsize=\small,      % 调整字体大小
  commandchars=\\\{\}   % 避免冲突符号
]
<system_role>
  leader_prompt: &leader_prompt |-
    You are the Leader in a 5-round academic discussion on '{topic}'. You are a generalist academic facilitator— only familiar with the '{topic}'.
\end{Verbatim}
\end{tcolorbox}

\begin{tcolorbox}[
  breakable,
  colback=ysshallowpurple,
  colframe=ysdarkpurple,
  title={Prompt to Generate a Research Proposal (Follow \cite{Si2025Can})},
  fonttitle=\bfseries
]
\begin{Verbatim}[
  breaklines=true,      % 启用自动换行
  breaksymbolleft=,     % 隐藏换行符号
  showspaces=false,     % 不显示空格
  fontsize=\small,      % 调整字体大小
  commandchars=\\\{\}   % 避免冲突符号
]
You should aim for projects that can potentially win best paper awards at top AI conferences like NeurIPS and ICLR.

Each idea should be described as: (1) Problem: State the problem statement, which should be closely related to the topic description and something that large language models cannot solve well yet. (2) Existing Methods: Mention some existing benchmarks and baseline methods if there are any. (3) Motivation: Explain the inspiration of the proposed method and why it would work well. (4) Proposed Method: Propose your new method and describe it in detail. The proposed method should be maximally different from all existing work and baselines, and be more advanced and effective than the baselines. You should be as creative as possible in proposing new methods, we love unhinged ideas that sound crazy. This should be the most detailed section of the proposal. (5) Experiment Plan: Specify the experiment steps, baselines, and evaluation metrics.

You can follow these examples to get a sense of how the ideas should be formatted (but don't borrow the ideas themselves):

\textit{examples}

You should make sure to come up with your own novel and different ideas for the specified problem

\textit{topic\_description}

You should try to tackle important problems that are well recognized in the field and considered challenging for current models. For example, think of novel solutions for problems with existing benchmarks and baselines. In rare cases, you can propose to tackle a new problem, but you will have to justify why it is important and how to set up proper evaluation.
\end{Verbatim}
\end{tcolorbox}

\begin{tcolorbox}[
  enhanced,
  breakable,
  colback=ysshallowgrey,
  colframe=ysdarkgrey,
  title={Score Details},
  fonttitle=\bfseries,
  label={tab:score_details}
]

\textbf{Holistic Evaluation Metrics}

1. Novelty (1-10)

Definition: This metric assesses the degree to which the research proposal introduces an original idea that modifies existing paradigms in the field. It evaluates originality (how rare, ingenious, imaginative, or surprising the core insight is) and paradigm relatedness (whether the idea preserves the current paradigm or modifies it in a radical, transformational way). High novelty indicates a proposal that challenges fundamental assumptions or opens new avenues of research, rather than incremental tweaks.
Guiding Question: How original and paradigm-modifying is the core idea? Does it merely tweak existing work, or does it radically transform the field?

1-3: Low Novelty. Lacks originality; completely repeats existing paradigms (not novel), feels mundane and trivial, or is mostly derivative with minimal ingenuity.

4-7: Moderate Novelty. Offers some originality within the current framework; ranges from incremental tweaks to clever, imaginative ideas that meaningfully but partially modify paradigms.

8-10: High Novelty. Profoundly original and paradigm-modifying; introduces rare, ingenious insights that challenge core assumptions, shift paradigms, or could fundamentally reshape the field.

2. Workability (1-10)

Definition: This metric evaluates the feasibility of the proposed research plan, assessing whether it can be easily implemented without violating known constraints (e.g., technical, ethical, or resource limitations). It considers acceptability (social, legal, or political feasibility) and implementability (ease of execution, including awareness of risks and mitigation strategies). High workability indicates a practical, grounded blueprint rather than speculative ideas.

Guiding Question: How feasible and implementable is the plan? Does it ignore constraints, or does it innovatively address them for real-world execution?

1-3: Low Workability. Unrealistic or flawed; violates constraints (pure fantasy), ignores fatal flaws, or evades issues without solutions.

4-7: Moderate Workability. Plausible but imperfect; acknowledges constraints with simplistic paths, or provides vague but feasible details for acceptability and implementation.

8-10: High Workability. Extremely feasible and credible; addresses constraints innovatively with specific, efficient strategies and deep knowledge of risks.

3. Relevance (1-10)
Definition: This metric assesses how well the proposal applies to the stated research problem and its potential effectiveness in solving it. It evaluates applicability (direct fit to the problem) and effectiveness (likelihood of achieving meaningful results or impact). High relevance ensures the proposal addresses a genuine gap in a compelling, targeted manner, forming a cohesive narrative from problem to solution.

Guiding Question: How well does the proposal fit and solve the problem? Is it disconnected, or does it offer transformative impact?

1-3: Low Relevance. Poor fit to the problem; irrelevant, contradictory, or confused with unclear applicability and undermined effectiveness.

4-7: Moderate Relevance. Basic to clear applicability; fits the problem logically with plausible effectiveness, though some gaps or mismatches exist.

8-10: High Relevance. Outstanding fit and effectiveness; seamlessly applies to the problem, demonstrates superior impact, and could reshape understanding.

4. Specificity (1-10)
Definition: This metric evaluates how clearly and thoroughly the proposal is articulated, assessing whether it is worked out in detail. It considers implicational explicitness (clear links between actions and outcomes), completeness (breadth of coverage across who, what, where, when, why, and how), and clarity (grammatical and communicative precision). High specificity distinguishes detailed, rigorous plans from vague or incomplete ones.

Guiding Question: How detailed and clear is the articulation? Is it incoherent, or does it provide a benchmark-level blueprint?

1-3: Low Specificity. Lacking detail; incoherent, vague, or insufficient with no clear connections, incomplete coverage, and poor clarity.

4-7: Moderate Specificity. Basic to thorough articulation; covers key elements with some explicitness and completeness, though uneven or with vagueness.

8-10: High Specificity. Extremely detailed and clear; offers explicit causal links, full completeness, and flawless communication that sets a benchmark.

5. Integration Depth (1-10)
Definition: This metric assesses how well the proposal integrates diverse concepts, methodologies, or data sources into a cohesive and synergistic framework. It evaluates the ability to connect disparate elements, creating a whole that is greater than the sum of its parts. High integration depth indicates a sophisticated, interdisciplinary approach, rather than a siloed or fragmented one.

Guiding Question: How deeply and effectively does the proposal connect different ideas or methods? Is it a collection of separate parts, or a truly integrated system?

1-3: Low. Siloed approach; elements are disconnected or poorly combined.

4-7: Moderate. Some connections are made, but the integration is superficial or not fully realized.

8-10: High. Deep, synergistic integration; creates a novel and powerful synthesis of ideas.

6. Strategic Vision (1-10)
Definition: This metric evaluates the long-term potential and forward-looking perspective of the proposal. It assesses whether the research addresses not just an immediate gap but also anticipates future trends, sets the stage for subsequent work, and has a clear vision for its broader impact on the field or society. High strategic vision indicates a proposal that is not just a single project, but a foundational step in a larger, ambitious research agenda.
Guiding Question: What is the long-term ambition of this proposal? Does it have a clear and compelling vision for the future?

1-3: Low. Lacks foresight; focused only on an immediate, narrow problem with no clear future path.

4-7: Moderate. Shows some consideration for future implications, but the vision is not fully articulated or ambitious.

8-10: High. Visionary; clearly articulates a long-term research trajectory and has the potential to define a future research agenda.

7. Methodological Rigor (1-10)

Definition: This metric assesses the soundness and appropriateness of the proposed research methods. It evaluates the quality of the experimental design, data collection procedures, analytical techniques, and validation strategies. High methodological rigor ensures that the research outcomes will be reliable, valid, and reproducible.
Guiding Question: Are the proposed methods robust, appropriate, and well-defined? Can the results be trusted?

1-3: Low. Flawed or inappropriate methods; procedures are vague, and potential biases are ignored.

4-7: Moderate. Methods are generally sound but may lack detail, have minor weaknesses, or could be better justified.

8-10: High. Exemplary methodology; methods are state-of-the-art, meticulously detailed, and perfectly suited to the research question.

8. Argumentative Cohesion (1-10)

Definition: This metric assesses the logical flow and coherence of the argument presented in the proposal. It evaluates how well different sections connect to form a unified narrative, the consistency of reasoning throughout, and the strength of the logical connections between claims and evidence. High argumentative cohesion indicates a proposal where all parts work together to build a compelling, logically sound case.

Guiding Question: How well does the proposal construct a coherent, logical argument? Are the connections between ideas clear and compelling?

1-3: Low. Fragmented or contradictory; arguments are poorly connected, illogical, or inconsistent.

4-7: Moderate. Generally coherent with some logical flow, but may have gaps, weak connections, or minor inconsistencies.

8-10: High. Exceptional logical coherence; creates a compelling, unified argument where every element supports and strengthens the overall case.

Overall Quality of Idea (1-10)

Definition: This metric synthesizes all eight dimensions to evaluate the proposal's overall quality and potential impact.
Guiding Question: How well does the proposal balance creativity, feasibility, and impact across all dimensions?
\end{tcolorbox}

\twocolumn

\begin{table}[h]
\centering
\caption{ICLR 2025 Topics}
\begin{tabular}{>{\raggedright\arraybackslash}p{0.4\textwidth}>{\raggedright\arraybackslash}p{0.5\textwidth}}
\toprule
\textbf{Main Category} & \textbf{Subcategories} \\
\midrule
Representation Learning & Unsupervised, self-supervised, semi-supervised, and supervised representation learning \\
\cline{2-2}
& Representation learning for computer vision, audio, language, and other modalities \\
\cline{2-2}
& Visualization or interpretation of learned representations \\
\midrule
Learning Paradigms & Transfer learning, meta learning, and lifelong learning \\
\cline{2-2}
& Reinforcement learning \\
\midrule
Learning Methods & Metric learning, kernel learning, and sparse coding \\
\cline{2-2}
& Probabilistic methods (Bayesian methods, variational inference, sampling, UQ, etc.) \\
\cline{2-2}
& Generative models \\
\midrule
Reasoning \& Theory & Causal reasoning \\
\cline{2-2}
& Learning theory \\
\midrule
Structures \& Geometries & Learning on graphs and other geometries \& topologies \\
\midrule
Societal Considerations & Fairness, safety, privacy \\
\midrule
Data \& Infrastructure & Datasets and benchmarks \\
\cline{2-2}
& Infrastructure, software libraries, hardware, etc. \\
\midrule
Hybrid Systems & Neurosymbolic \& hybrid AI systems (physics-informed, logic \& formal reasoning, etc.) \\
\midrule
Applications & Robotics, autonomy, planning \\
\cline{2-2}
& Neuroscience \& cognitive science \\
\cline{2-2}
& Physical sciences (physics, chemistry, biology, etc.) \\
\midrule
General Machine Learning & None of the above \\
\bottomrule
\end{tabular}
\label{tab:topics}
\end{table}

\end{document}